\documentclass[manuscript]{aastex6}   		
\usepackage[left=1.0in,bottom=0in,right=0.5in,top=1.5in]{geometry} 
\usepackage{graphicx}
\usepackage{subfloat}
\usepackage{amssymb}
\usepackage{amsmath}
\usepackage{hyperref}
\usepackage{epsfig, natbib}
\usepackage{color,soul}

\newcommand{\RNum}[1]{\uppercase\expandafter{\romannumeral #1\relax}}
\usepackage{afterpage}
\shorttitle{The RAMPS Pilot Survey}
\shortauthors{Hogge et al.}

\begin{document}

\title{The Radio Ammonia Mid-plane Survey (RAMPS) Pilot Survey}
\author{Taylor Hogge\altaffilmark{1}, James Jackson\altaffilmark{1,2}, Ian Stephens\altaffilmark{1,3}, Scott Whitaker\altaffilmark{4}, Jonathan Foster\altaffilmark{5}, Matthew Camarata\altaffilmark{1}, D. Anish Roshi\altaffilmark{6},  James Di Francesco\altaffilmark{7}, Steven Longmore\altaffilmark{8,9}, Robert Loughnane\altaffilmark{10}, Toby Moore\altaffilmark{9}, Jill Rathborne\altaffilmark{11}, Patricio Sanhueza\altaffilmark{12}, and Andrew Walsh\altaffilmark{13}}
\altaffiltext{1}{\itshape Institute for Astrophysical Research, Boston University, Boston, MA 02215, USA; thogge@bu.edu}
\altaffiltext{2}{\itshape School of Mathematical and Physical Sciences, University of Newcastle, University Drive, Callaghan NSW 2308, Australia}
\altaffiltext{3}{\itshape Harvard-Smithsonian Center for Astrophysics, 60 Garden Street, Cambridge, MA 02138, USA}
\altaffiltext{4}{\itshape Department of Physics, Boston University, Boston, MA 02215, USA}
\altaffiltext{5}{\itshape Yale Center for Astronomy and Astrophysics, New Haven, CT 06520, USA}
\altaffiltext{6}{\itshape National Radio Astronomy Observatory, Charlottesville, VA 22903, USA}
\altaffiltext{7}{\itshape National Research Council of Canada, Herzberg Institute of Astrophysics, Victoria, BC, V9E 2E7, Canada }
\altaffiltext{8}{\itshape School of Physics and Astronomy, University of Leeds, Leeds, UK}
\altaffiltext{9}{\itshape Astrophysics Research Institute, Liverpool John Moores University, Liverpool, UK}
\altaffiltext{10}{\itshape Instituto de Radioastronom\'{a} y Astrof\'{i}sica, Universidad Nacional Aut\'{o}noma de M\'{e}xico, Apdo. Postal 3-72, Morelia, Michoac\'{a}n, 58089, M\'{e}xico}
\altaffiltext{11}{\itshape CSIRO Astronomy and Space Science, Epping, Sydney, Australia}
\altaffiltext{12}{\itshape National Astronomical Observatory of Japan, National Institutes of Natural Sciences, 2-21-1 Osawa, Mitaka, Tokyo 181-8588, Japan}
\altaffiltext{13}{\itshape International Centre for Radio Astronomy Research, Curtin University, Perth, WA 6845, Australia}

\begin{abstract}
The Radio Ammonia Mid-Plane Survey (RAMPS) is a molecular line survey that aims to map a portion of the Galactic midplane in the first quadrant of the Galaxy ($l= 10^{\circ} - 40^{\circ}$, $\lvert b\rvert \leq 0.4^{\circ}$) using the Green Bank Telescope. We present results from the pilot survey, which has mapped approximately 6.5 square degrees in fields centered at $l=10^{\circ}, \ 23^{\circ},\  24^{\circ}, \ 28^{\circ}, \ 29^{\circ}, \ 30^{\circ}, \ 31^{\circ}, \ 38^{\circ}, \ 45^{\circ}, \ \mathrm{and} \ 47^{\circ}$. RAMPS observes the $\mathrm{NH_{3}}$ inversion transitions $\mathrm{NH_{3}(1,1) - (5,5)}$, the $\mathrm{H_{2}O} \ 6_{1,6}-5_{2,3}$ maser line at 22.235 GHz, and several other molecular lines. We present a representative portion of the data from the pilot survey, including $\mathrm{NH_{3}(1,1)}$ and $\mathrm{NH_{3}(2,2)}$ integrated intensity maps, $\mathrm{H_2O}$ maser positions, maps of $\mathrm{NH_{3}}$ velocity, $\mathrm{NH_{3}}$ line width, total $\mathrm{NH_{3}}$ column density, and $\mathrm{NH_{3}}$ rotational temperature. These data and the data cubes from which they were produced are publicly available on the RAMPS website\footnote{\url{http://sites.bu.edu/ramps/} \label{foot:ramps}}. 

\end{abstract}

\section{Introduction}
\label{sec:intro}

Although high-mass stars ($\mathrm{M > 8 \ M_{\odot}}$) are rare, they dominate the chemical and energetic input into the interstellar medium (ISM). Gaining a detailed understanding of the formation of high-mass stars is thus important for theories of stellar cluster formation and galactic evolution. The current theoretical picture of high-mass star formation is that high-mass molecular clumps ($\mathrm{M > 200 \ M_{\odot}, \ R\sim1 \ pc}$) are the nurseries of high-mass stars and star clusters. Density enhancements within clumps (here we define a $\lq \lq$clump" as a molecular clump), called cores \citep[$\mathrm{M \sim 1 - 10 \ M_{\odot}, \ R\sim0.05 \ pc}$; ][]{S17}, are initially devoid of stars, and are thus referred to as $\lq \lq$prestellar" cores. Their ensuing collapse forms deeply embedded, accreting $\lq \lq$protostellar" cores, where a high-mass star or multiple stellar system may form. High-mass protostars quickly enter the main sequence and ionize their surrounding material to form an H II region. Despite this broad theoretical understanding, the details of high-mass star formation are not well understood compared to the formation of low-mass stars, especially with regard to the early fragmentation history, turbulent support of cores, and to the physical and dynamical evolution of protostars, as well as their physical and dynamical evolution. This difference is in part due to the difficulty of observing high-mass star-forming regions (SFRs), especially at early evolutionary stages. In contrast to low-mass stars, high-mass stars are rarer, form more quickly, and form in regions that are more deeply embedded in gas and dust.

To make progress in the face of the observational challenges, large surveys are necessary to observe a statistically significant sample of high-mass SFRs. As high-mass stars form predominantly in the Galactic plane, surveys of high-mass SFRs typically focus their observations in the plane. Recently, continuum surveys of the Galactic plane, such as the 1.1 mm Bolocam Galactic Plane Survey \citep[BGPS;][]{2011ApJS..192....4A}, the 870 $\mathrm{\mu m}$ APEX Telescope Large Area Survey of the Galaxy \citep[ATLASGAL;][]{2009A&A...504..415S}, the 70 $-$ 500 $\mathrm{\mu m}$ $Herschel$ Infrared Galactic Plane Survey \citep[HiGAL;][]{2010A&A...518L.100M}, the Red MSX Source \citep[RMS; ][]{2009A&A...501..539U}, and the Coordinated Radio and Infrared Survey for High-Mass Star Formation \citep[CORNISH; ][]{2012PASP..124..939H}, have identified thousands of dense, high-mass, star-forming clumps from their dust emission. In addition to the position and structure of star-forming clumps, continuum surveys have contributed important information that helps characterize these clumps. In particular, modeling the dust continuum spectral energy distribution (SED) of a clump allows one to derive its dust temperature and column density. From the column density, one can estimate the dust mass of a clump at a known distance. With the dust-to-gas mass ratio, one can then determine the total mass of the clump. This information is crucial for determining whether a clump or core will go on to form high-mass stars and exactly how the clumps evolve.

Although continuum surveys are essential, they do have significant limitations. Continuum emission may be blended owing to multiple clumps or unrelated diffuse dust along the line of sight, both of which will hinder the estimation of parameters from the dust SED. In addition, assumptions about the dust-to-gas mass ratio, the dust emissivity coefficient $\kappa$, and the dust emissivity index $\beta$ are uncertain, with the combination of such uncertainties affecting the accuracy of derived column densities and temperatures. Furthermore, the derivation of temperatures from graybody dust SEDs usually assumes optically thin emission at all far-IR to millimeter wavelengths. While this assumption is reliable for the majority of high-mass SFRs, it may not be true for the densest, coldest clumps. Many of the limitations of dust continuum surveys can be overcome by a focused molecular line survey.

The main advantage of molecular line data is their ability to provide kinematic information, such as the velocity dispersion $\sigma$, a crucial parameter in all theories of high-mass star formation. The velocity dispersion, measured from the turbulent line width, sets the turbulent pressure $\mathrm{(\propto \rho \sigma^2)}$, the mass accretion rate (isothermal sphere: $\dot{M} \propto \sigma^3$ \citep{1980ApJ...241..637S}; Bondi-Hoyle: $\dot{M} \propto \sigma^{-3}$ \citep{1952MNRAS.112..195B}), the dynamical timescale $(\propto R / \sigma)$, and the virial parameter $(\alpha = M_{vir} / M \propto \sigma^2 R / G M)$. Using the kinematic distance method \citep{O58}, the velocity of a line can provide an estimate of distance, which is necessary to calculate the size, mass, luminosity, and Galactic location of a clump. Additionally, velocity fields can be used to separate multiple clumps along the line of sight and reveal bulk flows and rotation. Molecular line surveys that target transitions with large Einstein A-coefficients have an additional important advantage over continuum surveys. Such transitions have large critical densities, and thus they primarily trace regions with dense ($n > 10^3 \ \mathrm{cm^{-3}}$), star-forming gas, rather than unrelated diffuse gas along the line of sight.

Spectral lines can also provide a robust estimate of the gas temperature. In local thermodynamic equilibrium (LTE), the gas temperature of an emitting medium may be determined by observing spectral lines of the same species that are well separated in excitation energy. The excitation temperature sets the level populations, and the excitation temperature is equal to the gas temperature when the gas is sufficiently dense. In LTE, measuring the relative intensity of the lines thus provides the temperature of dense gas. 
In addition, spectral lines can help to determine optical depth by comparing two or more spectral lines that have a known intensity ratio. This estimation is often done with a molecule and its isotopic counterpart, since the ratio of their optical depths is equal to their relative abundance. A similar method is available for spectral transitions that exhibit hyperfine splitting. In LTE, the ratio of the optical depths in various hyperfine lines is proportional to the ratio of their quantum statistical weights, which are constant, unlike relative abundance. This feature allows for a more reliable determination of optical depth and can be accomplished by observing a single set of hyperfine lines.

The $\mathrm{H_{2}O}$ Southern Galactic Plane Survey \citep[HOPS;][]{2011MNRAS.416.1764W,P12} is a previous molecular line survey of dense gas. HOPS observed 100 deg$^{2}$ of the Galactic plane and primarily targeted several $\mathrm{NH_3}$ inversion lines and the 22.235 GHz $\mathrm{H_{2}O} \ 6_{1,6}-5_{2,3}$ maser line using the 22 m Mopra telescope. HOPS and similar surveys have provided a wealth of data for the high-mass star formation community. These data have helped advance our understanding of the complex kinematics, chemistry, and evolution of high-mass clumps \citep{2017MNRAS.470.1462L}. To further probe these SFRs, we must exploit new advancements in instrumentation. To this end, we are undertaking the Radio Ammonia Mid-Plane Survey (RAMPS). RAMPS is a new Galactic midplane molecular line survey, which employs the $K$-band Focal Plane Array on the Green Bank Telescope (GBT) to image several $\mathrm{NH_3}$ inversion lines and the 22.235 GHz $\mathrm{H_{2}O}$ line. In this paper, we describe the survey and highlight its first results. 

{We begin by discussing the survey and its observations (Section~\ref{sec:surv}). Subsequently, we present the results of the RAMPS pilot survey (Section~\ref{sec:results}). We then present a preliminary analysis of the data (Section~\ref{sec:analysis}) and a comparison of the features of the RAMPS survey to those of previous surveys (Section~\ref{sec:comparison}). Finally, we summarize our conclusions (Section~\ref{sec:conclusion}).

\section{The Survey}
\label{sec:surv}

RAMPS is a blind molecular line survey that targets a portion of the Galactic midplane in the first quadrant of the Galaxy. In this section, we describe in detail the survey and the processing of the data. In Section~\ref{subsec:line}, we discuss the observed lines. In Section~\ref{subsec:inst}, we describe the telescope, receiver, and spectrometer. In Section~\ref{subsec:obs}, we introduce our observing strategy. In Section~\ref{subsec:red}, we outline the data reduction pipeline. In Section~\ref{subsec:proc}, we describe the post-reduction processing of the data. Then, in Section~\ref{subsec:rel}, we detail the public release of the data. 

\subsection{Line Selection}
\label{subsec:line}

RAMPS observes 13 molecular transitions, which we present in Table~\ref{tab:lines}. The most frequently detected lines, and the lines we limit our focus to in the current paper, are $\mathrm{NH_3}$ (1,1), $\mathrm{NH_3}$ (2,2), and the $\mathrm{H_{2}O} \ 6_{1,6}-5_{2,3}$ maser line.

The $\mathrm{NH_3}$ inversion transitions near 23 GHz are particularly well suited to the study of high-mass stars. In addition to having a large critical density ($n_{\rm{crit}} \sim 3\times10^3 \ \mathrm{cm^{-3}}$) and revealing kinematic information, the $\mathrm{NH_3}$ inversion transitions provide a robust estimate of the gas temperature and the column density. The excitation temperature (also called the rotational temperature) representing a series of $\mathrm{NH_3}$ rotational transitions for an observed source of emission is set by the $\mathrm{NH_3}$ level populations. For gas with a density well above the critical density, the rotational temperature is equal to the gas temperature. Thus, in LTE one can determine the gas temperature from the brightness ratios of the inversion lines. We can measure column density from the relative intensities of the nuclear quadrupole hyperfine lines since the intensity ratios of the satellite hyperfine lines to the main hyperfine line are set by the optical depth. 

The collisionally pumped $\mathrm{H_{2}O}$ maser line at 22.235 GHz \citep{1989ApJ...346..983E} is useful because it is known to trace active star formation. Although the exact evolutionary stage or stages probed by $\mathrm{H_{2}O}$ masers in star-forming clumps remain uncertain \citep{2010MNRAS.405.2471V}, $\mathrm{H_{2}O}$ masers are frequently found in high-mass SFRs. They are, however, also seen toward low-mass SFRs. Given that $\mathrm{H_{2}O}$ can be one of the brightest spectral lines emitting from low-mass SFRs, these masers can help us detect low-mass SFRs at much larger distances than continuum surveys. $\mathrm{H_{2}O}$ masers are also associated with asymptotic giant branch (AGB) stars, which can be observed using VLBI techniques to study the dynamics of their atmospheres and winds \citep{1997PASP..109.1286M,2008PASJ...60.1077S}.
Furthermore, masers are well suited for parallax measurements \citep{2014ApJ...783..130R} since they are extremely luminous compact sources. Consequently, $\mathrm{H_{2}O}$ masers are particularly useful for measuring accurate distances to SFRs throughout the Galaxy. 

The RAMPS spectral setup also includes two shock-excited $\mathrm{CH_{3}OH}$ lines and high-density tracing lines of $\mathrm{HC_{5}N}$, $\mathrm{HC_{7}N}$, and HNCO, as well as CCS, which is found in SFRs that are in an early evolutionary state \citep{1992ApJ...392..551S}. 

\begin{deluxetable}{ccccc}

\tablecolumns{5}
\tablecaption{The 13 molecular lines observed by RAMPS.\label{tab:lines}}

\tablehead{\colhead{Molecule}&
		  \colhead{Transition}&
		  \colhead{Frequency}&
		  \colhead{$\mathrm{E_{upper} / k}$}&
		  \colhead{Number of} \\
		  \colhead{}&
		  \colhead{}&
		  \colhead{(MHz)}&
		  \colhead{(K)}&
		  \colhead{Receivers}}
\startdata
$\mathrm{NH_{3}}$ & ($J,K$) = (1,1) & 23694.47 & 23 & 7 \\
$\mathrm{NH_{3}}$ & ($J,K$) = (2,2) & 23722.60 & 64 & 7 \\
$\mathrm{NH_{3}}$ & ($J,K$) = (3,3) & 23870.08 & 124 & 7 \\
$\mathrm{NH_{3}}$ & ($J,K$) = (4,4) & 24139.35 & 201 & 7 \\
$\mathrm{NH_{3}}$ & ($J,K$) = (5,5) & 24532.92 & 295 & 7 \\
$\mathrm{CH_{3}OH}$ & $\mathrm{J_{K_p}}$ = $10_1$ -- $9_2 \ A^{-}$ & 23444.78 & 143 & 7 \\
$\mathrm{HC_{5}N}$ & $J$ = 9 -- 8 & 23963.90 & 6 & 7 \\
$\mathrm{HC_{5}N}$ & $J$ = 8 -- 7 & 21301.26 & 5 & 1 \\
$\mathrm{HC_{7}N}$ & $J$ = 19 -- 18 & 21431.93 & 10 &1 \\
$\mathrm{CH_{3}OH}$ & $\mathrm{J_{K_p}}$ = $12_2$ -- $11_1 \ A^{-}$ & 21550.34 & 479  & 1 \\
HNCO & $\mathrm{J_{K_p,K_o}}$ = $1_{0,1} - 0_{0,0}$ & 21981.57 & 1 & 1 \\
$\mathrm{H_{2}O}$ & $\mathrm{J_{K_p,K_o}}$ = $6_{1,6} - 5_{2,3}$ & 22235.08 & 644 & 1 \\
CCS  & $J$ = 2 -- 1 & 22344.03 & 2 & 1 \\
\enddata
\tablecomments{The quantum numbers given in the $\lq \lq$Transition" column are $J$, the rotational quantum number, $K$, the projection of $J$ along the molecular axis of symmetry, $K_{\rm{p}}$, the value of $K$ in the limiting case of a prolate spheroid molecule, and $K_{\rm{o}}$, the value of $K$ in the limiting case of an oblate spheroid molecule. $\mathrm{CH_{3}OH}$ $12_2-11_1 \ A^{-}$ is a rotational transition within the first vibrationally excited state, i.e. $v=1$.}
\end{deluxetable}

\subsection{Instrumentation}
\label{subsec:inst}

We performed observations for RAMPS using the 100 m diameter Robert C. Byrd GBT \citep{P09} at the NRAO,\footnote{The National Radio Astronomy Observatory (NRAO) is a facility of the National Science Foundation operated under cooperative agreement by Associated Universities, Inc.}, which operates in a nearly continuous frequency range of 0.29 $-$ 115 GHz. The GBT is the most sensitive fully steerable single-dish telescope in the world, which allows us to observe a large area with high spatial resolution. RAMPS uses the $K$-band Focal Plane Array \citep[KFPA; ][]{2008ursi.confE...2M}, which is a seven-element receiver array that operates in a frequency range of 18-27.5 GHz. Each receiver has a beam pattern that is well represented by a Gaussian with a $32\arcsec$ FWHM at the rest frequency of $\mathrm{NH_3}$(1,1) and a beam-to-beam distance of approximately $95 \arcsec$ (Figure~\ref{fig:beam}). The receivers feed into the VErsatile GBT Astronomical Spectrometer \citep[VEGAS; ][]{2012arXiv1202.0938A}, a spectrometer equipped for use with focal plane arrays. VEGAS is capable of processing up to 1.25 GHz bandwidth from eight spectrometer banks, each with eight dual polarized sub-bands. 

\begin{figure}[!hbtp]
\centering
\includegraphics[height=\linewidth, angle=0, scale=0.7]{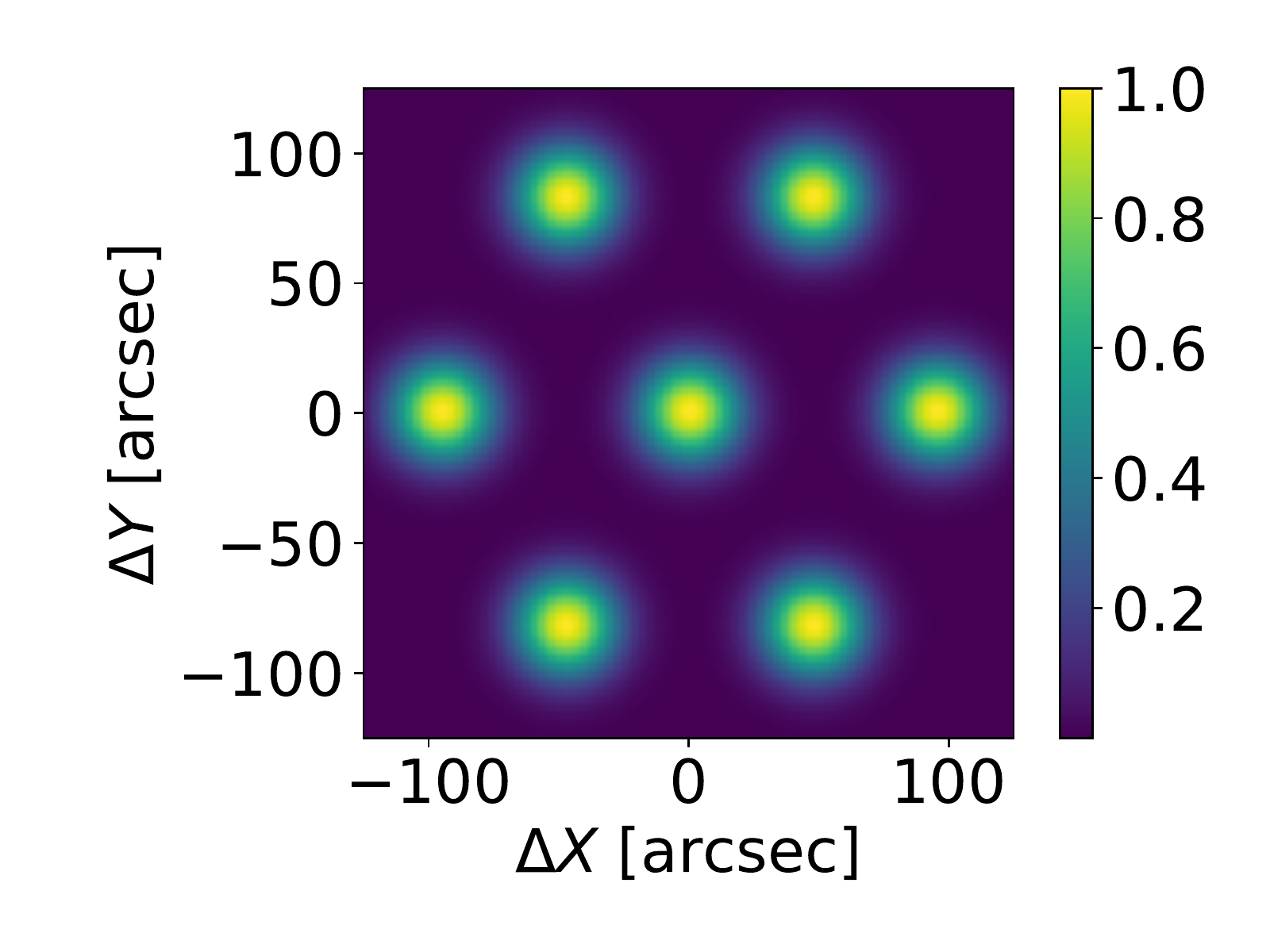}
\caption{Beam pattern of the KFPA. The color represents the sensitivity as a function of angle relative to the sensitivity at beam center. Each receiver has a Gaussian beam shape with an FWHM of $32\arcsec$ at the rest frequency of $\mathrm{NH_3}$(1,1), and the beam-to-beam distance is $\sim 95 \arcsec$.}
\label{fig:beam}
\end{figure}

\subsection{Observations}
\label{subsec:obs}

In 2014, RAMPS was awarded 210 hr of observing time on the GBT for a pilot survey. The purpose of the pilot survey was to test the feasibility of the RAMPS project and to help commission VEGAS. We performed observations for the RAMPS pilot study between 2014 March 16 and 2015 January 22. We used all seven of the KFPA's receivers, with 13 dual polarized sub-bands and 23 MHz bandwidth per sub-band. We observed with the $\lq \lq$medium" spectral resolution, providing a channel width of 1.4 kHz ($\sim0.018 \ \mathrm{km \ s^{-1}}$). We performed Doppler tracking using the $\mathrm{NH_{3}}$(1,1) rest frequency. 

While the KFPA has seven available receivers, the VEGAS back end supports eight spectrometer banks. Hence, six of the seven KFPA receivers each feed into an individual spectrometer bank, while the central receiver feeds into two spectrometer banks. We observed the $\mathrm{NH_{3}}$ inversion transitions, $\mathrm{NH_{3}}$(1,1)$-$(5,5), with all seven receivers to achieve better sensitivity for the $\mathrm{NH_{3}}$ data. 
We observed the 22.235 GHz $\mathrm{H_2O}$ maser line with only the central receiver. Although this significantly reduced the sensitivity of our $\mathrm{H_2O}$ observations, $\mathrm{H_2O}$ masers are typically bright, and thus the GBT frequently detected this line. As discussed in Section~\ref{subsec:line}, RAMPS also observed several other lines; the numbers of receivers used to observe each of these spectral lines are indicated in Table~\ref{tab:lines}.

The proposed RAMPS region extends from Galactic longitude $10^{\circ}$ to $40^{\circ}$ and from Galactic latitude $-0^{\circ}.4$ to $+0^{\circ}.4$.  The survey region is broken up into $1^{\circ} \times 0^{\circ}.8$ $\lq \lq$fields" centered on integer-valued Galactic latitudes and $0^{\circ}$ Galactic longitude. We also observed a portion of two additional fields centered on Galactic longitudes $45^{\circ}$ and $47^{\circ}$, due to the presence of several infrared dark clouds of interest. For the first two fields observed in the pilot survey, centered at $l = \mathrm{10^{\circ} \ and \ 30^{\circ}}$, we tested two different mapping schemes. The first of these divides the mapped field into rectangular $\lq \lq$tiles" of size $0^{\circ}.25 \times 0^{\circ}.20$, and the second divides the field into $\lq \lq$strips" of size $1^{\circ} \times 0^{\circ}.058$. The two schemes differ considerably in the quality of the resulting maps, mainly due to gain variations caused by differing elevations and weather conditions. Due to the long, thin shape of the strips, clumps are often too large to fit completely within a single strip. A clump that was observed in two separate strips was thus observed in different weather conditions and at different elevations. Once the separate observations were combined to create a larger map, this resulted in $\lq \lq$striping" artifacts in the mapping direction. Given that clumps usually fit completely within tiles, gain variations were less problematic for the tile division scheme. Consequently, we chose to map the rest of the survey region with tiles. After the initial tests of the tiling scheme, we adjusted the parameters for the size and position of the tiles to optimize the sensitivity in the overlap regions between adjacent tiles and fields. Specifically, we increased the tile size to $0^{\circ}.26 \times 0^{\circ}.208$ and performed additional observations at the overlap regions between the fields already observed.

We observed in on-the-fly mapping mode, scanning in Galactic longitude, with 4 integrations beam$^{-1}$, 1 s integrations, and $0^{\circ}.008$ between rows. Due to these mapping parameters, the sampling of a tile is uneven. In addition, the sampling pattern is dependent on the angle of the KFPA with respect to the Galactic plane. The uneven sampling pattern and its dependence on the array angle are displayed in Figures~\ref{fig:int30} and \ref{fig:int0}, which show the expected integration time for each spectrum in a data cube assuming the KFPA configuration displayed in the lower left corner of each map. The angle of the array depends on the target position; thus, different tiles may be mapped with the array at a different angle. Observing an individual tile takes approximately 1 hr. Before mapping a tile, we adjust the pointing and focus of the telescope by observing a known calibrator with flux greater than 3 Jy in the $K$ band. This meets the suggested pointing calibration frequency of once per hour and provides a typical pointing error of $\sim 5 \arcsec$. Before observing a new field, we also perform a single pointed observation ($\lq \lq$on/off") toward one of the brightest BGPS 1.1 mm sources in the field. This observation serves as a test to ensure that the receiver and back end are configured correctly, as well as a way to evaluate system performance and repeatability over the observing season. A reference $\lq \lq$off" observation is taken at an offset of $+1^{\circ}$ in Galactic latitude from the tile center immediately before and after mapping in order to subtract atmospheric emission. Although we did not check the ``off'' positions for emission, we found no evidence of a persistent negative amplitude signal in any of the data.

\begin{figure}[!hbtp]
\centering
\includegraphics[height=\linewidth, angle=0, scale=0.7]{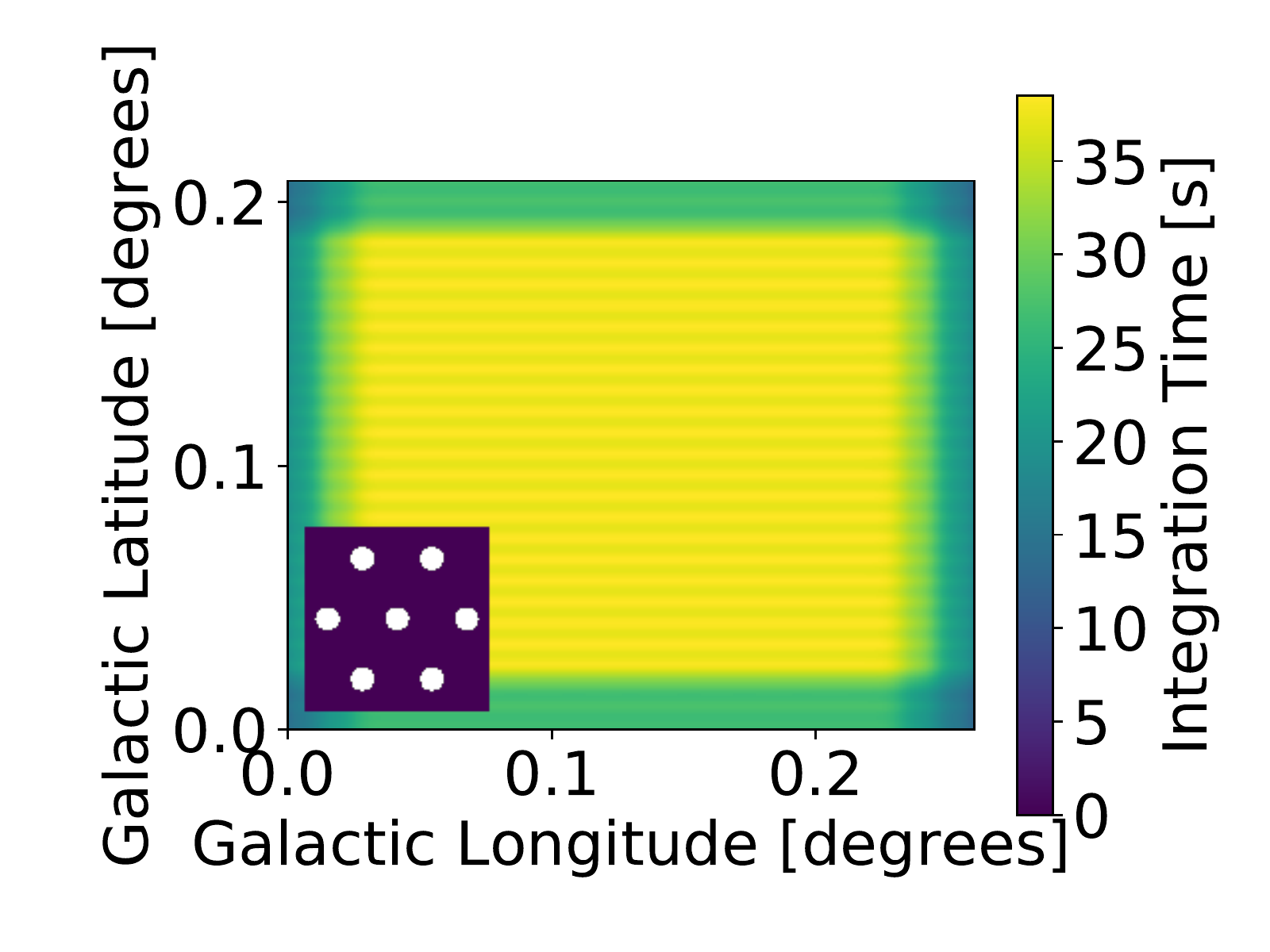}
\caption{The color shows the expected integration time for each spectrum in a data cube assuming the use all seven beams and the KFPA configuration given in the box to the lower left. The angle of the KFPA here provides the least uniform sampling.}
\label{fig:int30}
\end{figure}

\begin{figure}[!hbtp]
\centering
\includegraphics[height=\linewidth, angle=0, scale=0.7]{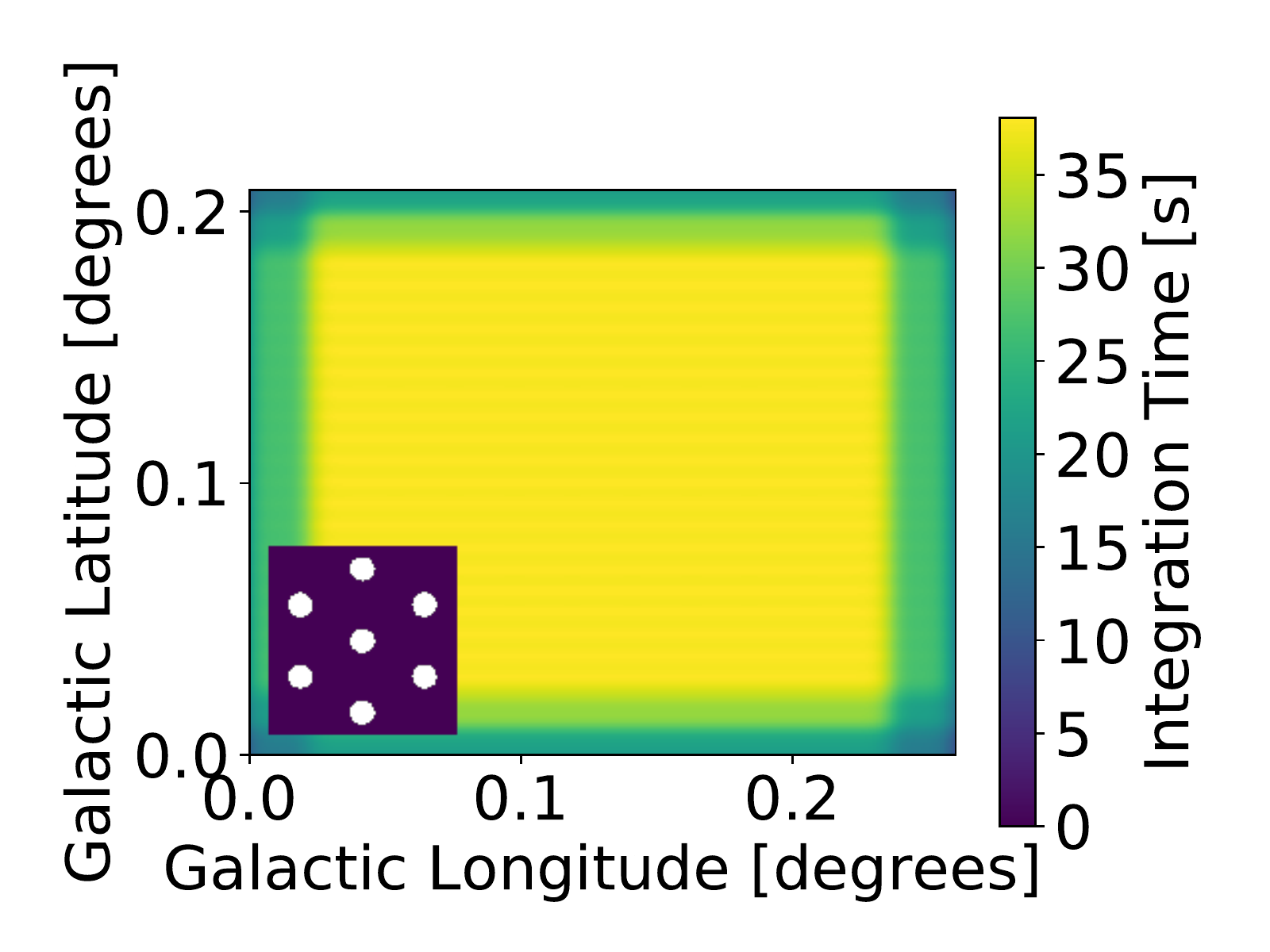}
\caption{The color shows the expected integration time for each spectrum in a data cube assuming the use all seven beams and the KFPA configuration given in the box to the lower left. The angle of the KFPA here provides more uniform sampling than in Figure~\ref{fig:int30}, but the sampling is still coarser in the Galactic latitude direction.}
\label{fig:int0}
\end{figure}

During the first 210 hr of GBT observing, RAMPS mapped approximately 6.5 square degrees in total for fields centered at $l=10^{\circ}, 23^{\circ}, 24^{\circ}, 28^{\circ}, 29^{\circ}, 30^{\circ}, 31^{\circ}, 38^{\circ}, 45^{\circ}, \mathrm{and} \ 47^{\circ}$. Due to the success of the pilot survey and the legacy nature of the RAMPS dataset, RAMPS has been awarded additional observing time to extend the survey. Our goal is to map completely the 24 square degree survey region.

\subsection{Data Reduction}
\label{subsec:red}

We have reduced RAMPS data cubes in a standard manner 
using the GBT Mapping Pipeline \citep{2011ASPC..442..127M} and the \texttt{gbtgridder}\footnote{https://github.com/nrao/gbtgridder}. The reduction process calibrates and grids the KFPA data to produce $l$,$b$,$v$ data cubes (i.e. an array of data with two spatial axes in Galactic coordinates, $l$ and $b$, and one velocity axis, $v$). The mapping pipeline calibrates and processes the raw data into FITS files for each array receiver, sub-band, and polarization, and the \texttt{gbtgridder} grids the spectra using a Gaussian kernel. 
We grid the data cubes with a pixel size of $6\arcsec$ and a channel width of $1.43 \ \mathrm{kHz} \ (\sim0.018 \ \mathrm{km \ s^{-1}})$, where the central channel is at $V = 0$ in the local standard of rest (LSR) frame.
For each spectrum, the \texttt{gbtgridder} determines a zeroth-order baseline from the average of a group of channels near the edges of the band. It then generates a baseline-subtracted data cube that we use for further analysis.

\subsection{Data Processing}
\label{subsec:proc}

We cropped the data cubes along both spatial and spectral axes. We performed the spatial cropping to remove pixels with no spectral data. We did this using PySpecKit \citep{2011ascl.soft09001G}, a Python spectral analysis and reduction toolkit. Specifically, we used the \texttt{subcube} function of the \texttt{Cube} class. We also cropped the data cubes on their spectral axis, and we did so for two reasons: to remove artifacts due to low gain at the edges of the passband, and to remove a portion of the $\mathrm{NH_{3}}$ spectra at large negative velocities. The edge of a spectrometer sub-band is less sensitive than at its center and can also exhibit a steep cusp if the baselines are not steady. We cropped all spectra by $\sim3\%$ at each edge to remove this feature. After baseline fitting, we performed additional cropping on the $\mathrm{NH_{3}}$ spectra to remove unnecessary channels at large negative velocities. At the Galactic longitudes that RAMPS observes ($l= 10^{\circ} - 40^{\circ}$), CO source velocities range from $-60 \ \mathrm{km \ s^{-1}}$ to  $160 \ \mathrm{km \ s^{-1}}$ \citep{2001ApJ...547..792D}. For the $\mathrm{NH_{3}}$ spectra, cropping the channels at velocities less than $-60 \ \mathrm{km \ s^{-1}}$ should not remove real signal. 

After cropping the edge channels, we regridded and combined adjacent cubes using the MIRIAD \citep{STW95} tasks \texttt{REGRID} (version 1.17) and \texttt{IMCOMB} (version 1.11), respectively. This process resulted in data cubes of the L10, L23, 24, L28, L29, L30, and 31 fields, as well as portions of the L38, L45, and L47 fields. We also combined adjacent data cubes to create multifield maps of the L23-24 and L28-31 fields. Next, we applied a median filter to the spectra to increase signal-to-noise ratio (S/N), as well as to remove any anomalously large channel-to-channel variations. The original channel width of the RAMPS data cubes is $0.018 \ \mathrm{km \ s^{-1}}$. We smoothed the $\mathrm{NH_{3}}$ data cubes along their spectral axis using a median filter with a width of 11 channels, which resulted in a new channel width of $0.2 \ \mathrm{km \ s^{-1}}$. We chose this channel width to resolve in at least five spectral channels the typical line width found in high-mass SFRs \citep{2016PASA...33...30R} and infrared dark clouds \citep{2012ApJ...756...60S}. We smoothed the $\mathrm{H_{2}O}$ data cubes using a median filter with a width of seven channels, which resulted in a new channel width of $0.12 \ \mathrm{km \ s^{-1}}$. We smoothed the $\mathrm{H_{2}O}$ data with a smaller filter, in part because $\mathrm{H_{2}O}$ maser lines are generally bright and have larger S/Ns than the typical $\mathrm{NH_{3}}$ lines, as well as the need for higher spectral resolution to avoid blending multiple velocity components. 

Next, we subtracted a polynomial baseline to remove any remaining passband shape. Before fitting for a baseline, we attempted to mask any spectral lines present in the spectra, since these would influence the baseline fit if left unmasked. To perform this masking in an automated manner, we masked groups of spectral channels that had a larger-than-average standard deviation, since these channels likely contained spectral lines. For each channel we calculated the standard deviation of the nearest 40 channels, which we will refer to as a channel's $\lq \lq$local standard deviation." We then masked channels that had a local standard deviation larger than 1.5 times the median of the local standard deviations of all channels in the spectrum. Channels with a large local standard deviation were likely the result of a spectral line, while, on the other hand, a slowly varying baseline shape would result in channels with a smaller local standard deviation. This method reliably masked the majority of lines but was prone to miss broad-line wings. To mitigate this, we also masked channels that were within 10 channels of a masked channel. Next, we fit spectra for a polynomial baseline of up to second order, where the order is chosen such that the fit results in the smallest reduced $\chi^{2}$. We then subtracted the baseline function from the original spectrum and smoothed the baseline-subtracted spectrum as described above.

After subtracting a baseline, we attempted to test the quality of the fit in an automated manner. Our method involved comparing the true noise in a spectrum to the rms in the line-free regions of the spectrum. To estimate the true noise, we calculated the noise using the average channel-to-channel difference. We refer to the channel-to-channel noise as $\sigma_{\rm{diff}}$, where $\sigma_{\rm{diff}} = \sqrt{\frac{1}{2} \langle(T_{i}-T_{i+1})^2\rangle_{i}}$, where $T_i$ is the intensity of the $i^{th}$ channel and $\langle(T_{i}-T_{i+1})^2\rangle_{i}$ is the mean value of the square of the channel-to-channel differences. While the rms is influenced by both the true noise and any baseline present in the spectrum, $\sigma_{\rm{diff}}$ is relatively unaffected by the presence of both a signal and a baseline, as long as they are slowly varying compared to the channel spacing \citep{2016PASA...33...30R}. Thus, if the rms and $\sigma_{\rm{diff}}$ of the line-free portion of a spectrum are very different, there is likely a significant residual baseline present.

To test this, we simulated $10^5$ synthetic spectra with 15,384 channels, the size of unsmoothed RAMPS spectra after cropping. The synthetic spectra consisted of random Gaussian noise with a known standard deviation. We then smoothed the spectra with a median filter to match the real data since the $\mathrm{H_{2}O}$ data were smoothed with a seven-channel filter and the $\mathrm{NH_{3}}$ data were smoothed with an 11-channel filter. Next, we calculated the relative difference ($R$) between the rms and $\sigma_{\rm{diff}}$, given by $R=1-\frac{\sigma_{\rm{diff}}}{rms}$, for each synthetic spectrum. In Figure~\ref{fig:rdiff_nosig} we present two histograms of the distribution of $R$. The left panel shows the distribution of $R$ for the synthetic spectra smoothed with a filter width of seven channels, while right panel shows the distribution of $R$ for the synthetic spectra smoothed with a filter width of 11 channels. The two histograms have a mean of $\mu_R \sim 0$ and standard deviations of $\sigma_R  \sim 0.01$ or $1\%$. Thus, $\sigma_{\rm{diff}}$ is a reliable estimator of the true rms for Gaussian noise. Next, we added a Gaussian signal to each synthetic spectrum to determine how $\sigma_{\rm{diff}}$ responds to the presence of signal. We gave the Gaussian signals uniform random values for both their line widths and S/Ns, where the line widths ranged from 0 to 10 channels and the S/Ns ranged from 0 to 100. For each synthetic spectrum of noise plus signal, we calculated $R$ and binned the values as a function of the amplitude and standard deviation of the synthetic signal, which is shown in Figure~\ref{fig:rdiff_sig}. Thus, $\sigma_{\rm{diff}}$ is also a reliable estimate of the noise when signal is present, except in spectra that contain very strong signals with relatively small line widths. This is not a problem for the $\mathrm{NH_{3}}$ data because the $\mathrm{NH_{3}}$ lines in the RAMPS dataset have S/N $< 100$ and line widths of $\sigma > 1$ channel. On the other hand, the $\mathrm{H_{2}O}$ masers in the RAMPS dataset can have S/N $> 1000$ and line widths of $\sigma \sim 2$ channels, which adds a large source of error to $\sigma_{\rm{diff}}$. Hence, bright, narrow lines must be masked in order for $\sigma_{\rm{diff}}$ to accurately represent the true noise in a spectrum.

\begin{figure}[!hbtp]
\centering
\includegraphics[height=\linewidth, angle=0, scale=0.7]{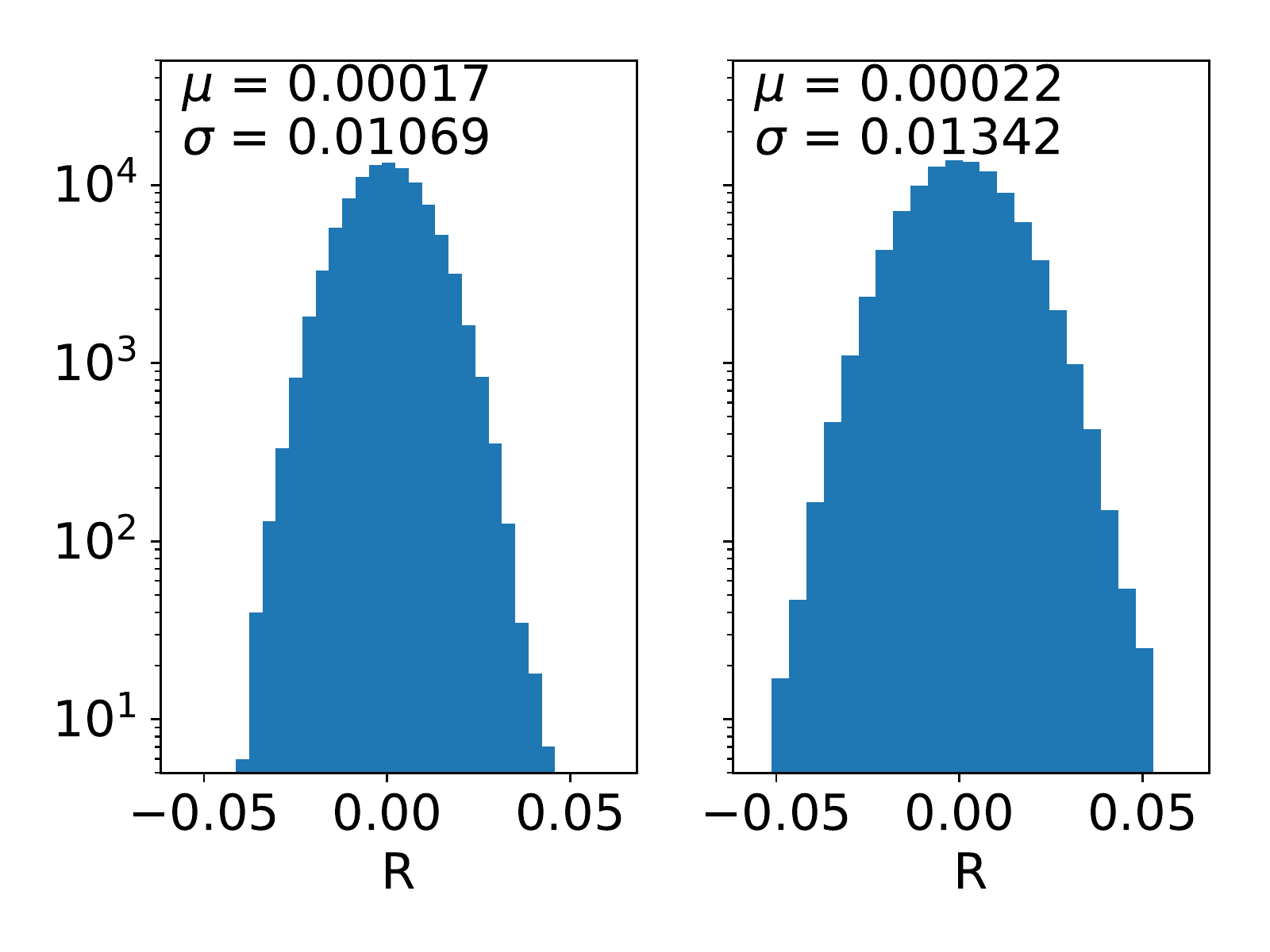}
\caption{Histograms of the relative difference $R$ between the rms and $\sigma_{\rm{diff}}$ for $10^5$ synthetic spectra of Gaussian noise, where the relative difference is given by $R=1-\frac{\sigma_{\rm{diff}}}{rms}$. Our noise estimate is given by $\sigma_{\rm{diff}} = \sqrt{\frac{1}{2} \langle(T_{i}-T_{i+1})^2}\rangle_{i}$, where $T_i$ is the intensity of the $i^{\rm{th}}$ channel and $\langle(T_{i}-T_{i+1})^2\rangle_{i}$ is the mean of the square of all channel-to-channel differences. Listed in each panel are the mean ($\mu$) and standard deviation ($\sigma$) of each distribution. Left: distribution of $R$ for the synthetic spectra smoothed with a median filter width of seven channels.  Right: distribution of $R$ for the synthetic spectra smoothed with a median filter width of 11 channels. Thus, for pure Gaussian noise, our noise estimate is a reliable estimator of the rms, with $\mu \sim 0$ and $\sigma \sim 0.01$ or 1\%.}
\label{fig:rdiff_nosig}
\end{figure}

\begin{figure}[!hbtp]
\centering
\includegraphics[height=\linewidth, angle=0, scale=0.7]{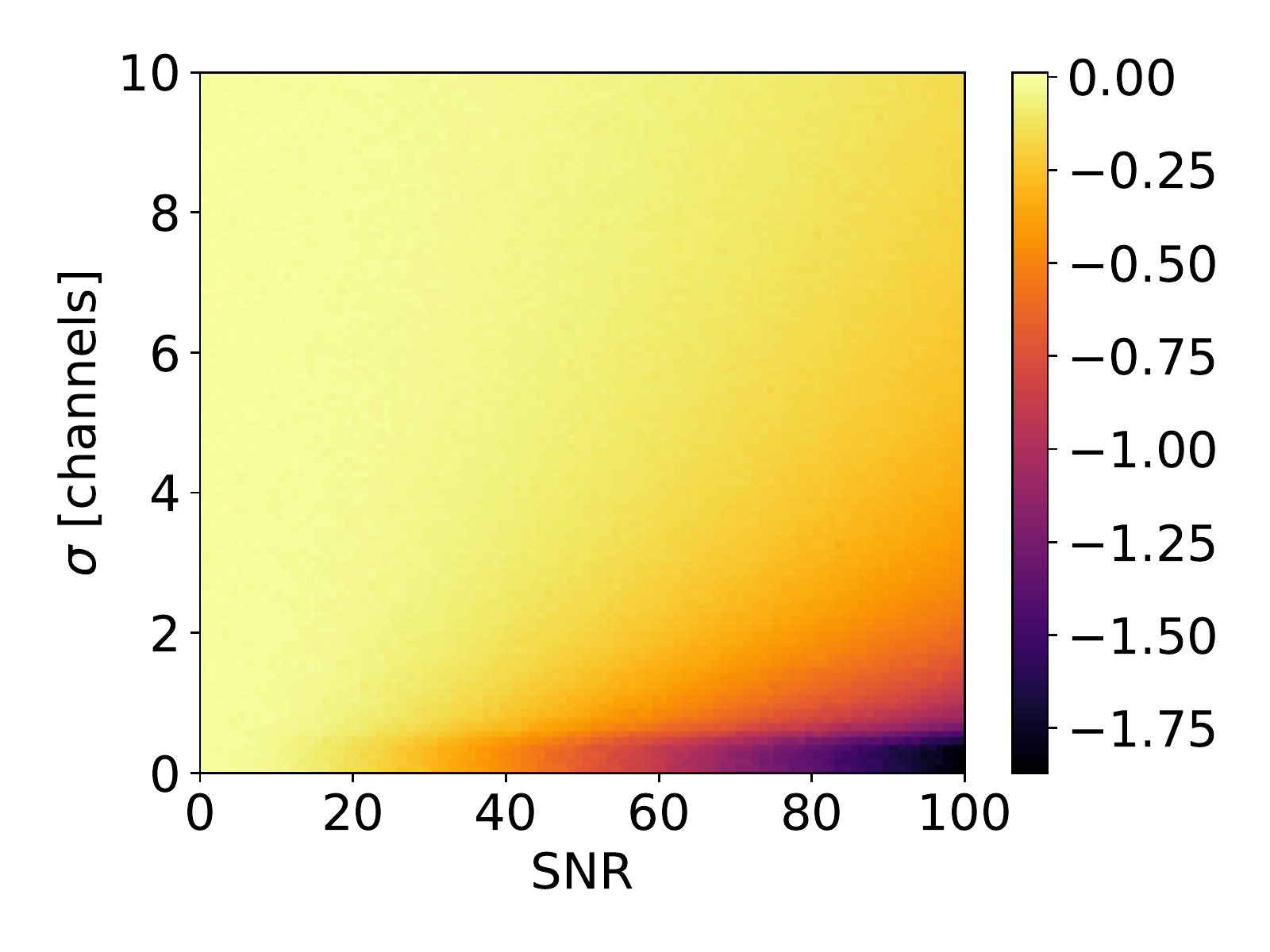}
\caption{Color corresponds to the relative difference $R$ between the rms and $\sigma_{\rm{diff}}$ for $10^5$ synthetic spectra. We added synthetic Gaussian signals of varying width and amplitude and calculated $R=1-\frac{\sigma_{\rm{diff}}}{rms}$ for each spectrum. We binned the values of $R$ according to the line width $\sigma$ and the S/N to show the effect on $\sigma_{\rm{diff}}$ caused by the presence of signal. This analysis shows that only bright, narrow lines significantly affect the accuracy of $\sigma_{\rm{diff}}$.}
\label{fig:rdiff_sig}
\end{figure}

Because bright lines add error to our estimate of the true noise, we masked each spectrum before comparing the rms to $\sigma_{\rm{diff}}$. As a first estimate of the true noise, we calculated $\sigma_{\rm{diff}}$ for the unmasked spectrum. We then masked channels with an intensity greater than $3\sigma_{\rm{diff}}$, as well as channels that were within 10 channels of a masked channel. Because bright $\mathrm{H_{2}O}$ masers add a large source of error to $\sigma_{\rm{diff}}$, we also measure $\sigma_{\rm{diff}}$ for the masked spectrum, which does not include very bright lines. We then used this new measurement of $\sigma_{\rm{diff}}$ to again mask channels with an intensity greater than $3\sigma_{\rm{diff}}$, as well as channels that were within 10 channels of a masked channel. We then calculated $R$ for this masked spectrum and used this value of $R$ to test the quality of the baseline fit. We also recorded the rms of the spectra, which we used as our estimate of the noise for later analysis. In Figure~\ref{fig:ex_spec} we give a few examples of RAMPS $\mathrm{NH_{3}}$(1,1) spectra and their associated values of $R$, which show that a poor baseline fit generally results in a larger value of $R$. A poor baseline fit can occur for spectra in which the spectral mask did not exclude all of the signal, as well as for spectra with a baseline shape more complicated than second order. While our spectral mask was reliable for the majority of $\mathrm{NH_{3}}$ lines in the RAMPS dataset, some lines where broader than a typical $\mathrm{NH_{3}}$ line and were not well masked. To better fit spectra of this class, we attempted a second fit on spectra with $R > 3 \sigma_{R}$ using a slightly different mask. To mask broader lines more effectively, we employed the same masking technique as for the initial fit but this time used a 120-channel, rather than 40-channel, window to calculate the array of local standard deviations. Due to the larger window size, this mask was more sensitive to broader spectral features, and so it more successfully masked broad lines. We performed another baseline fit using this masked spectrum and once again calculated $R$. If the spectrum is well fit by the second fit, $R$ will likely be low, but if there is a residual baseline shape more complicated than second order, $R$ will still be large. Low-amplitude signal that was not well masked may also increase the measured value of $R$. In either case, a poor fit has the potential to alter line amplitude ratios, which would change the parameter values calculated from future fits to the data. To mitigate this potential problem, if a spectrum had $R > 3 \sigma_{R}$ after the second fit, we performed a third, more conservative fit. We used the mask from the second fit and forced a zeroth-order baseline fit, which is less likely to change the line amplitude ratios. In Figure~\ref{fig:R_real}, we show histograms of $R$ for all of the baseline fits of the $\mathrm{NH_{3}}$(1,1), $\mathrm{NH_{3}}$(2,2), and $\mathrm{H_{2}O}$ spectra. The distributions show a Gaussian component centered at $R \sim 0$, with long tails out to larger values of $R$. The Gaussian portions of each distribution match relatively well with the distributions found for synthetic Gaussian noise. The long tails in the distributions represent the poor baseline fits that were fit with a zeroth-order baseline. The vertical magenta line corresponds to $R=3 \sigma_R$, which shows the approximate threshold between good and bad baselines expected from the analysis of the synthetic data. Significantly bad baselines are rare in this dataset, with the percent of spectra with $R>3 \sigma_R$ for the $\mathrm{NH_{3}}$(1,1), $\mathrm{NH_{3}}$(2,2), and $\mathrm{H_{2}O}$ data equal to $3.1\%$, $2.4\%$, and $2.5\%$, respectively. While the majority of the data are of a high quality, there are spectra in the dataset that require higher-order baseline fitting and more careful masking than our automated techniques can provide. Because we intend to create a catalog of molecular clumps from the RAMPS dataset, we will look in more detail at each detected clump. For those clumps with poorly fit spectra, we will attempt another baseline fit with a more carefully chosen spectral mask and baseline polynomial order.

\begin{figure}[!hbtp]
\centering
\includegraphics[height=\linewidth, angle=0, scale=0.7]{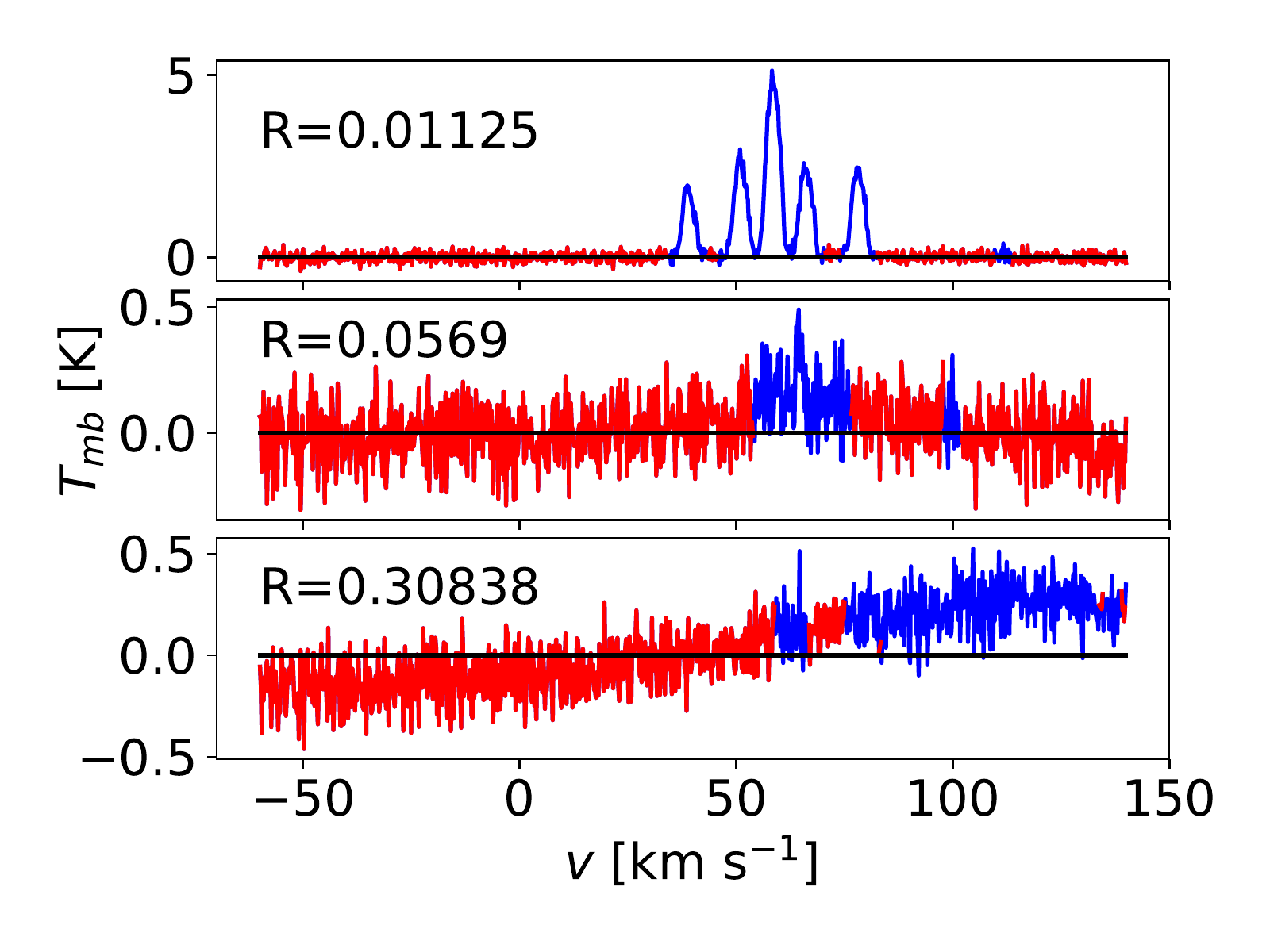}
\caption{Examples of RAMPS $\mathrm{NH_{3}}$(1,1) spectra and their associated values of $R$. We fit spectra that have a large value of $R$ in a more conservative manner in order to preserve the line shape of any signal present. We show the full spectra in blue, overplot the masked spectra in red, and also show a fiducial line at $T_{mb}=0$ K. Top: this spectrum exhibits a bright $\mathrm{NH_{3}}$(1,1) line and shows little evidence for a residual baseline. The line is well masked, and consequently the spectrum has a low value of $R$. Middle: this spectrum exhibits a weak $\mathrm{NH_{3}}$(1,1) line and shows evidence for a moderate residual baseline. The weak line is relatively well masked, but the residual baseline results in a moderately high value for $R$. Bottom: this spectrum does not contain an obvious $\mathrm{NH_{3}}$(1,1) line but shows evidence for a significant residual baseline, resulting in a large value for $R$.}
\label{fig:ex_spec}
\end{figure}

\begin{figure}[!hbtp]
\centering
\includegraphics[height=\linewidth, angle=0, scale=0.5]{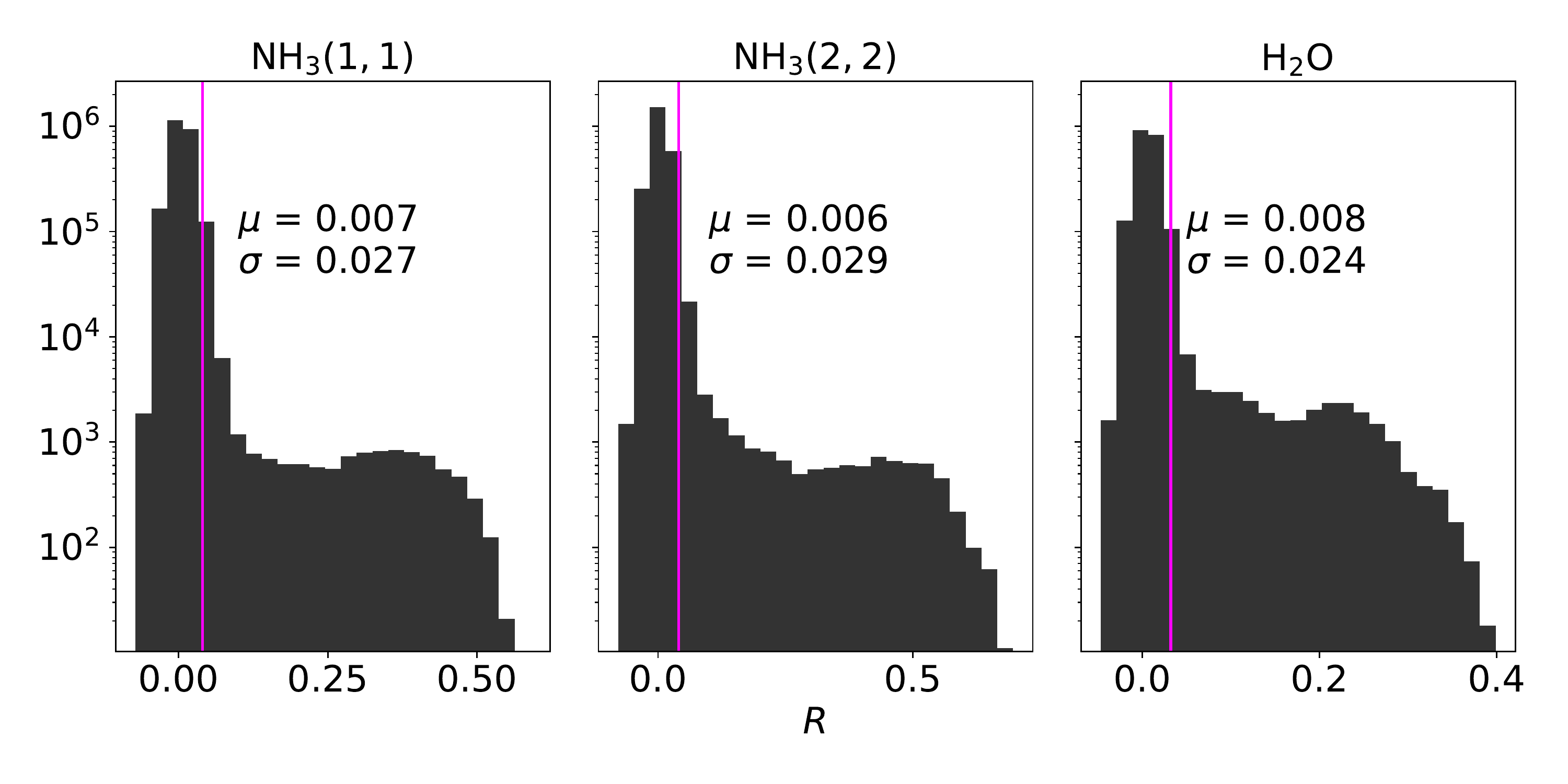}
\caption{Histograms of $R$ for all RAMPS spectra, separated by line. The distributions each have a Gaussian component expected from the analysis of simulated spectra, but they also have a long tail to higher values of $R$, indicating poorly fit spectra. The magenta lines show the $R=3 \sigma_R$ threshold used to restrict the complexity of the baseline fits.}
\label{fig:R_real}
\end{figure}

We used the rms, calculated in the manner described above, as our estimate of the noise in a spectrum. This estimate includes a contribution from the true noise, as well as from any residual baseline that is present. After calculating the noise, we determined the integrated intensity and first moment of each spectrum. First, we masked each channel with a value less than five times the rms. If there was only one unmasked channel, we masked the entire spectrum. Otherwise, we summed over the unmasked channels to obtain the integrated intensity in units of $\mathrm{K \ km \ s^{-1}}$. Using the same spectral mask, we determined the first moment using the formula $\langle v \rangle = \frac{\Sigma v_{i} T_{i}}{\Sigma T_{i}}$, where $T_i$ and $v_i$ are the intensity and velocity of the $i^{th}$ channel, respectively.

\subsection{Data Release}
\label{subsec:rel}

RAMPS data that are currently released to the public consist of $\mathrm{NH_{3}}$(1,1), $\mathrm{NH_{3}}$(2,2), and $\mathrm{H_{2}O}$ data cubes and their corresponding noise maps, as well as maps of $\mathrm{NH_{3}}$(1,1) and $\mathrm{NH_{3}}$(2,2) integrated intensity, $\mathrm{NH_{3}}$ velocity field, rotational temperature, total $\mathrm{NH_{3}}$ column density, $\mathrm{NH_{3}}$(1,1) optical depth, and $\mathrm{NH_{3}}$ line width. We present the integrated intensity and velocity field maps in Section \ref{sec:results} and the maps of rotational temperature, total $\mathrm{NH_{3}}$ column density, $\mathrm{NH_{3}}$ line width, and $\mathrm{H_{2}O}$ maser positions in Section \ref{sec:analysis}. RAMPS is an ongoing observing project, with the derived data being released annually upon verification. These data from the pilot survey are available at the RAMPS website (see footnote \ref{foot:ramps}).

\section{Results}
\label{sec:results}

Figure~\ref{fig:noise_hist} shows three histograms of the noise in the smoothed RAMPS Pilot spectra, one histogram each for $\mathrm{NH_{3}}$(1,1), $\mathrm{NH_{3}}$(2,2), and $\mathrm{H_{2}O}$ $6_{1,6} - 5_{2,3}$. Since we use seven receivers to observe the $\mathrm{NH_{3}}$ lines as compared to the single receiver we use to observe the $\mathrm{H_{2}O}$ maser line, the integration times per pixel are longer for the $\mathrm{NH_{3}}$ spectra. Thus, the $\mathrm{NH_{3}}$ spectra have much lower noise than the $\mathrm{H_{2}O}$ spectra. To show the spatial variations in the noise, we also present noise maps of all RAMPS fields observed during the pilot survey. Figures~\ref{fig:L10_11_noise}$-$\ref{fig:L47_11_noise} show the $\mathrm{NH_{3}(1,1)}$ noise maps, Figures~\ref{fig:L10_22_noise}$-$\ref{fig:L47_22_noise} show the $\mathrm{NH_{3}(2,2)}$ noise maps, and Figures~\ref{fig:L10_H2O_noise}$-$\ref{fig:L47_H2O_noise} show the $\mathrm{H_{2}O}$ maser noise maps. Since spectra from tiles observed in poor weather or at low elevations have much higher noise, the noise often varies significantly from tile to tile. Although several tiles show significantly higher noise than the average noise within their fields, we intend to reobserve only those tiles that show evidence of emission in the BGPS maps, so as not to waste future observing time. There is also evidence for noise variations within tiles due to the nonuniform integration time across a tile (Figures~\ref{fig:int30} and \ref{fig:int0}), changes in weather or source elevation over the course of an observation, and the stitching together of partial observations of a single tile. As shown in Figures~\ref{fig:L10_11_noise}-\ref{fig:L47_H2O_noise}, these variations are generally small, but they can be significant in certain tiles. For the rare circumstances where the noise variations within a tile are a significant detriment to our analysis of the data, we intend to reobserve.

\begin{figure}[!hbtp]
\centering
\includegraphics[angle=0,trim={3cm 0 3cm 0}, scale=0.6]{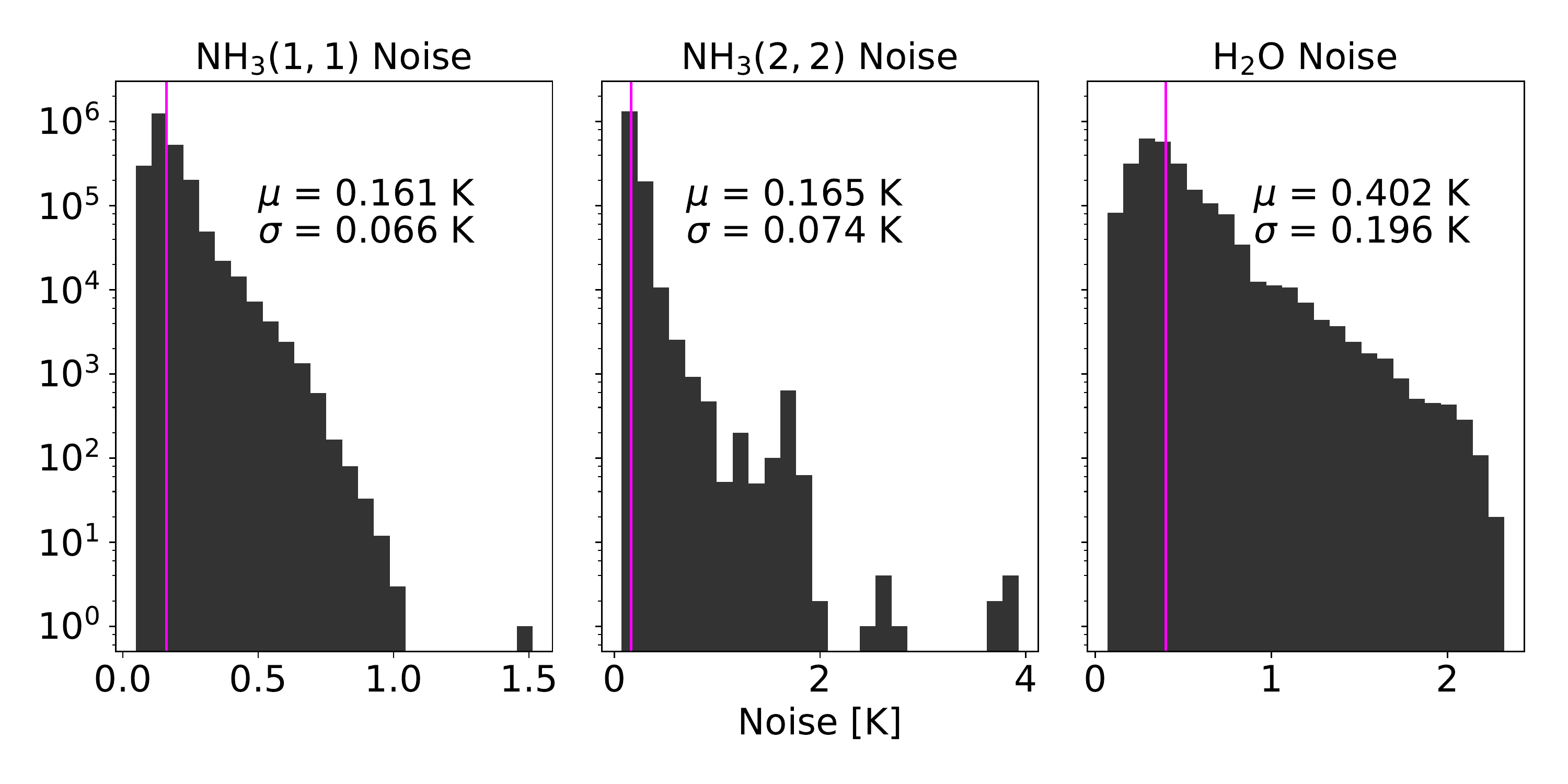}
\caption{Histograms of the rms in each spectrum, separated by line. The mean value $\mu$ of each noise distribution is indicated by a magenta line and is printed on each plot along with the standard deviation $\sigma$ of each distribution.}
\label{fig:noise_hist}
\end{figure}

\begin{subfigures}
\begin{figure}[!hbtp]
\centering
\includegraphics[scale=0.6]{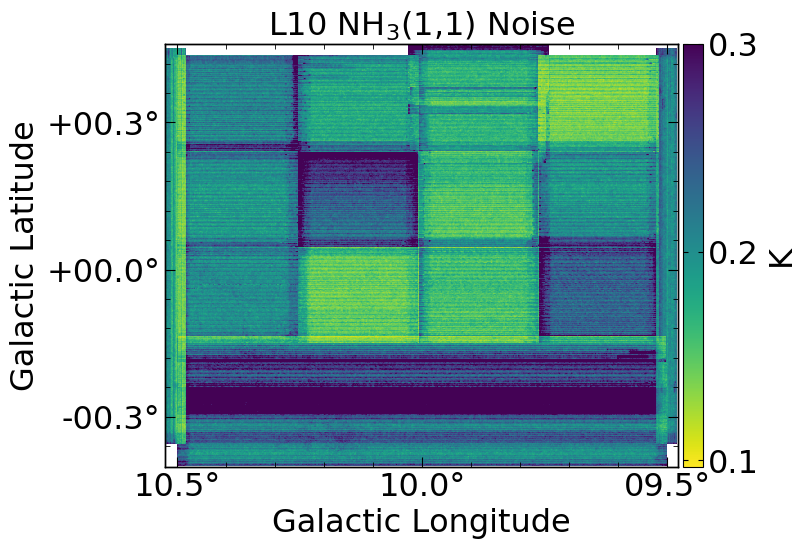}
\caption{$\mathrm{NH_{3}(1,1)}$ noise map of the L10 field. We used the L10 and L30 fields as test cases for our mapping scheme. Consequently, we mapped L10 in 12 $0^{\circ}.25 \times 0^{\circ}.20$ $\lq \lq$tiles" and eight $1^{\circ} \times 0^{\circ}.058$ $\lq \lq$strips". The white areas on the map represent regions that we have not observed.}
\label{fig:L10_11_noise}
\end{figure}

\begin{figure}[!hbtp]
\centering
\includegraphics[scale=0.6]{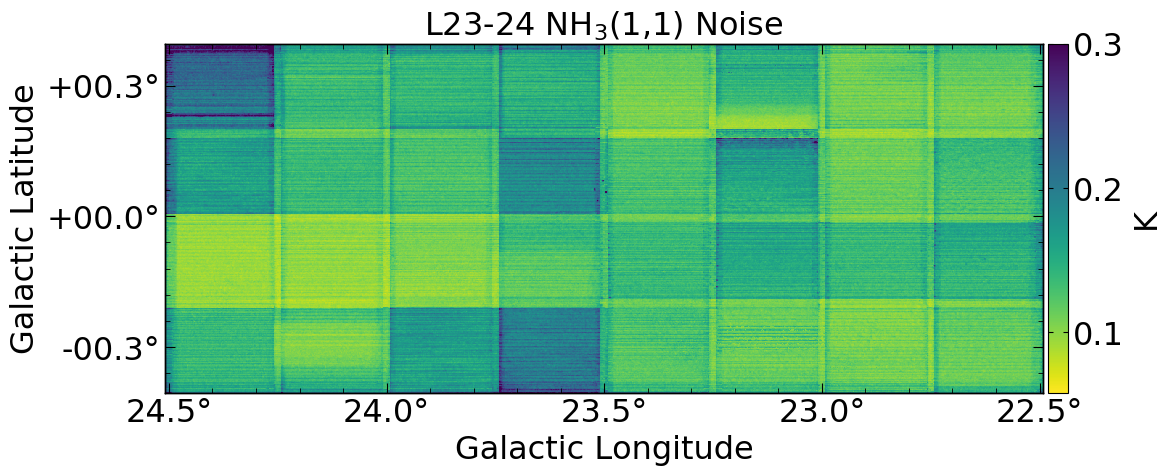}
\caption{$\mathrm{NH_{3}(1,1)}$ noise map of the L23 and L24 fields. The L23 and L24 fields were each mapped in sixteen $0.26^{\circ} \times 0.208^{\circ}$ rectangular tiles.}
\label{fig:L23-24_11_noise}
\end{figure}

\begin{figure}[!hbtp]
\centering
\includegraphics[scale=0.34]{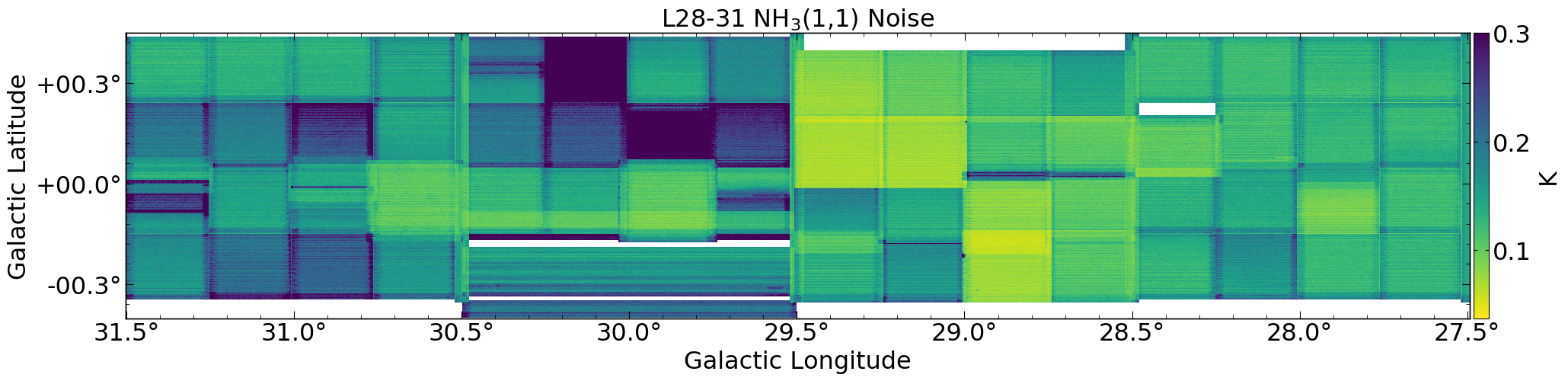}
\caption{$\mathrm{NH_{3}(1,1)}$ noise map of the L28, L29, L30, and L31 fields. The L28 and L31 fields were each mapped in sixteen $0.25^{\circ} \times 0.20^{\circ}$ rectangular tiles. The L29 field was mapped in sixteen $0.26^{\circ} \times 0.208^{\circ}$ rectangular tiles. The L10 and L30 field as test cases for our mapping scheme. Consequently, we mapped L30 in twelve $0.25^{\circ} \times 0.20^{\circ}$ $\lq \lq$tiles" and eight $1^{\circ} \times 0.058^{\circ}$ $\lq \lq$strips".}
\label{fig:L28-31_11_noise}
\end{figure}

\begin{figure}[!hbtp]
\centering
\includegraphics[scale=0.7]{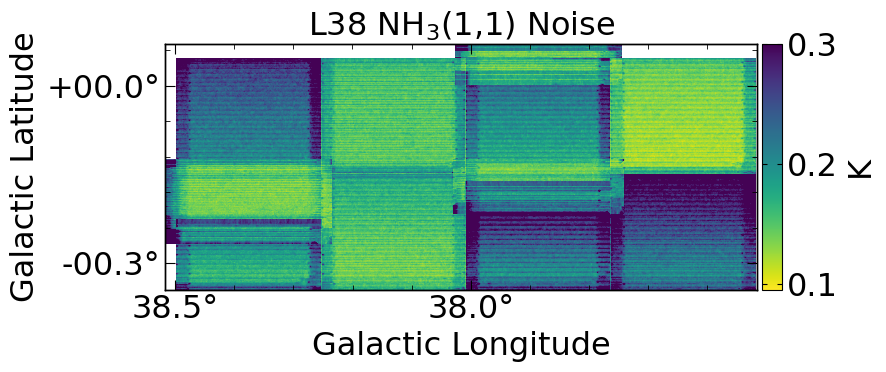}
\caption{$\mathrm{NH_{3}(1,1)}$ noise map of a portion of the L38 field. The L38 field was mapped in eight $0.25^{\circ} \times 0.20^{\circ}$ rectangular tiles.}
\label{fig:L38_11_noise}
\end{figure}

\begin{figure}[!hbtp]
\centering
\includegraphics[scale=0.65]{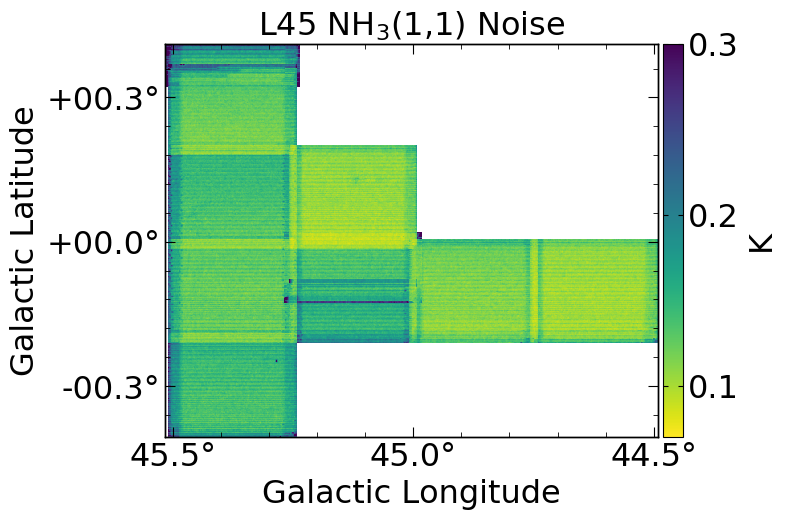}
\caption{$\mathrm{NH_{3}(1,1)}$ noise map of a portion of the L45 field. The L45 field was mapped in eight $0.26^{\circ} \times 0.208^{\circ}$ rectangular tiles.}
\label{fig:L45_11_noise}
\end{figure}

\begin{figure}[!hbtp]
\centering
\includegraphics[scale=0.7]{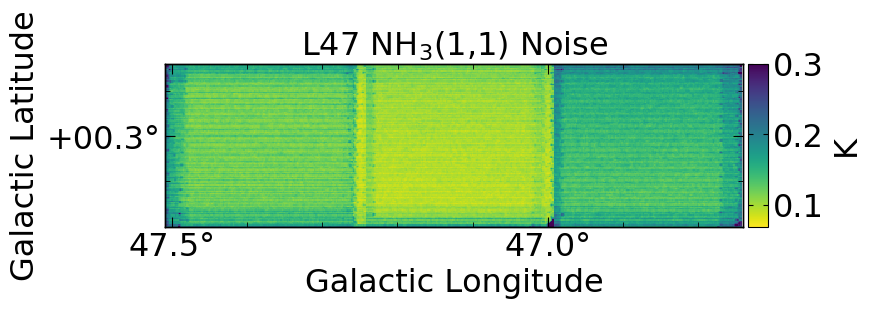}
\caption{$\mathrm{NH_{3}(1,1)}$ noise map of a portion of the L47 field. The L47 field was mapped in three $0.26^{\circ} \times 0.208^{\circ}$ rectangular tiles.}
\label{fig:L47_11_noise}
\end{figure}
\end{subfigures}

\begin{subfigures}
\begin{figure}[!hbtp]
\centering
\includegraphics[scale=0.6]{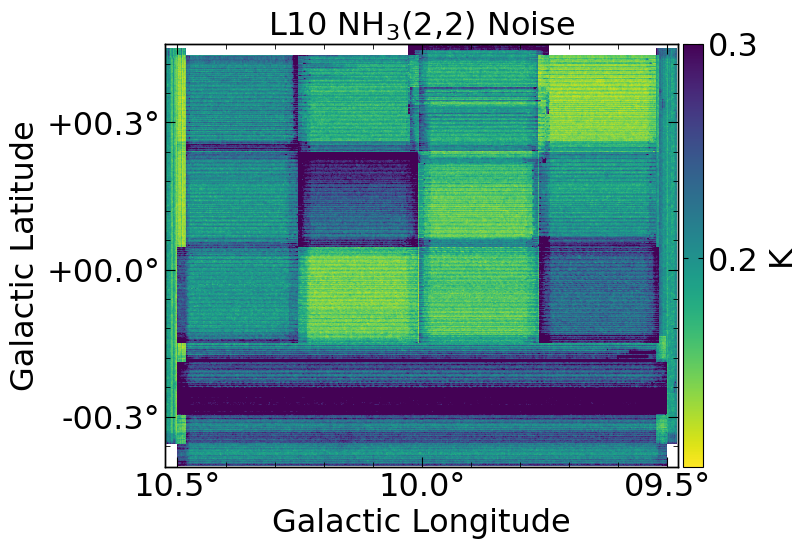}
\caption{$\mathrm{NH_{3}(2,2)}$ noise map of the L10 field.}
\label{fig:L10_22_noise}
\end{figure}

\begin{figure}[!hbtp]
\centering
\includegraphics[scale=0.6]{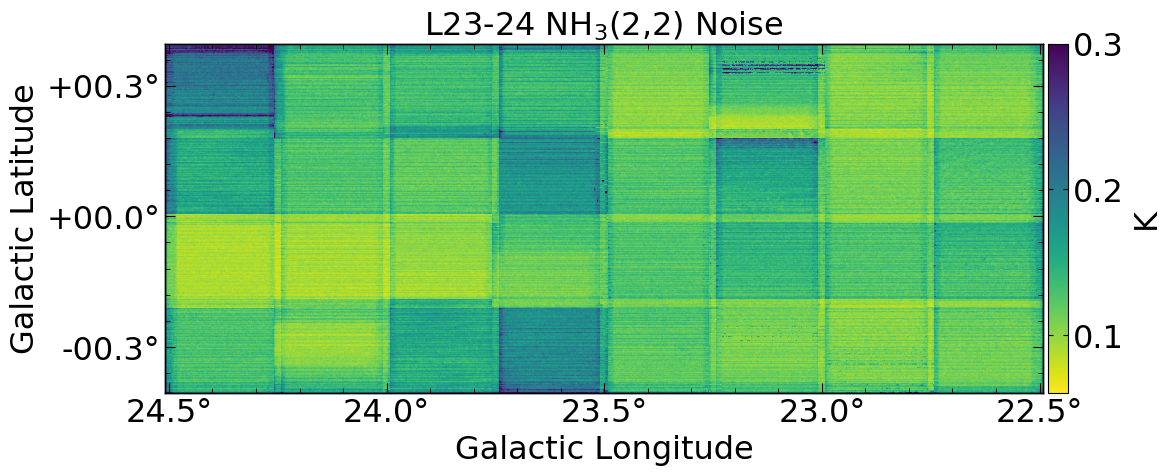}
\caption{$\mathrm{NH_{3}(2,2)}$ noise map of the L23 and L24 fields.}
\label{fig:L23-24_22_noise}
\end{figure}

\begin{figure}[!hbtp]
\centering
\includegraphics[scale=0.34]{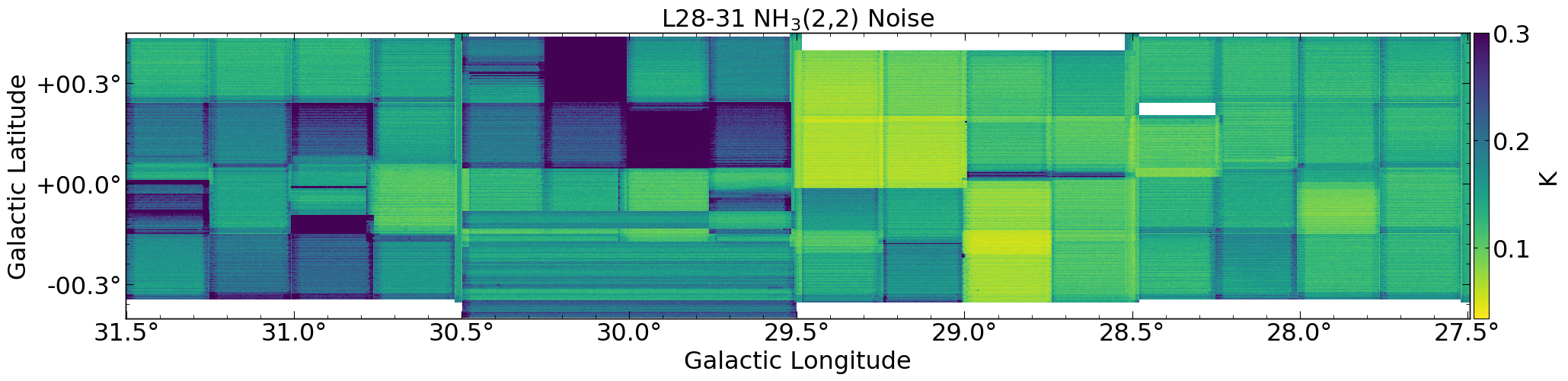}
\caption{$\mathrm{NH_{3}(2,2)}$ noise map of the L28, L29, L30, and L31 fields.}
\label{fig:L28-31_22_noise}
\end{figure}

\begin{figure}[!hbtp]
\centering
\includegraphics[scale=0.7]{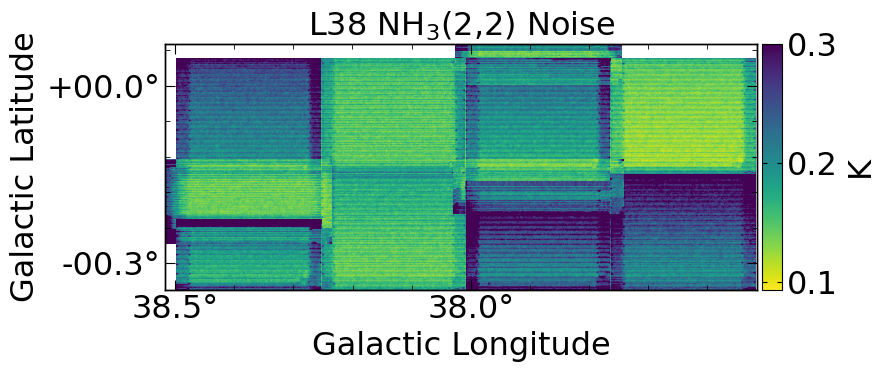}
\caption{$\mathrm{NH_{3}(2,2)}$ noise map of a portion of the L38 field.}
\label{fig:L38_22_noise}
\end{figure}

\begin{figure}[!hbtp]
\centering
\includegraphics[scale=0.65]{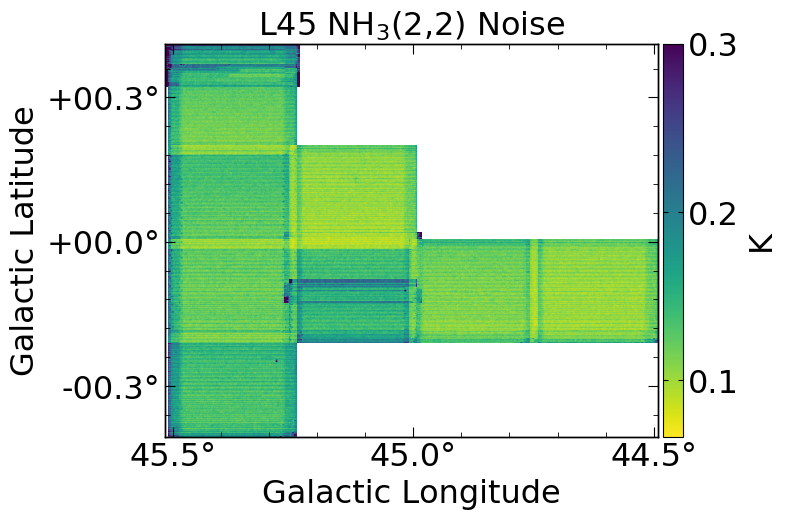}
\caption{$\mathrm{NH_{3}(2,2)}$ noise map of a portion of the L45 field.}
\label{fig:L45_22_noise}
\end{figure}

\begin{figure}[!hbtp]
\centering
\includegraphics[scale=0.7]{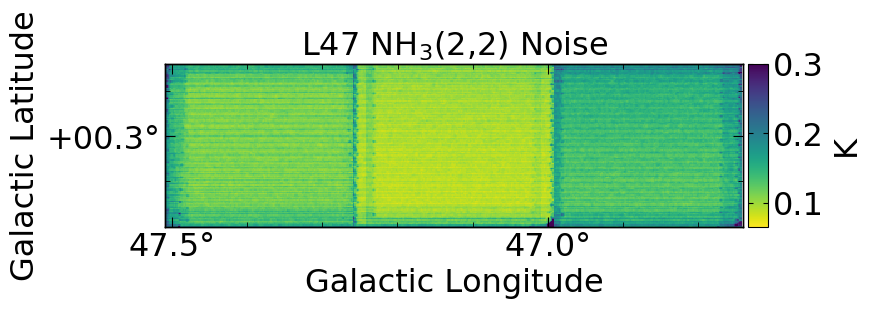}
\caption{$\mathrm{NH_{3}(2,2)}$ noise map of a portion of the L47 field.}
\label{fig:L47_22_noise}
\end{figure}
\end{subfigures}

\begin{subfigures}
\begin{figure}[!hbtp]
\centering
\includegraphics[scale=0.6]{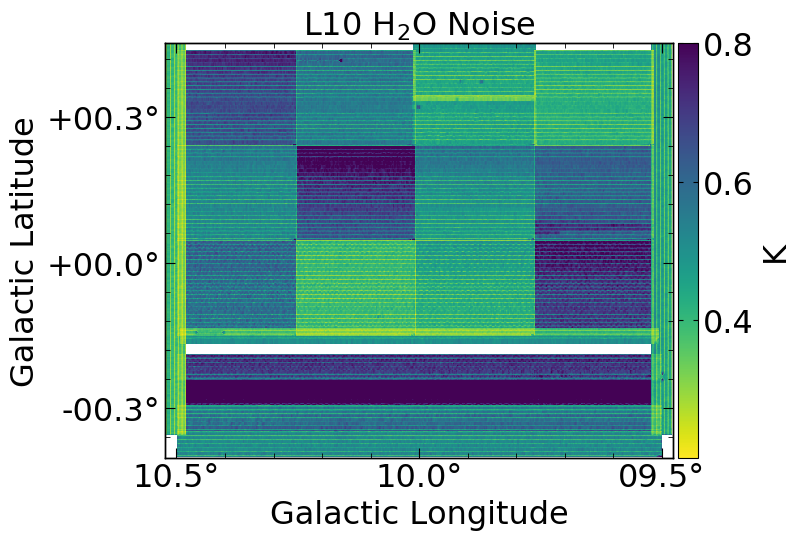}
\caption{$\mathrm{H_{2}O}$ noise map of the L10 field.}
\label{fig:L10_H2O_noise}
\end{figure}

\begin{figure}[!hbtp]
\centering
\includegraphics[scale=0.6]{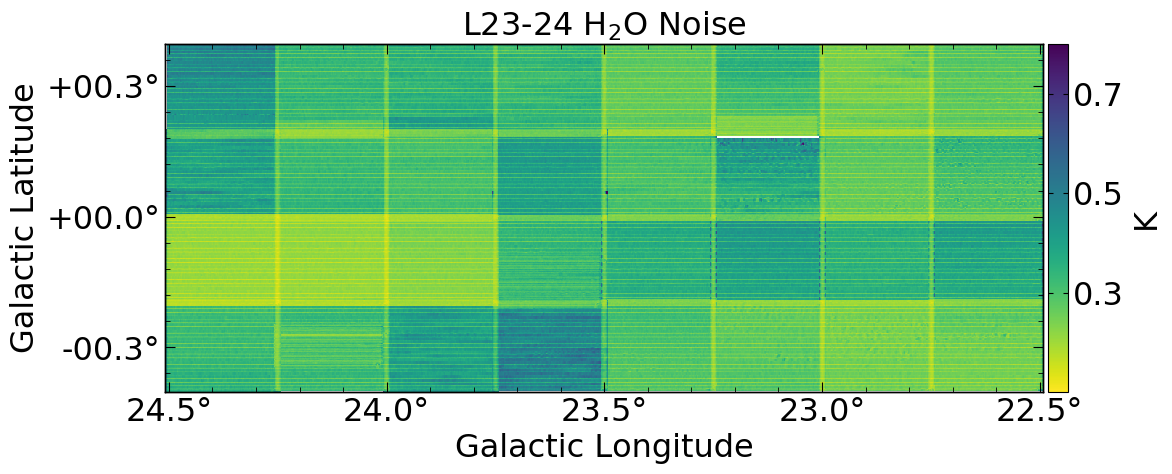}
\caption{$\mathrm{H_{2}O}$ noise map of the L23 and L24 fields.}
\label{fig:L23-24_H2O_noise}
\end{figure}

\begin{figure}[!hbtp]
\centering
\includegraphics[scale=0.34]{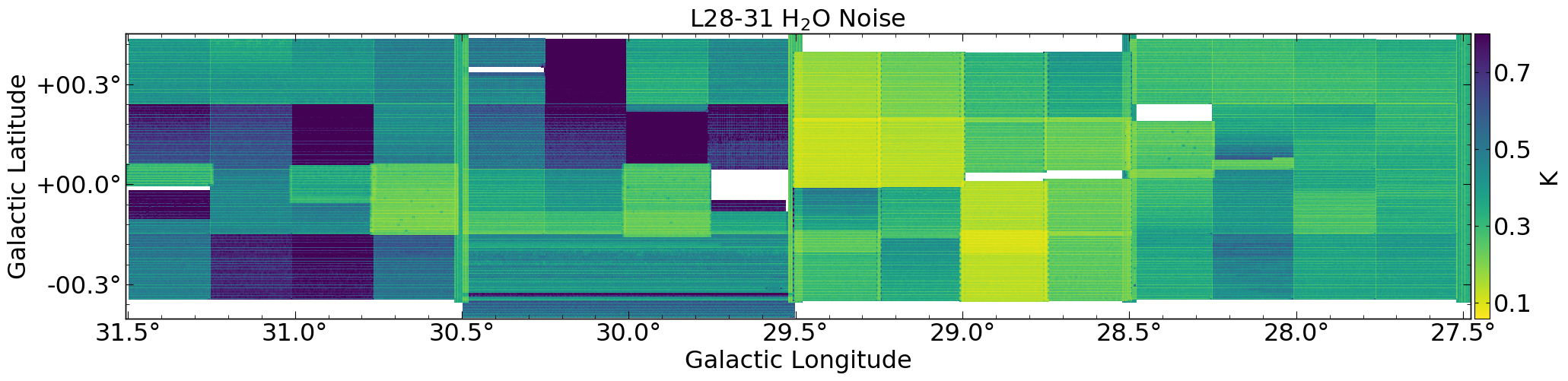}
\caption{$\mathrm{H_{2}O}$ noise map of the L28, L29, L30, and L31 fields.}
\label{fig:L28-31_H2O_noise}
\end{figure}

\begin{figure}[!hbtp]
\centering
\includegraphics[scale=0.7]{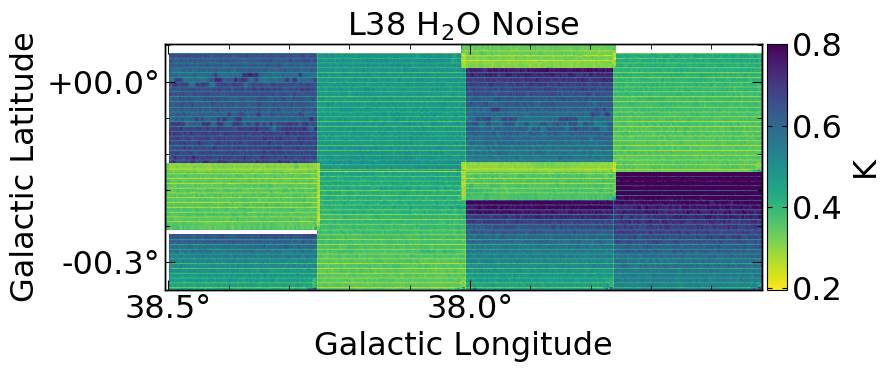}
\caption{$\mathrm{H_{2}O}$ noise map of a portion of the L38 field.}
\label{fig:L38_H2O_noise}
\end{figure}

\begin{figure}[!hbtp]
\centering
\includegraphics[scale=0.65]{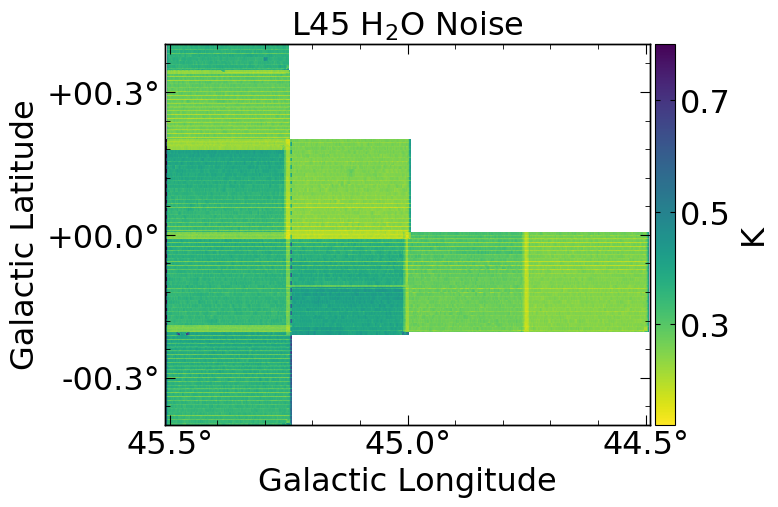}
\caption{$\mathrm{H_{2}O}$ noise map of a portion of the L45 field.}
\label{fig:L45_H2O_noise}
\end{figure}

\begin{figure}[!hbtp]
\centering
\includegraphics[scale=0.7]{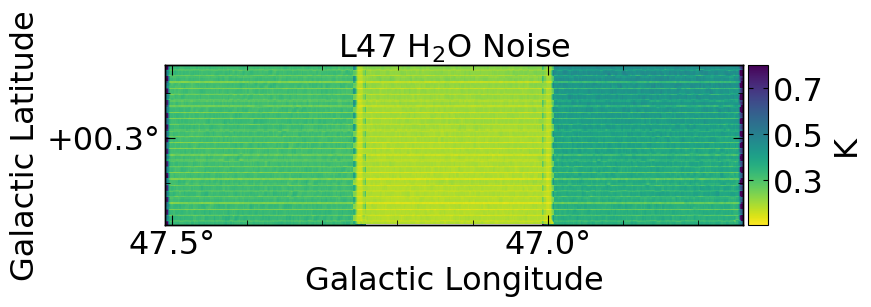}
\caption{$\mathrm{H_{2}O}$ noise map of a portion of the L47 field.}
\label{fig:L47_H2O_noise}
\end{figure}
\end{subfigures}

Figures \ref{fig:L10_11_mom0}$-$\ref{fig:L47_11_mom0} show the $\mathrm{NH_{3}(1,1)}$ integrated intensity maps of the RAMPS fields observed during the pilot survey. We also present our $\mathrm{NH_{3}(2,2)}$ integrated intensity maps in Figures \ref{fig:L10_22_mom0}$-$\ref{fig:L45_22_mom0}, where we did not plot the $\mathrm{NH_{3}(2,2)}$ integrated intensity map for the L47 field since it did not include andy significant emission. We detected significant $\mathrm{NH_{3}(1,1)}$ and $\mathrm{NH_{3}(2,2)}$ emission in $20.8\%$ and $5.4\%$, respectively, of the mapped area. Furthermore, we found that $20.7\%$ of pixels with a significant $\mathrm{NH_{3}(1,1)}$ detection also had a significant $\mathrm{NH_{3}(2,2)}$ detection, while there were no pixels with a significant $\mathrm{NH_{3}(2,2)}$ detection and no significant $\mathrm{NH_{3}(1,1)}$ detection. The integrated intensity maps reveal molecular clumps of various shapes and angular sizes. While a portion of the detected clumps seem to be grouped together in large complexes, many clumps appear to be more isolated and spread somewhat uniformly across the survey region. We also present a map of clump velocities in Figure~\ref{fig:L23-24_11_mom1}. In this map, we have detected clumps over a velocity range of $\sim20 - 140 \ \mathrm{km \ s^{-1}}$. There are several groupings of clumps with similar velocities. Although we have not performed a quantitative analysis of the positions and velocities of the detected clumps, one could use the RAMPS dataset to advance our understanding of Galactic structure.

\begin{subfigures}
\begin{figure}[!hbtp]
\centering
\includegraphics[scale=0.6]{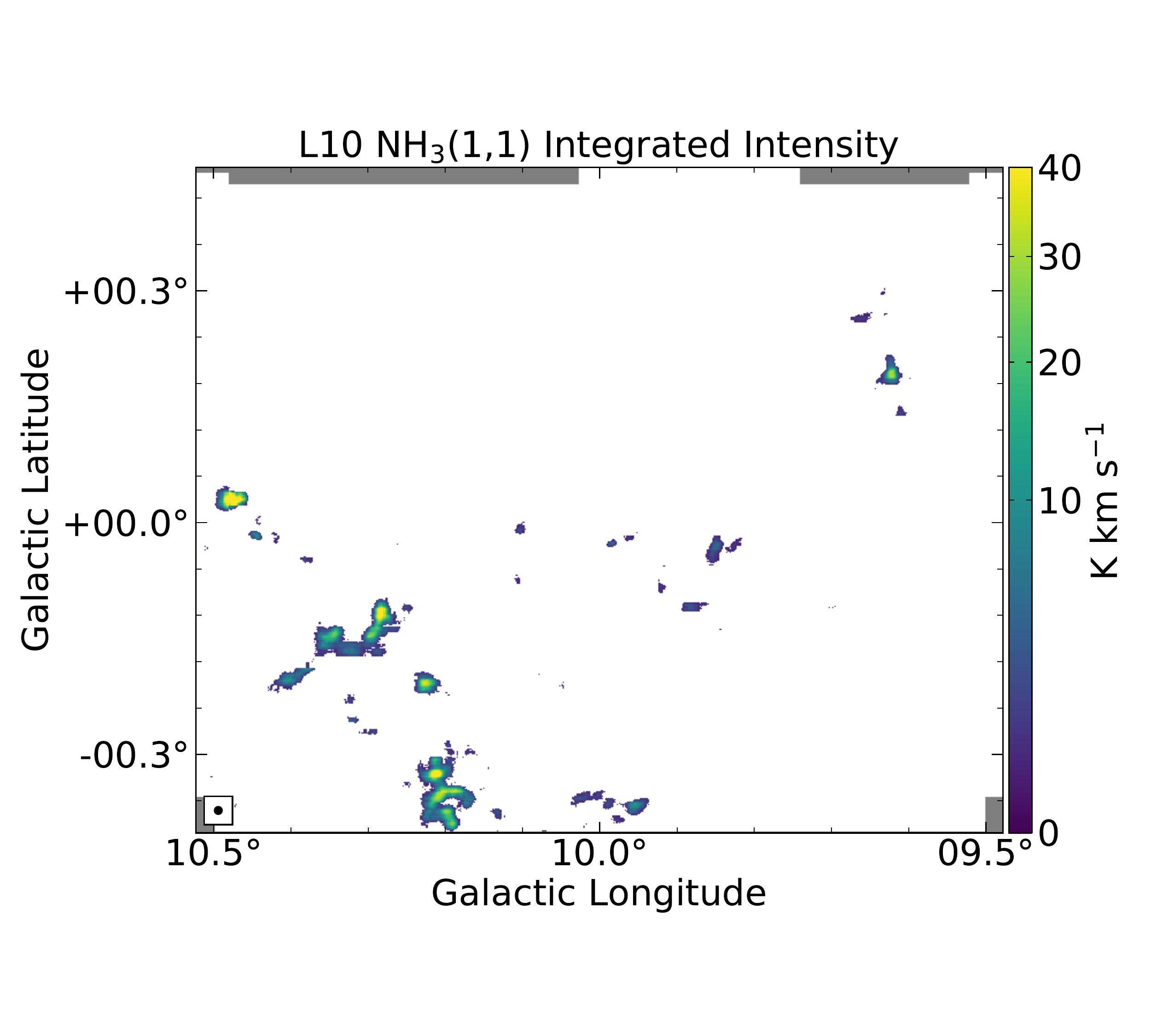}
\caption{$\mathrm{NH_{3}(1,1)}$ integrated intensity map of the L10 field. The beam size is shown in the box at the lower left corner of the map. The gray parts of the map represent regions that were not observed.}
\label{fig:L10_11_mom0}
\end{figure}

\begin{figure}[!hbtp]
\centering
\includegraphics[scale=0.7,trim={0 3cm 0 3cm},clip]{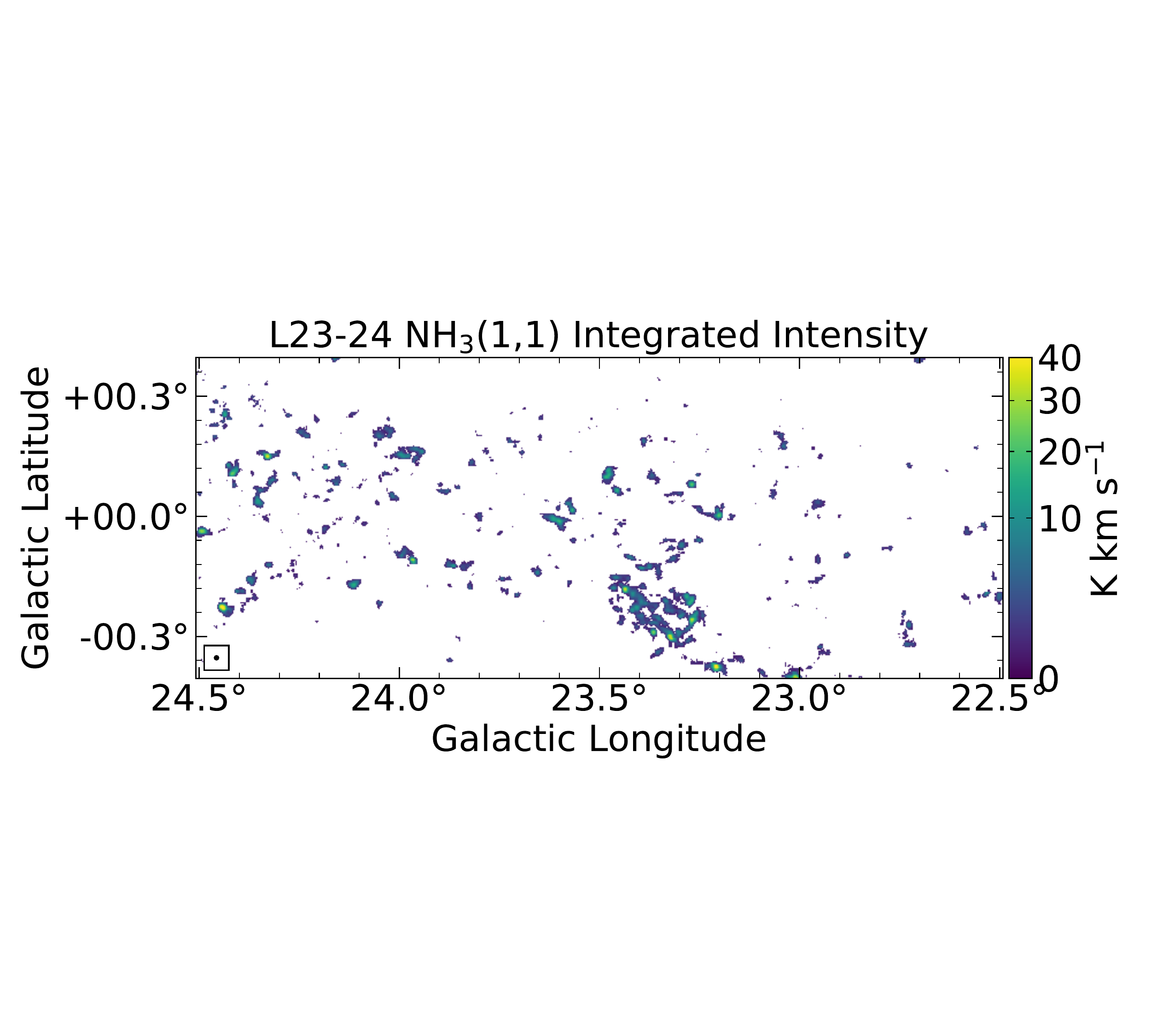}
\caption{Combined $\mathrm{NH_{3}(1,1)}$ integrated intensity map of the L23 and L24 fields.}
\label{fig:L23-24_11_mom0}
\end{figure}

\begin{figure}[!hbtp]
\centering
\includegraphics[scale=0.7,trim={0 5cm 0 5cm},clip]{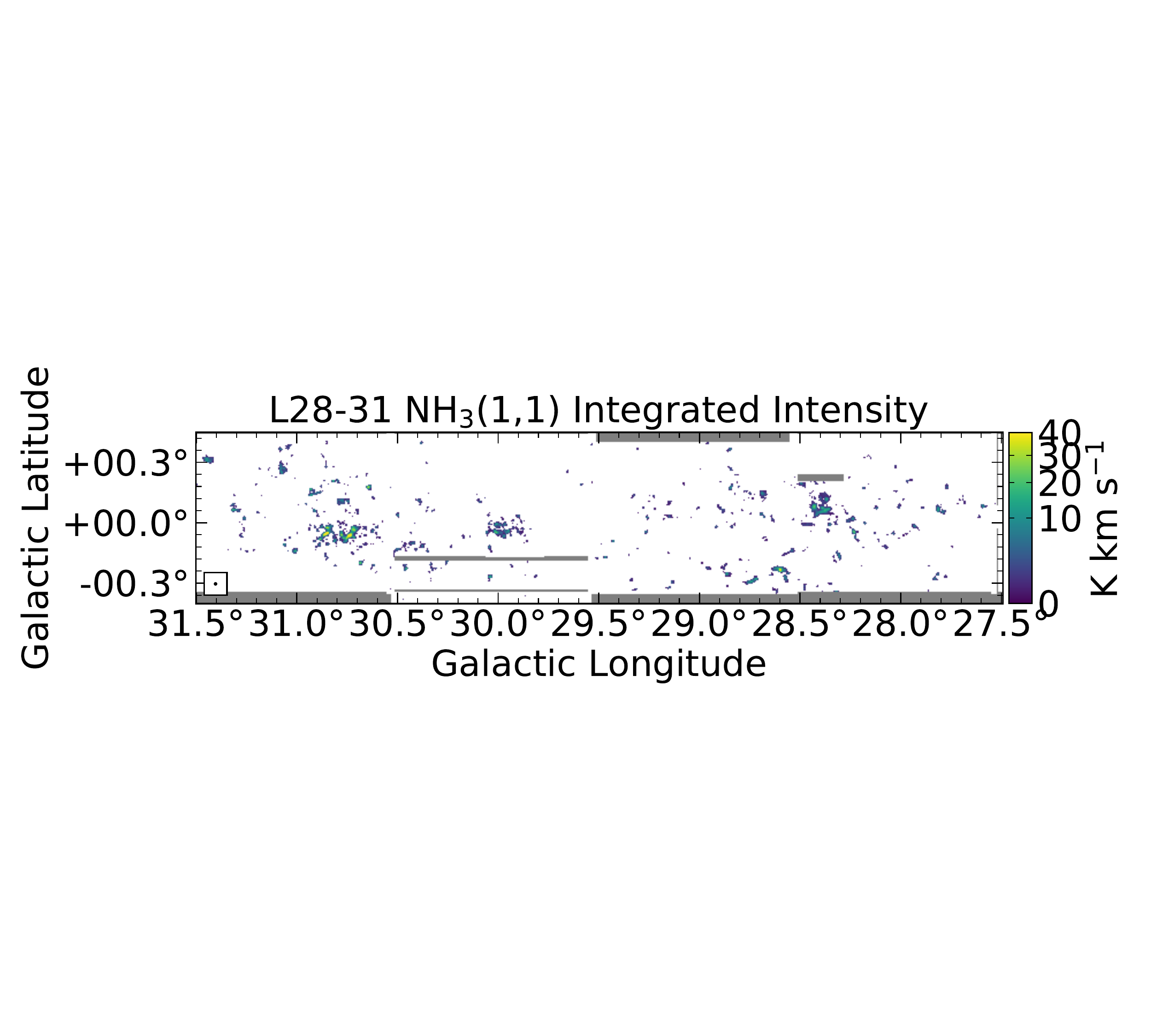}
\caption{Combined $\mathrm{NH_{3}(1,1)}$ integrated intensity map of the L28, L29, L30, and L31 fields.}
\label{fig:L28-31_11_mom0}
\end{figure}

\begin{figure}[!hbtp]
\centering
\includegraphics[scale=0.7,trim={0 3cm 0 3cm},clip]{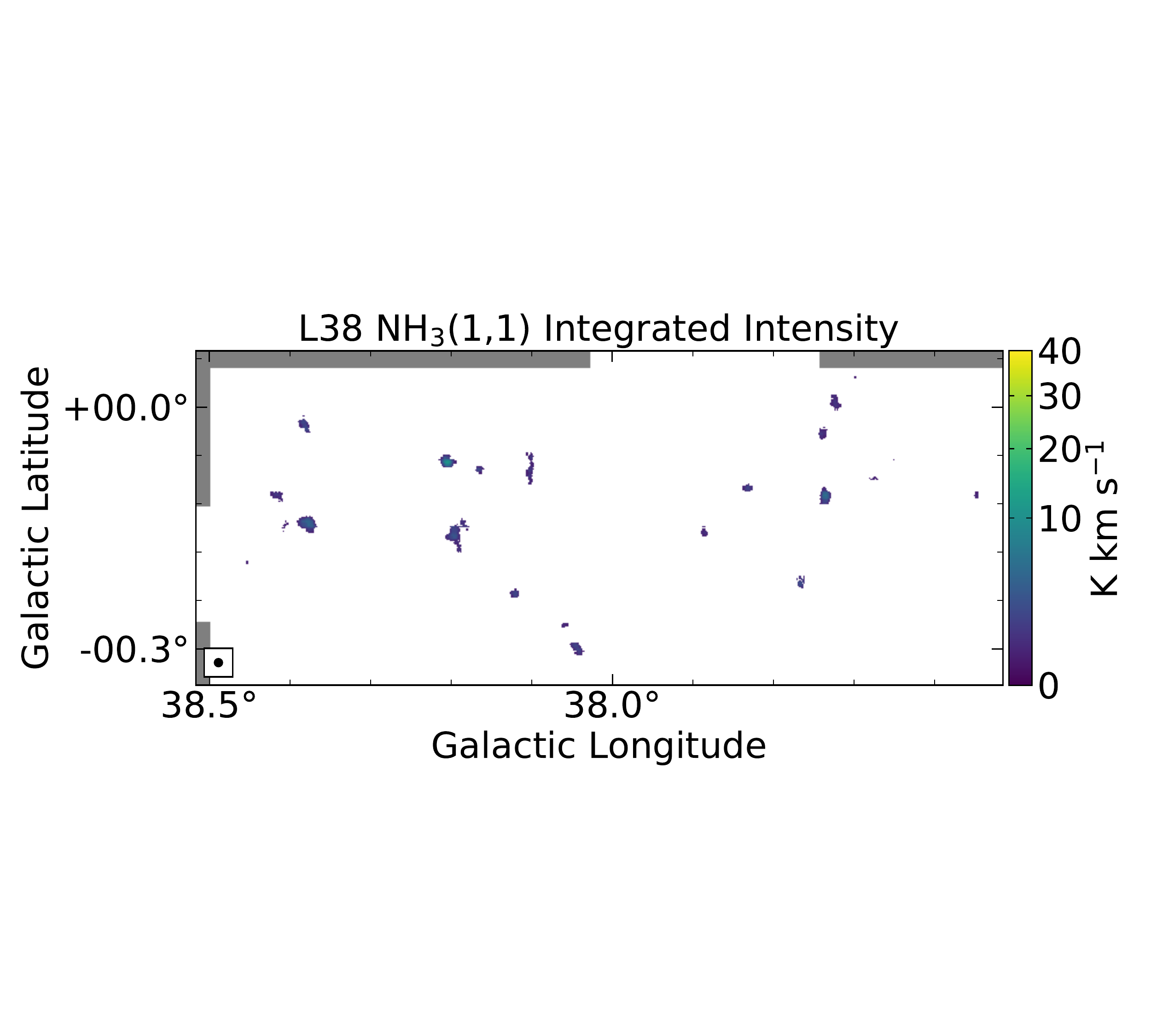}
\caption{$\mathrm{NH_{3}(1,1)}$ integrated intensity map of the L38 field.}
\label{fig:L38_11_mom0}
\end{figure}

\begin{figure}[!hbtp]
\centering
\includegraphics[scale=0.65]{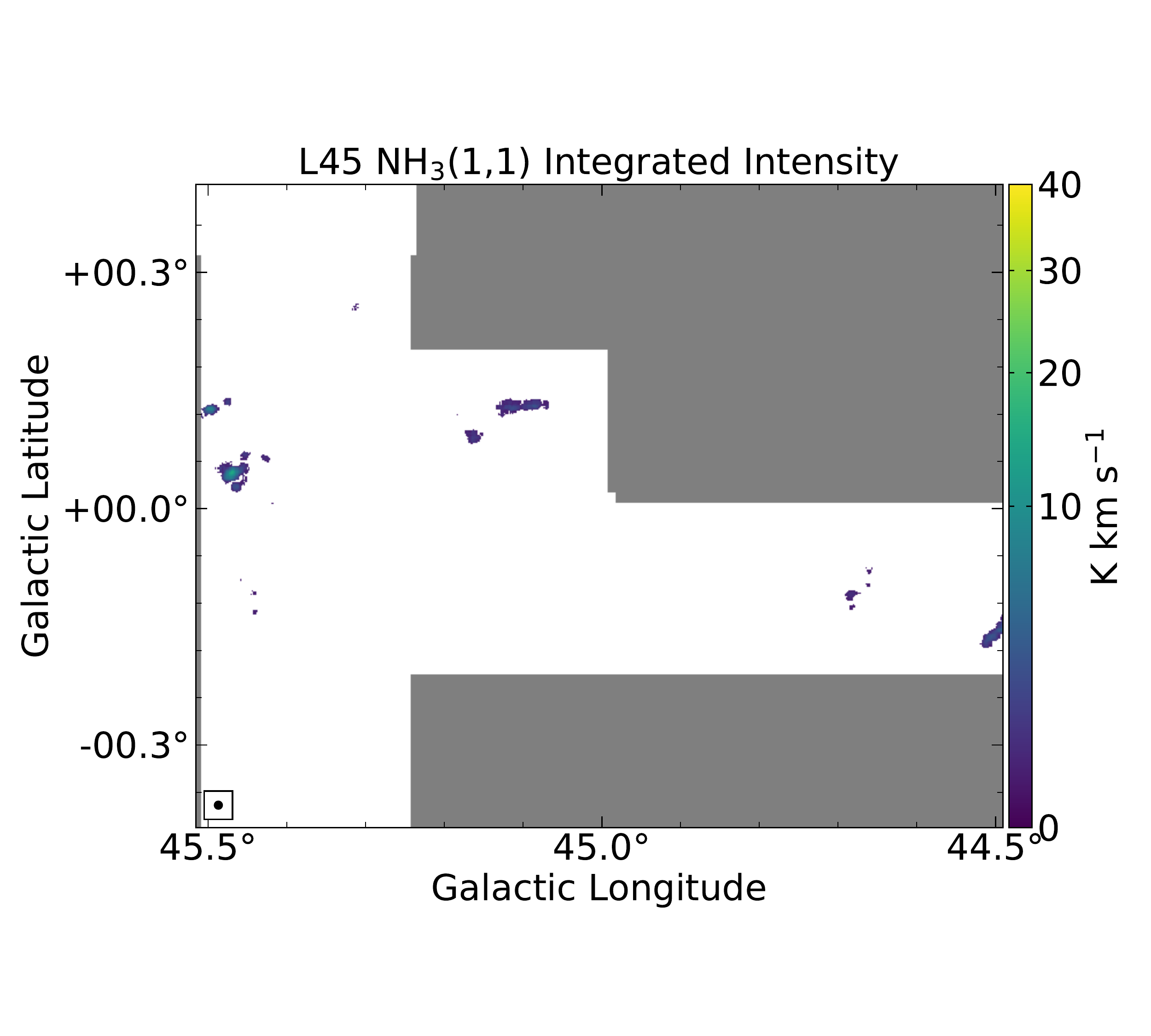}
\caption{$\mathrm{NH_{3}(1,1)}$ integrated intensity map of the L45 field.}
\label{fig:L45_11_mom0}
\end{figure}

\begin{figure}[!hbtp]
\centering
\includegraphics[scale=0.75,trim={0 4cm 0 4cm},clip]{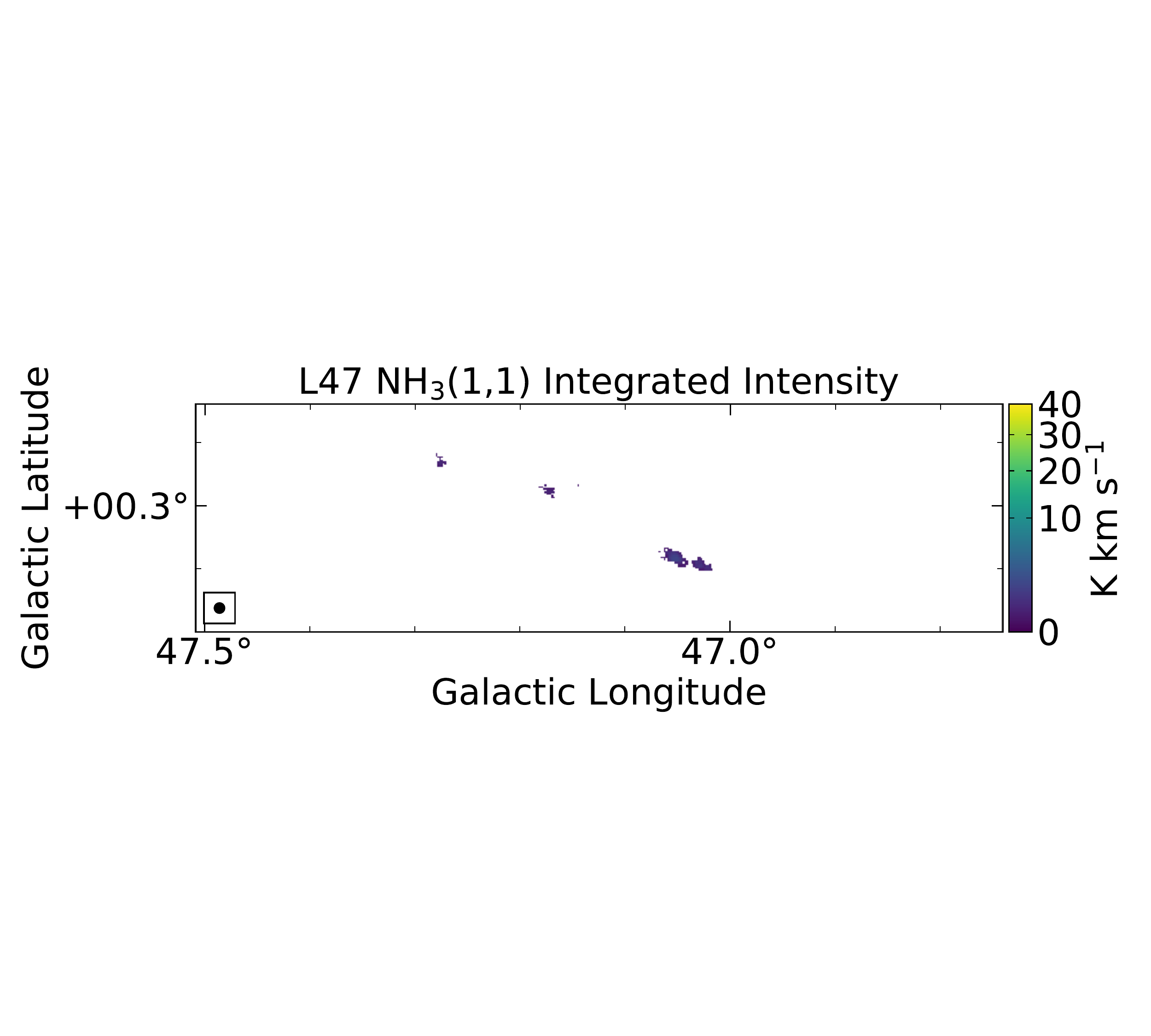}
\caption{$\mathrm{NH_{3}(1,1)}$ integrated intensity map of the L47 field.}
\label{fig:L47_11_mom0}
\end{figure}
\end{subfigures}

\begin{subfigures}
\begin{figure}[!hbtp]
\centering
\includegraphics[scale=0.7]{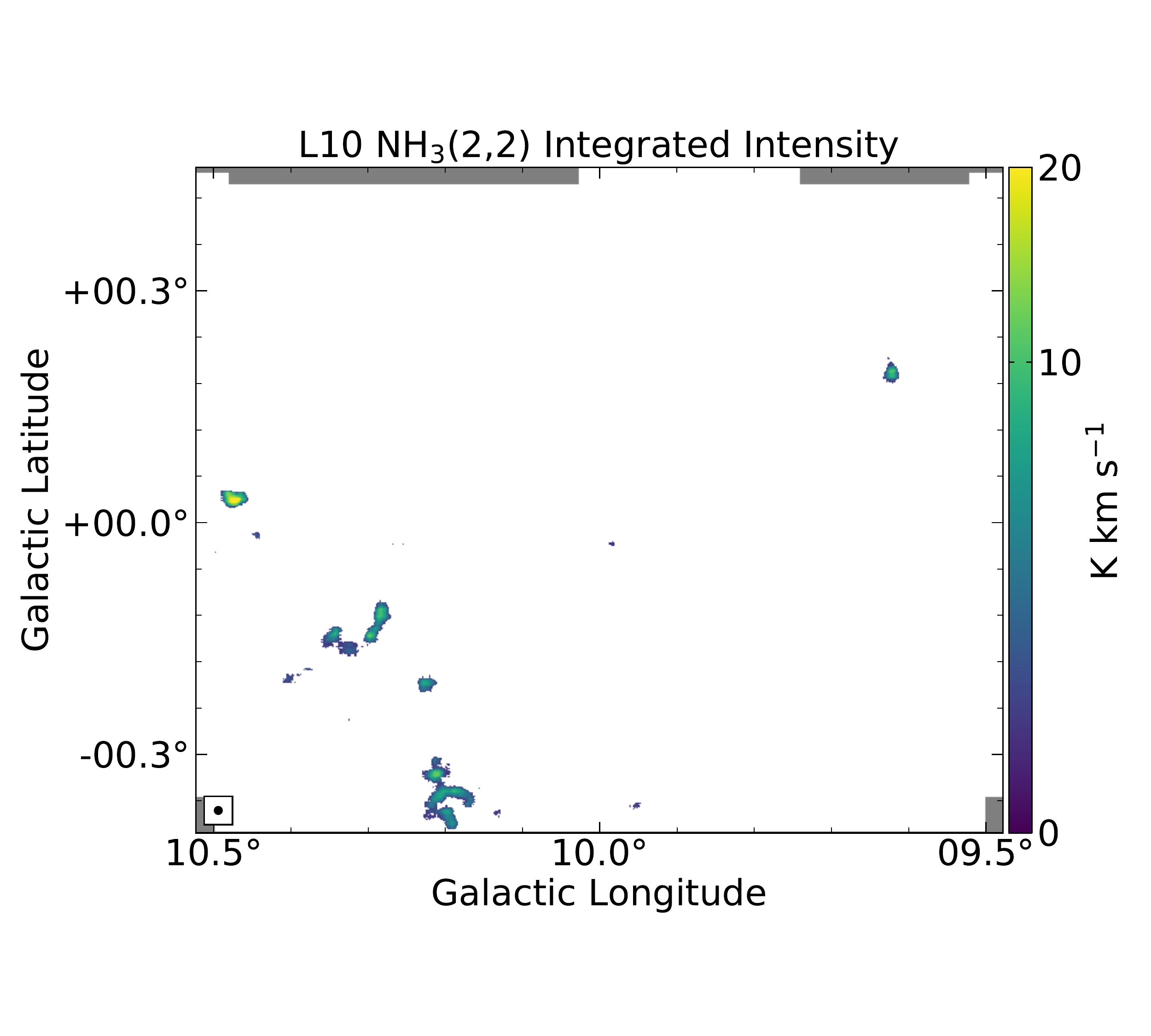}
\caption{$\mathrm{NH_{3}(2,2)}$ integrated intensity map of the L10 field.}
\label{fig:L10_22_mom0}
\end{figure}

\begin{figure}[!hbtp]
\centering
\includegraphics[scale=0.7,trim={0 3cm 0 3cm},clip]{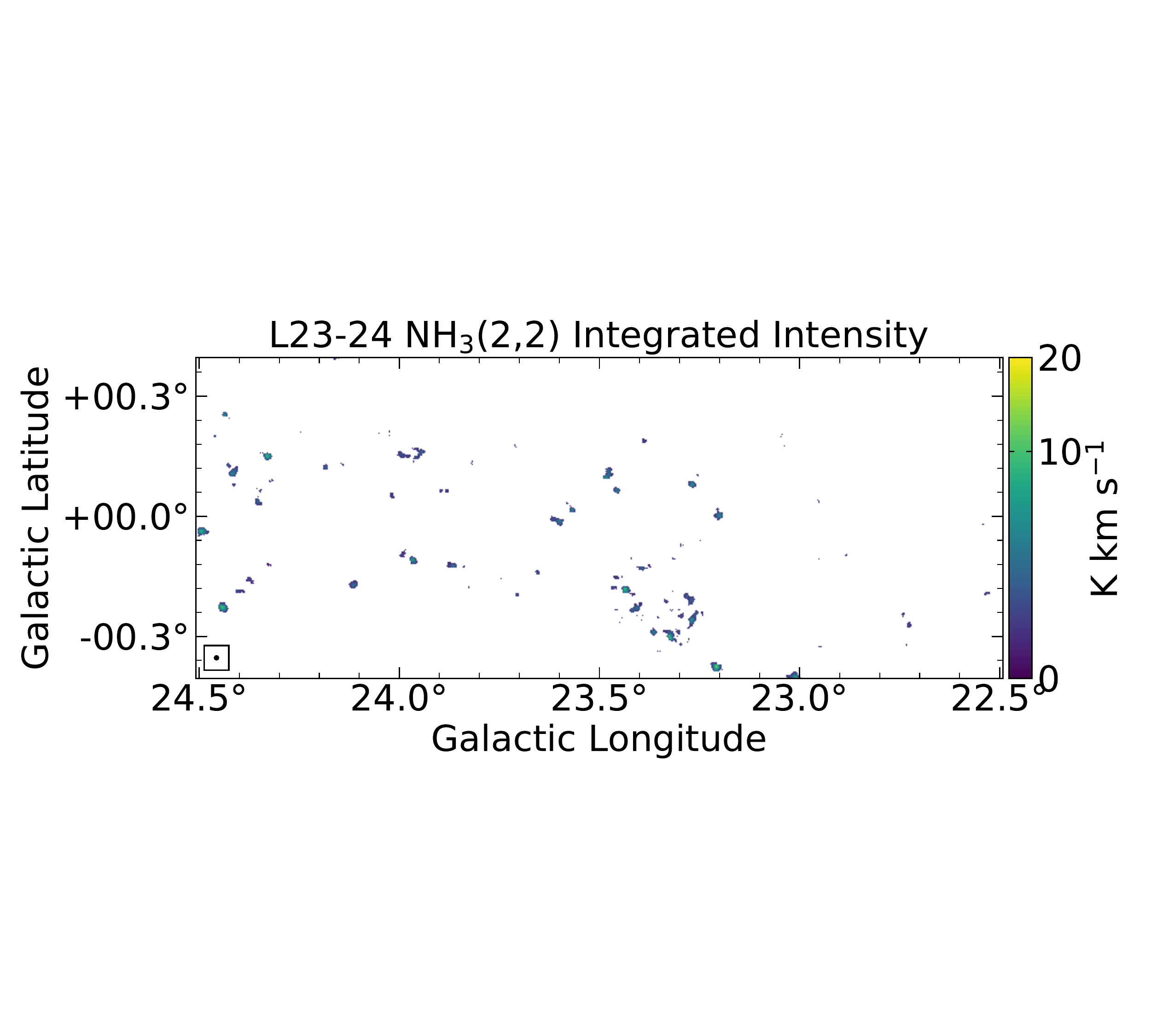}
\caption{Combined $\mathrm{NH_{3}(2,2)}$ integrated intensity map of the L23 and L24 fields.}
\label{fig:L23-24_22_mom0}
\end{figure}

\begin{figure}[!hbtp]
\centering
\includegraphics[scale=0.7,trim={0 5cm 0 5cm},clip]{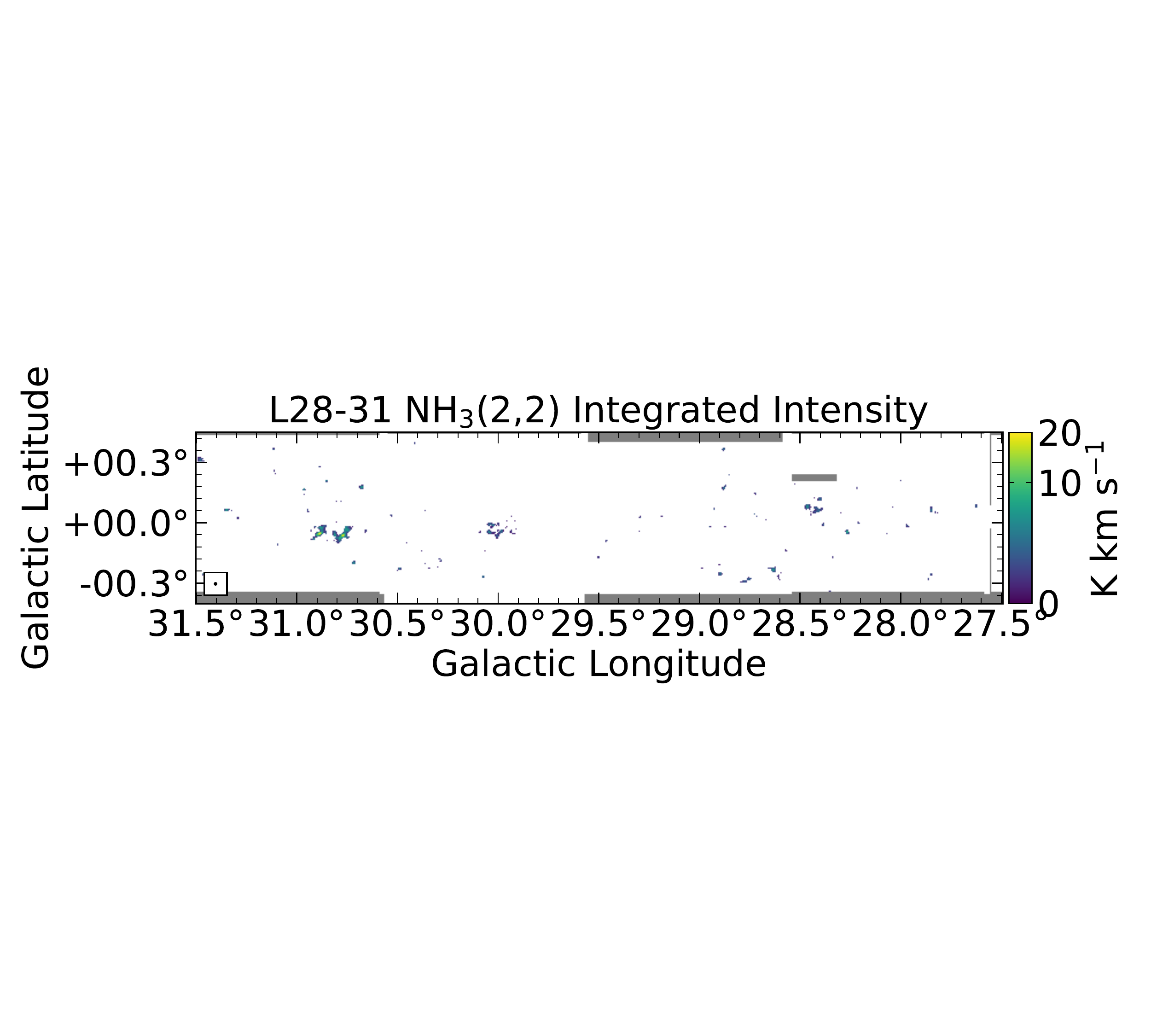}
\caption{Combined $\mathrm{NH_{3}(2,2)}$ integrated intensity map of the L28, L29, L30, and L31 fields.}
\label{fig:L28-31_22_mom0}
\end{figure}

\begin{figure}[!hbtp]
\centering
\includegraphics[scale=0.7,trim={0 3cm 0 3cm},clip]{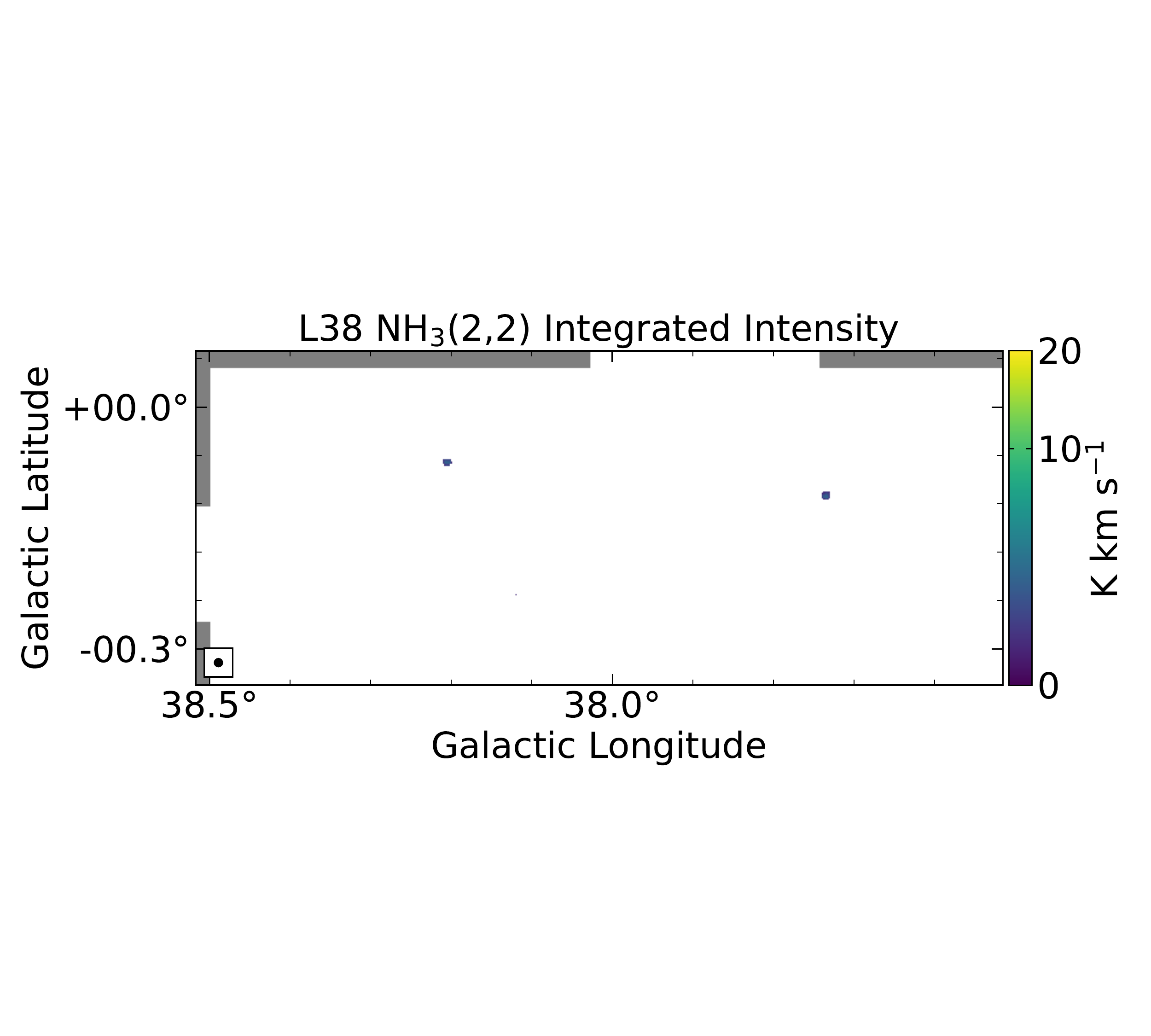}
\caption{$\mathrm{NH_{3}(2,2)}$ integrated intensity map of the L38 field.}
\label{fig:L38_22_mom0}
\end{figure}

\begin{figure}[!hbtp]
\centering
\includegraphics[scale=0.65]{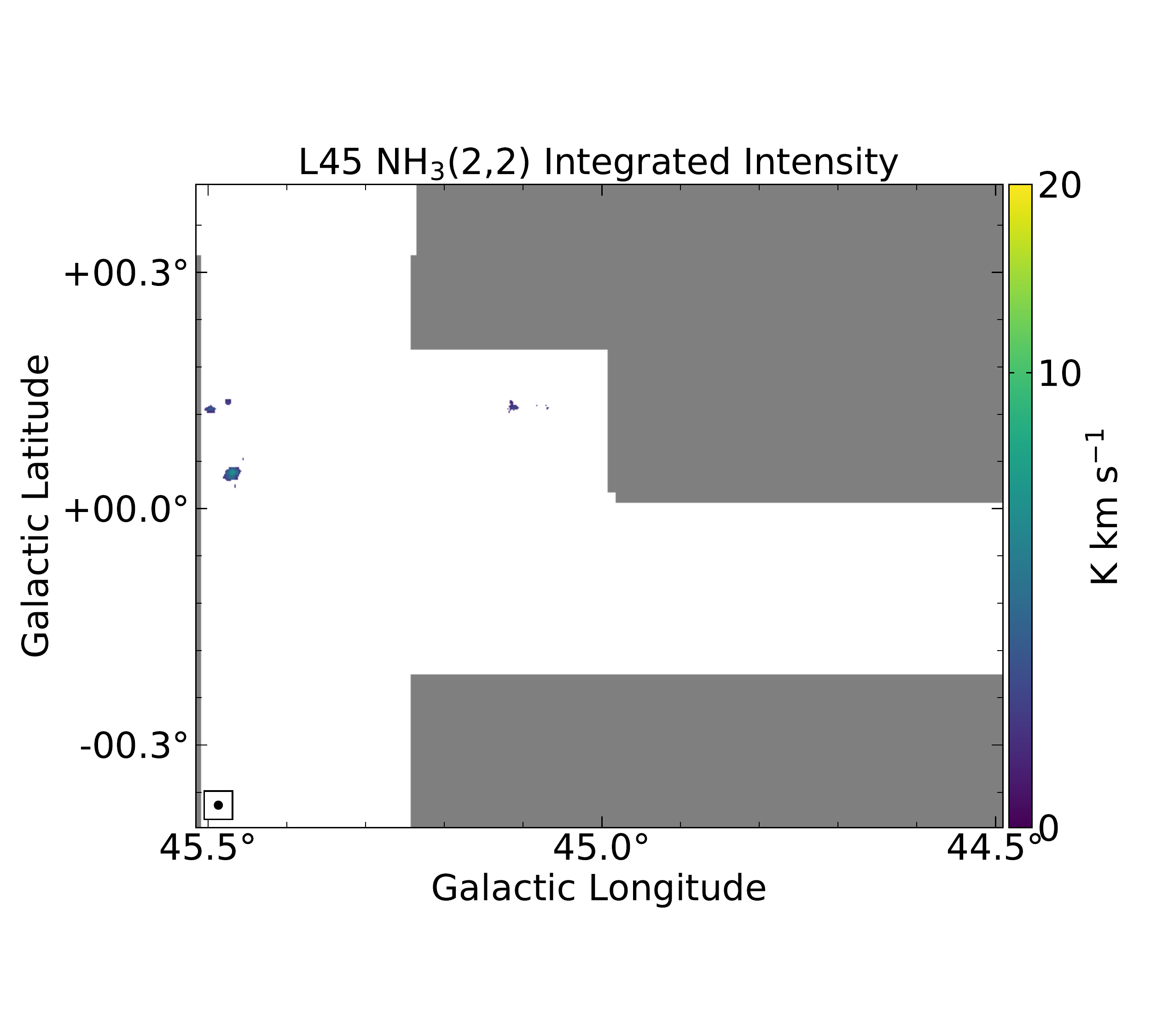}
\caption{$\mathrm{NH_{3}(2,2)}$ integrated intensity map of the L45 field.}
\label{fig:L45_22_mom0}
\end{figure}

\end{subfigures}

\begin{figure}[!hbtp]
\centering
\includegraphics[scale=0.7,trim={0 3cm 0 3cm},clip]{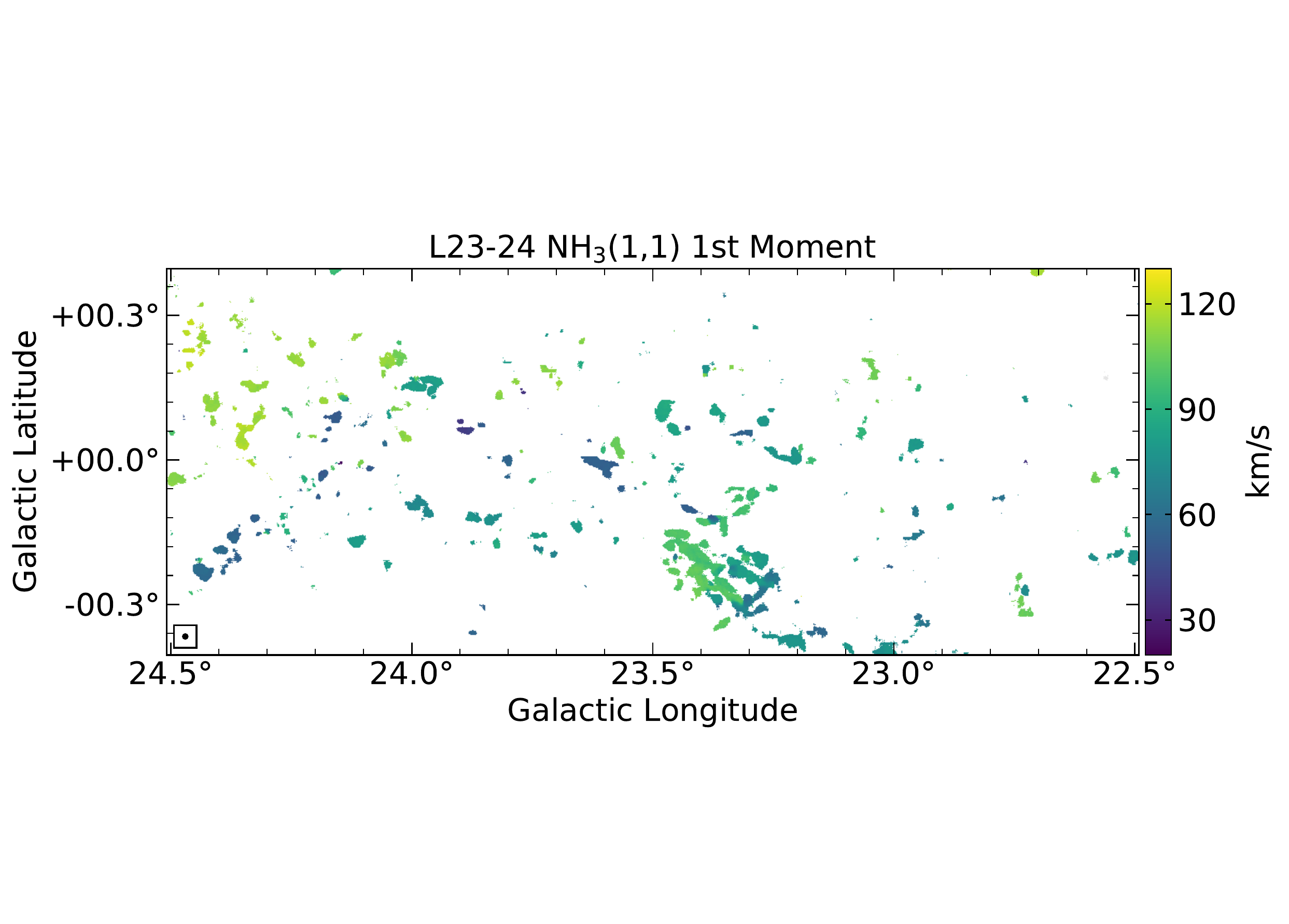}
\caption{$\mathrm{NH_{3}(1,1)}$ velocity field of the L23 and L24 fields.}
\label{fig:L23-24_11_mom1}
\end{figure}

\section{Analysis}
\label{sec:analysis}

In the following section, we report on some preliminary analysis of the RAMPS data. In Section~\ref{subsec:NH3}, we describe our methods for creating maps of $\mathrm{NH_{3}}$ rotational temperature, column density, line width, and velocity. In Section~\ref{subsec:H2O}, we discuss the $\mathrm{H_{2}O}$ data and describe our method for determining $\mathrm{H_{2}O}$ maser positions.

\subsection{$\mathrm{NH_{3}}$}
\label{subsec:NH3}

For our preliminary analysis of the RAMPS $\mathrm{NH_{3}}$ data, we fit the $\mathrm{NH_{3}}$(1,1) and $\mathrm{NH_{3}}$(2,2) spectra to determine $\mathrm{NH_{3}}$ rotational temperatures, total $\mathrm{NH_{3}}$ column densities, $\mathrm{NH_{3}}$(1,1) optical depths, $\mathrm{NH_{3}}$ line widths, and $\mathrm{NH_{3}}$ velocities. We calculated these quantities using a PySpecKit $\mathrm{NH_{3}}$ inversion line model and fitting routine. Before fitting RAMPS data with the PySpecKit $\mathrm{NH_{3}}$ model, we applied a simpler line-fitting routine to the $\mathrm{NH_{3}}$(1,1) spectra. The purpose of this initial fit was to measure the central velocity and line width for each $\mathrm{NH_{3}}$(1,1) line in order to provide more accurate initial parameters for the PySpecKit fitting routine. For the initial fit, we used our $\mathrm{NH_{3}}$(1,1) integrated intensity maps as masks for the fitting procedure in such a way that we did not attempt to fit a spectrum if there was no significant $\mathrm{NH_{3}}$(1,1) signal detected. We fit RAMPS $\mathrm{NH_{3}}$(1,1) spectra using the \texttt{optimize.curve\char`_fit} function from the \texttt{scipy} library. The \texttt{curve\char`_fit} function is a nonlinear least-squares method used to fit a function to data. As its input arguments, \texttt{curve\char`_fit} takes a model function, data, initial parameters, and parameter boundaries. We used a simple model function consisting of five Gaussians of equal width placed at the respective spacings of the main and satellite lines, where $\lq \lq$satellite line" refers to one of the four nuclear quadrupole hyperfine lines. Our model required that the two inner satellite components be equal in amplitude and likewise for the two outer satellite components. These equal intensities are expected if the hyperfine lines all have the same excitation temperature. Thus, the model contains five free parameters: the amplitude of the main line, the amplitude of the inner satellite pair, the amplitude of the outer satellite pair, the velocity of the main line, and the line width. We set the initial guess for the main line amplitude to the intensity of the brightest channel, while we set the initial guess for the inner and outer satellite amplitudes to half this value. We set the initial line width parameter to 1 $\mathrm{km \ s^{-1}}$ and the velocity parameter to the velocity of the brightest channel for each spectrum. We set sensible boundaries for each of the other parameters, which we determined from a preliminary fit to a subset of the data. The amplitude, the line width, and the velocity parameters were free to lie within the ranges $0-10$ K, $0.1-10 \ \mathrm{km \ s^{-1}}$, and $0-160 \ \mathrm{km \ s^{-1}}$, respectively. We then ran the fit routine with these initial parameters. Some of the RAMPS spectra contain two sets of lines at different velocities, which represent two different clumps along the line of sight. After the initial fit, we tested for the existence of a second velocity component by calculating the integrated intensity of the residual. If the residual satisfied our threshold for a significant detection (described in Section~\ref{sec:results}), then we attempted a two-component fit on the spectrum. If either of the main line amplitudes in the two-component fit was less than three times the noise, we used the single-component fit as the best-fit model of the spectrum. Otherwise, we calculated the reduced $\chi^2$ of the single- and two-component fits and selected the fit with the reduced $\chi^2$ closest to 1 as the best-fit model of the spectrum.

After the initial fit of the $\mathrm{NH_{3}}$(1,1) spectra, we employed the PySpecKit $\mathrm{NH_{3}}$ fitting routine. Our code utilizes a PySpecKit function called \texttt{fiteach}, which takes in $\mathrm{NH_{3}}$ inversion transition data cubes and fits them with an $\mathrm{NH_{3}}$ model. We use an LTE $\mathrm{NH_{3}}$ model; thus, we use a single rotational temperature to set all of the level populations. To create model spectra, the function uses the rotational temperature $T_{\rm{rot}}$, total $\mathrm{NH_{3}}$ column density $N_{\rm{tot}}$, line width $\sigma$, velocity $v$, the beam filling fraction $\phi$, and the ortho fraction, or fraction of $\mathrm{NH_{3}}$ in an ortho state, as input parameters. The beam filling factor is a scaling factor between 0 and 1. If the telescope beam is smaller than the smallest angular scale of the emitting source, the beam filling factor equals 1. If the emitting source is smaller than the beam, the filling factor equals the solid angle of the source divided by the solid angle of the beam. Ortho-$\mathrm{NH_{3}}$ states are those with $K = 3 n$, where $n$ is an integer including 0, while para-$\mathrm{NH_{3}}$ states have $K \neq 3 n$.  With these input parameters, the code calculates the optical depth from the transition using the equation 
\begin{equation}
\tau = \frac{N_{\rm{tot}} g}{Q_{\rm{tot}}} \frac{A_{ul}c^2}{8\pi\nu_0^2} \frac{c}{\sigma \nu_0 \sqrt{2 \pi}} \frac{1- e^{\frac{-h \nu_0}{k_B T_{\rm{rot}}}}} {1  + e^{\frac{-h \nu_0}{k_B T_{\rm{rot}}}}},
\end{equation}
where $\tau$ is the optical depth, $g$ is the statistical weight of the upper state, $Q_{\rm{tot}}$ is the molecular partition function, $A_{ul}$ is the Einstein A-coefficient, $c$ is the speed of light, $\nu_0$ is the rest frequency of the transition, $h$ is Planck's constant, and $k_B$ is the Boltzmann constant. The code uses $\tau$ to calculate $\tau(\nu)$, the optical depth profile of the magnetic hyperfine lines, by using the known statistical weights and assuming Gaussian line widths. We assume that the line widths of each of the magnetic hyperfine lines are equal. We also calculate and report $\tau_0$, the $\mathrm{NH_{3}}$(1,1) main line optical depth. The code then creates the model spectrum using the equation 
\begin{equation}
I_{\nu} =  \phi \frac{h \nu}{k_B} (1 - e^{- \tau(\nu)}) \left(\frac{1}{e^{\frac{h \nu}{k_B T_{\rm{rot}}}} - 1} - \frac{1}{e^{\frac{h \nu}{k_B T_{bkg}}} - 1}\right),
\label{eq:int}
\end{equation}
where $I_{\nu}(\nu)$ is the intensity as a function of frequency, $\phi$ is the beam filling factor, $\nu$ is the frequency, and $T_{bkg}$ is the temperature of the cosmic microwave background.

The fitting routine performs Levenberg$-$Marquardt least-squares minimization to find the best-fit parameters. We have modified the PySpecKit model to include an additional fit parameter. The original PySpecKit \texttt{ammonia\char`_model} class does not include the beam filling fraction as a fit parameter. We have modified the PySpecKit model to include the beam filling fraction as a fit parameter since we fit sources of various angular size. We determined sensible starting values for the rotational temperature $T_{\rm{rot}}$, the column density $N_{\rm{tot}}$, and the beam filling fraction $\phi$ using an initial fit on a subset of the data. We found that reasonable starting values for $T_{\rm{rot}}$, $N_{\rm{tot}}$, and $\phi$ were 18 K, $10^{15} \ \mathrm{cm^{-2}}$, and 0.1, respectively. The initial values for $\sigma$ and $v$ were best-fit parameters from the preliminary fit to the $\mathrm{NH_{3}}$(1,1) spectra. Just as with the preliminary fit, we have chosen to limit the parameter space. $T_{\rm{rot}}$, $N_{\rm{tot}}$, $\sigma$, $v$, and $\phi$ vary within the ranges $5-200$ K, $10^{12} - 10^{17} \ \mathrm{cm^{-2}}$, $0.05 - 10 \ \mathrm{km \ s^{-1}}$, $0 - 160 \ \mathrm{km \ s^{-1}}$, and $0 - 1$, respectively. For this preliminary analysis, we have fit only $\mathrm{NH_{3}}$(1,1) and $\mathrm{NH_{3}}$(2,2) spectra, both of which are para-$\mathrm{NH_{3}}$ transitions that give us no information on the ortho transitions. Consequently, we fixed the ortho fraction parameter to its equilibrium value of 0.5, although deviations from this value have been observed \citep{1999ApJ...525L.105U}. 

To ensure that we fit only pixels containing significant signal, we masked pixels that did not have a significant detection of both $\mathrm{NH_{3}}$(1,1) and (2,2). Next, we performed a single-component fit on each unmasked spectrum. In addition, we used our initial fit of the $\mathrm{NH_{3}}$(1,1) data to determine whether or not to attempt a two-component fit. For the spectra fit with a two-component model, if either of the $\mathrm{NH_{3}}$(2,2) line amplitudes was less than three times the noise, then we accepted the single-component fit as the best-fit model. Otherwise, we accepted the model with the reduced $\chi^2$ closest to 1 as the best fit. From these fits we created model $\mathrm{NH_{3}}$(1,1) and (2,2) data cubes, as well as maps of $T_{\rm{rot}}$, $N_{\rm{tot}}$, $\sigma$, $v$, $\phi$, and $\tau_0$ and their associated errors. In Figure~\ref{fig:L23-24_par_maps}, we show maps of the five fit parameters for the L23 and L24 fields. We also present a few examples of typical fit results in Figure~\ref{fig:spec_ex}.

\begin{figure}[!hbtp]
\centering
\includegraphics[scale=1.05,trim={0cm 0 0cm 0}]{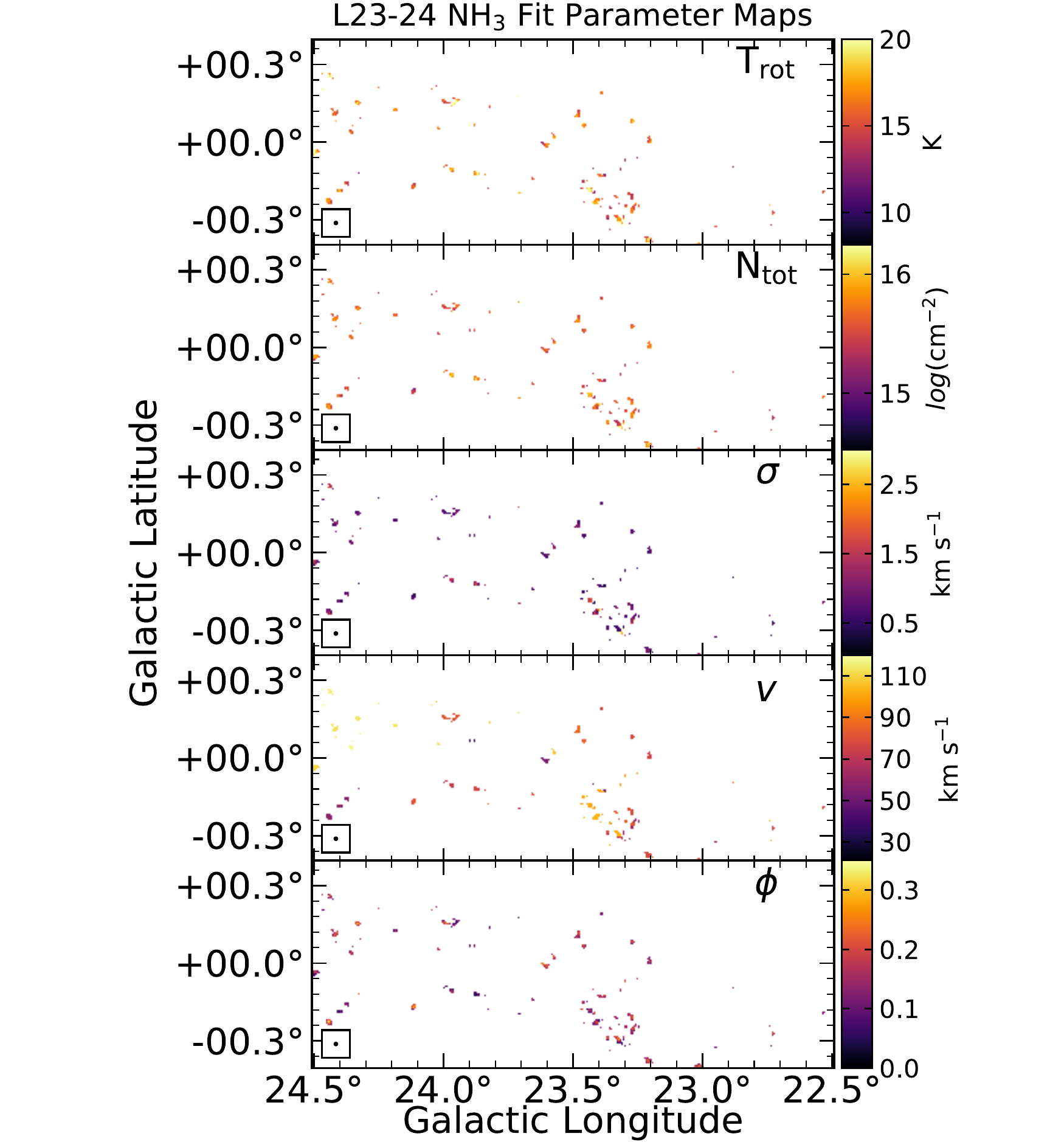}
\caption{Maps of the $\mathrm{NH_{3}}$ fit parameters for the L23 and L24 fields. We performed the fits using the $\mathrm{NH_{3}}$(1,1) and $\mathrm{NH_{3}}$(2,2) data. From top to bottom, the maps show the rotational temperature, the $\mathrm{NH_{3}}$ column density, the line width, the velocity, and the beam filling fraction. The beam size is shown in the box at the lower left corner of each map.}
\label{fig:L23-24_par_maps}
\end{figure}

\begin{figure}[!hbtp]
\centering
\includegraphics[scale=1]{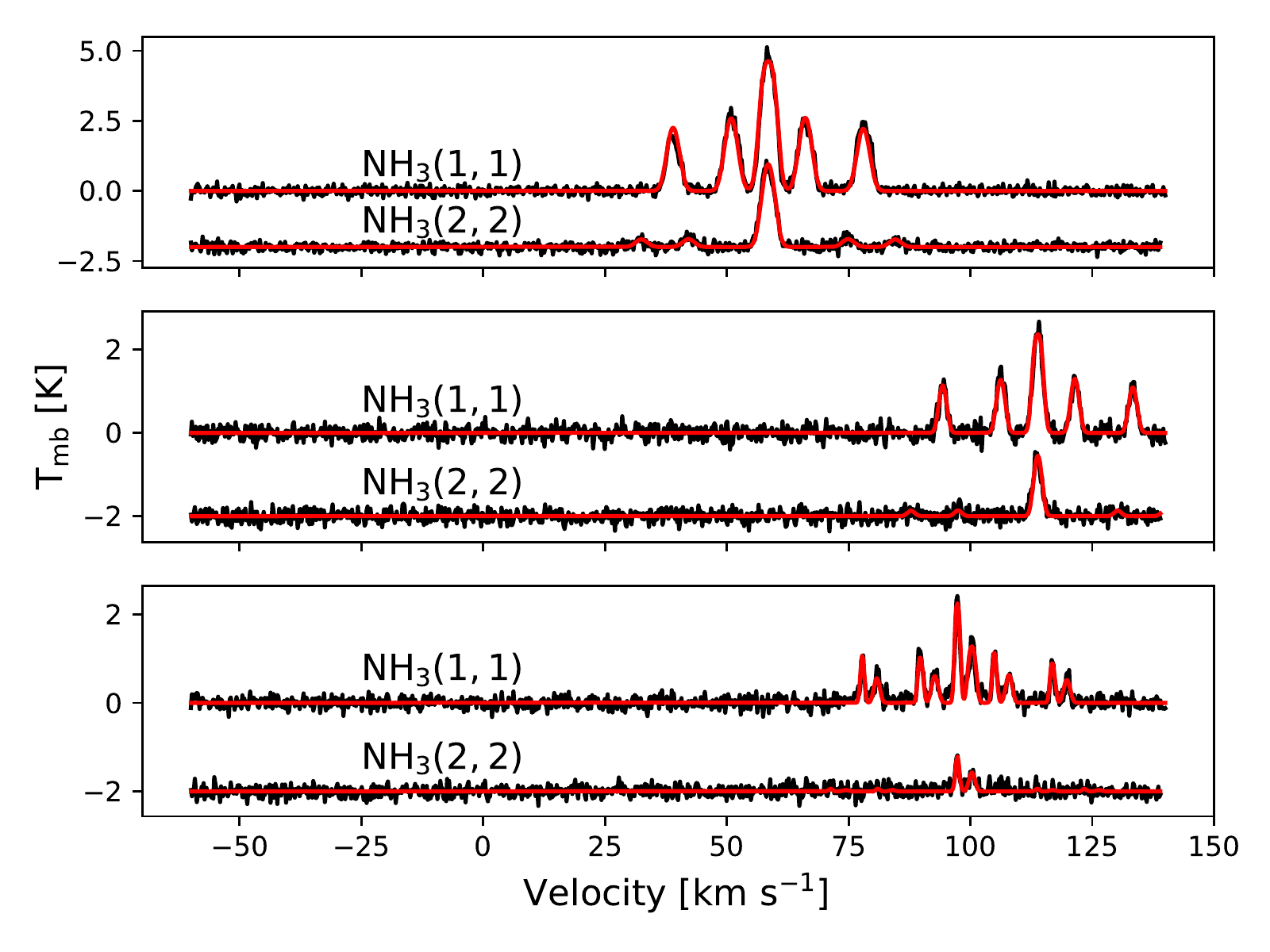}
\caption{Example fit results for three typical spectra. The $\mathrm{NH_{3}}$(1,1) and (2,2) spectra are shown in black, while the models are overplotted in red.}
\label{fig:spec_ex}
\end{figure}

We also present histograms of the best-fit parameters for all RAMPS $\mathrm{NH_{3}}$ fits in Figure~\ref{fig:par_hists}. From left to right, Figure~\ref{fig:par_hists} shows histograms of the rotational temperature, column density, line width, velocity, and beam filling fraction, with the mean of each distribution shown with a magenta line. The temperature distribution peaks at $\sim18$ K, with some fits having $T_{\rm{rot}}<10$ K and a small number of fits having $T_{\rm{rot}}>30$ K. The column density distribution appears roughly Gaussian, with a peak near $5 \times 10^{15} \mathrm{cm^{-2}}$. The line width distribution peaks at $\sim1 \ \mathrm{km \ s^{-1}}$, with another small population near $7 \ \mathrm{km \ s^{-1}}$. The velocity distribution shows several peaks, with a mean of $\sim80 \ \mathrm{km \ s^{-1}}$. The distribution of the filling fraction peaks at $\sim 0.1$, exhibits a small tail out to larger values, and has another peak at the parameter's upper limit of $\phi=1$. 

\begin{figure}[!hbtp]
\centering
\includegraphics[scale=0.35]{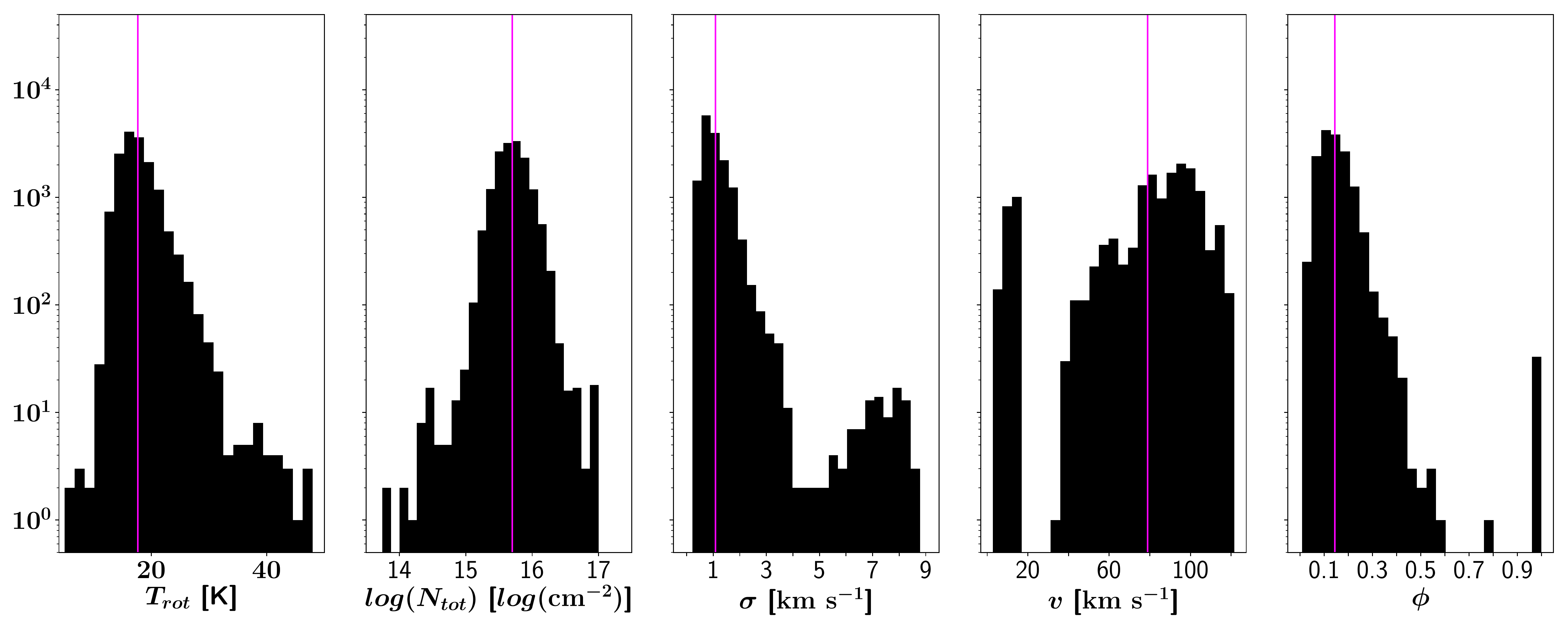}
\caption{Histograms of the $\mathrm{NH_{3}}$ fit parameters for the entire RAMPS pilot survey. The mean of each distribution is shown with a magenta line.}
\label{fig:par_hists}
\end{figure}
  
To help determine the reliability of the fit results, we plotted the 1$\sigma$ error in the parameters against the parameter values in Figures~\ref{fig:trot_err}$-$\ref{fig:tau_err}. The color scale shows the S/N of the model $\mathrm{NH_{3}}$(2,2) main line, and the overlaid dashed lines show the median of the parameter values and parameter errors. These plots reveal regions of parameter space populated by fits that do not accurately represent the data. Fits with large parameter errors or parameter values pegged at their limit are generally indicative of a failure with our fitting routine. Note that parameter values that are equal to the upper or lower limit of that parameter have no meaningful errors and are thus excluded in these plots. Figure~\ref{fig:trot_err} shows that fits with very low or very high temperatures have large errors, while the rest of the fits have temperature errors $< 3$ K. The general trend is that spectra with lower S/Ns have larger errors, although there are some exceptions to this trend at both low and high temperatures. Further investigation of these fit results revealed two relatively rare situations that can result in a poor fit. The first occurred when two molecular clumps along the line of sight were close in velocity, resulting in significant overlap of their line emission. This can add a large uncertainty to the two-component fit, especially for the fainter line component. This issue is largely responsible for the small group of fits with large errors and high S/Ns. The next issue we found was also the result of overlapping lines. In rare cases the velocity difference between two clumps was such that their satellite lines overlapped, causing the overlapping satellite lines to appear brighter than the main line. Due to the assumptions made in our fit routine, the bright satellite line was fit as if it were the main line of a single-component fit. These false main lines had no $\mathrm{NH_{3}}$(2,2) line associated with them, resulting in low best-fit temperatures and large errors on the fit parameters. Figure~\ref{fig:ntot_err} shows that at low column densities the fits separate into two distinct groups, those with small errors and those with large errors on $N_{\rm{tot}}$. This behavior at low column densities is likely the result of a degeneracy in our model, which is due to the dependence on $N_{\rm{tot}}$ and $\phi$ on the modeled line intensity. Equation~\ref{eq:int} shows that $I_{\nu} \propto \phi (1-e^{- \tau})$, but in the optically thin limit ($\tau \ll 1$) the dependence on $\tau$ becomes linear. Figure~\ref{fig:ntot_err} shows that the fits start to become degenerate near $N_{\rm{tot}} \sim 2 \times 10^{15} \mathrm{cm^{-2}}$, so fits above this limit have reliable values for $N_{\rm{tot}}$. Although fits below this limit have unreliable values for $N_{\rm{tot}}$, the beam-averaged column density can be obtained by taking the product of the best-fit $N_{\rm{tot}}$ and $\phi$. The error in line widths is plotted in Figure~\ref{fig:sigma_err}, which shows that fits with large line widths also tend to have larger errors on those line widths. The fits with large errors on their line widths and large S/Ns are mostly fits that have attempted to perform a single-component fit on two velocity components that are close in velocity, resulting in a larger error in $\sigma$. There is also a group of fits with $\sigma \sim 7 \ \mathrm{km \ s^{-1}}$, which corresponds to a particular source, G23.33-0.30 (Figure~\ref{fig:G23_ex}), that was previously observed by HOPS. Although this source also appears to consist of two velocity components that have been fit by a single-component model, the line widths are intrinsically much broader than the typical $\mathrm{NH_{3}}$ line and are thus well separated from the rest of the fits. High angular resolution observations are needed to determine the nature of the large line widths. Figure~\ref{fig:vel_err} shows that errors on the velocities are generally small compared to the typically measured line widths. Measuring accurate clump velocities is necessary to determine their kinematic distances and to resolve the kinematic distance ambiguity \citep{2017AJ....154..140W}. We plot the error in the filling fraction in Figure~\ref{fig:fillfrac_err}, which shows that the filling fraction is not well constrained for a small portion of the fits. The fits with poorly constrained $\phi$ are the same fits with small values of $N_{\rm{tot}}$ that are also poorly constrained owing to the degeneracy between $\phi$ and $N_{\rm{tot}}$. There are also a handful of fits with both low $\phi$ and low error in $\phi$. These were the result of spectra with similar $\mathrm{NH_{3}}$(1,1) and $\mathrm{NH_{3}}$(2,2) line amplitudes, potentially indicating a hot component. Since these lines were not very bright, it forced the filling fraction to be low to account for the small line amplitudes. Figure~\ref{fig:tau_err} shows a few distinct groups of main line optical depth ($\tau_0$) values. Most of the fits return $\tau_0 \sim 1-3$, but there are a group of fits with small $\tau_0$ and small error in $\tau_{0}$, as well as several fits with a large $\tau_0$ and a large error in $\tau_{0}$. The fits with very small $\tau$ are the degenerate fits that have $\phi = 1$.

\begin{subfigures}
\begin{figure}[!hbtp]
\centering
\includegraphics[scale=0.6]{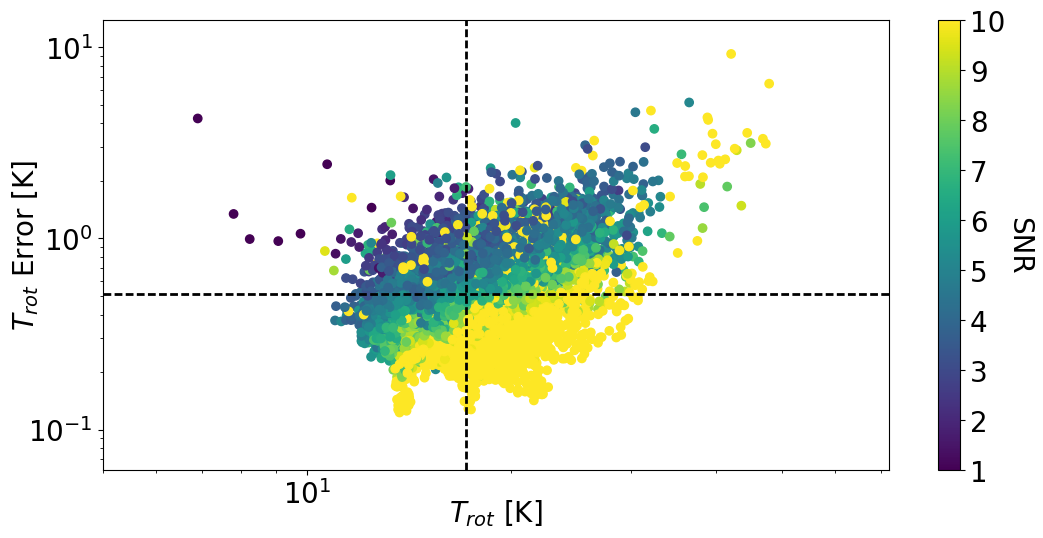}
\caption{Plot of the error in the rotational temperature against the rotational temperature for each fit. The median of the parameter value and the error in the parameter are represented by the dashed lines. The color corresponds to the S/N of the model $\mathrm{NH_{3}}$(2,2) main line.}
\label{fig:trot_err}
\end{figure}
 
\begin{figure}[!hbtp]
\centering
\includegraphics[scale=0.6]{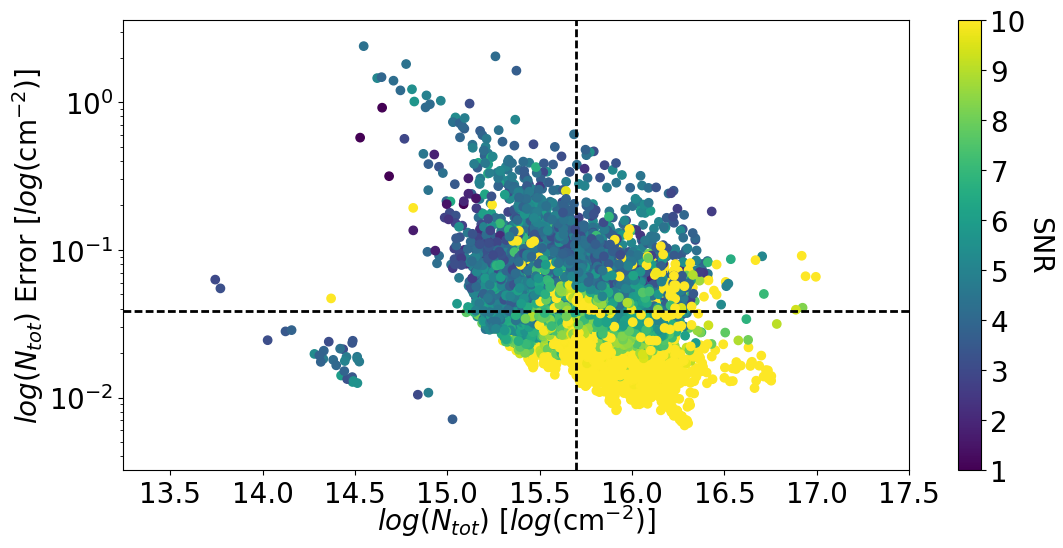}
\caption{Plot of the error in the column density against the column density for each fit. The median of the parameter value and the error in the parameter are represented by the dashed lines. The color corresponds to the S/N of the model $\mathrm{NH_{3}}$(2,2) main line. There are two relatively distinct groups of fits, which are discussed in Section~\ref{subsec:NH3}.}
\label{fig:ntot_err}
\end{figure}
 
\begin{figure}[!hbtp]
\centering
\includegraphics[scale=0.6]{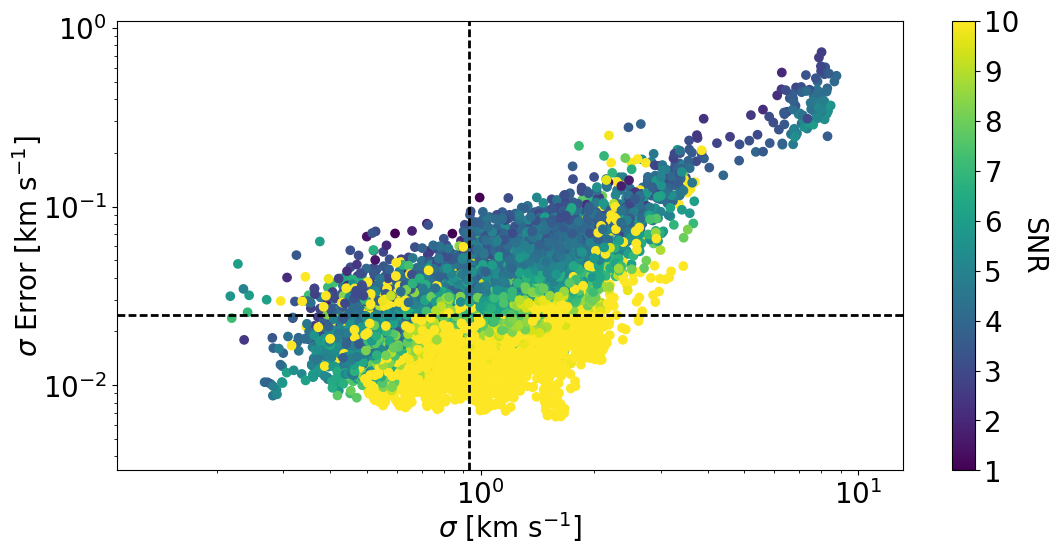}
\caption{Plot of the error in the line width against the line width for each fit. The median of the parameter value and the error in the parameter are represented by the dashed lines. The color corresponds to the S/N of the model $\mathrm{NH_{3}}$(2,2) main line.}
\label{fig:sigma_err}
\end{figure}
 
\begin{figure}[!hbtp]
\centering
\includegraphics[scale=0.6]{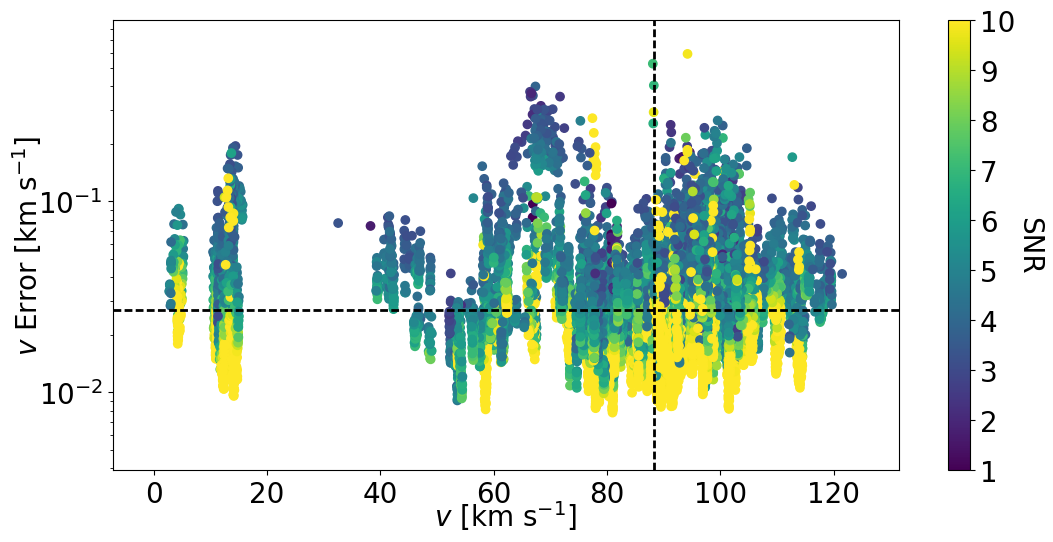}
\caption{Plot of the error in the velocity against the velocity for each fit. The median of the parameter value and the error in the parameter are represented by the dashed lines. The color corresponds to the S/N of the model $\mathrm{NH_{3}}$(2,2) main line.}
\label{fig:vel_err}
\end{figure}
 
\begin{figure}[!hbtp]
\centering
\includegraphics[scale=0.6]{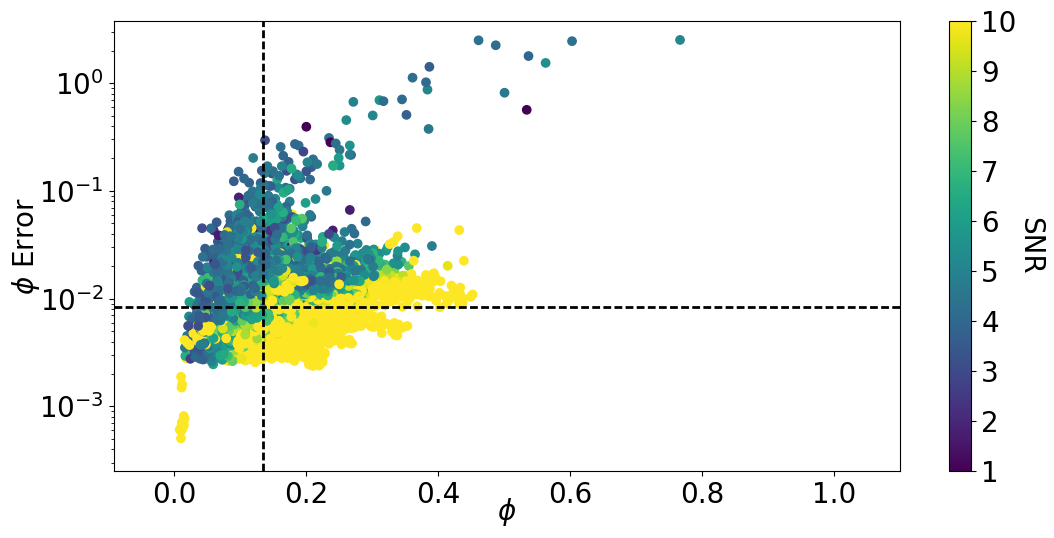}
\caption{Plot of the error in the filling fraction against the filling fraction for each fit. The median of the parameter value and the error in the parameter are represented by the dashed lines. The color corresponds to the S/N of the model $\mathrm{NH_{3}}$(2,2) main line.}
\label{fig:fillfrac_err}
\end{figure}

\begin{figure}[!hbtp]
\centering
\includegraphics[scale=0.6]{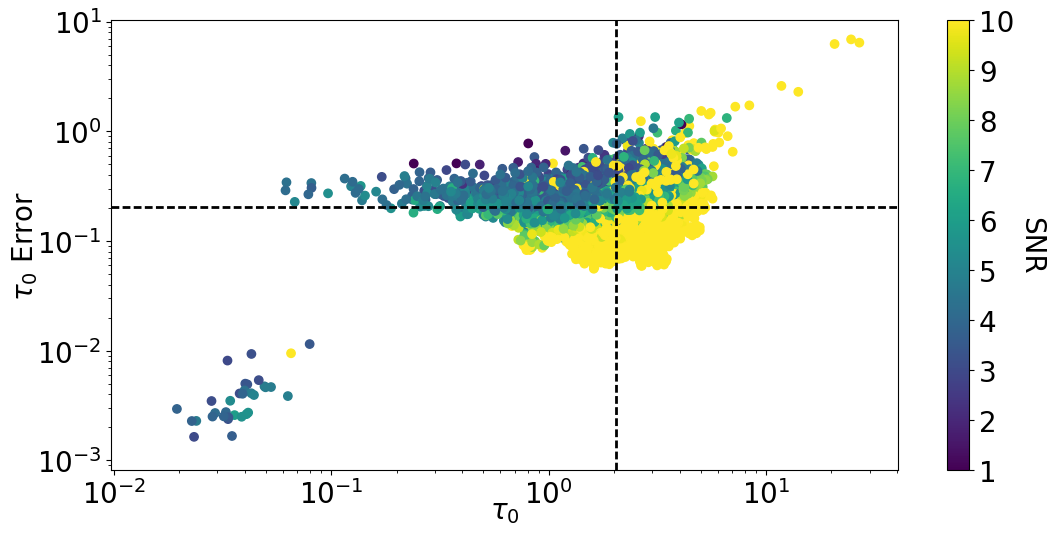}
\caption{Plot of the error in the $\mathrm{NH_{3}}$(1,1) main line optical depth against the $\mathrm{NH_{3}}$(1,1) main line optical depth for each fit. The median of the parameter value and the error in the parameter are represented by the dashed lines. The color corresponds to the S/N of the model $\mathrm{NH_{3}}$(2,2) main line.}
\label{fig:tau_err}
\end{figure}
\end{subfigures}

\begin{figure}[!hbtp]
\centering
\includegraphics[scale=1]{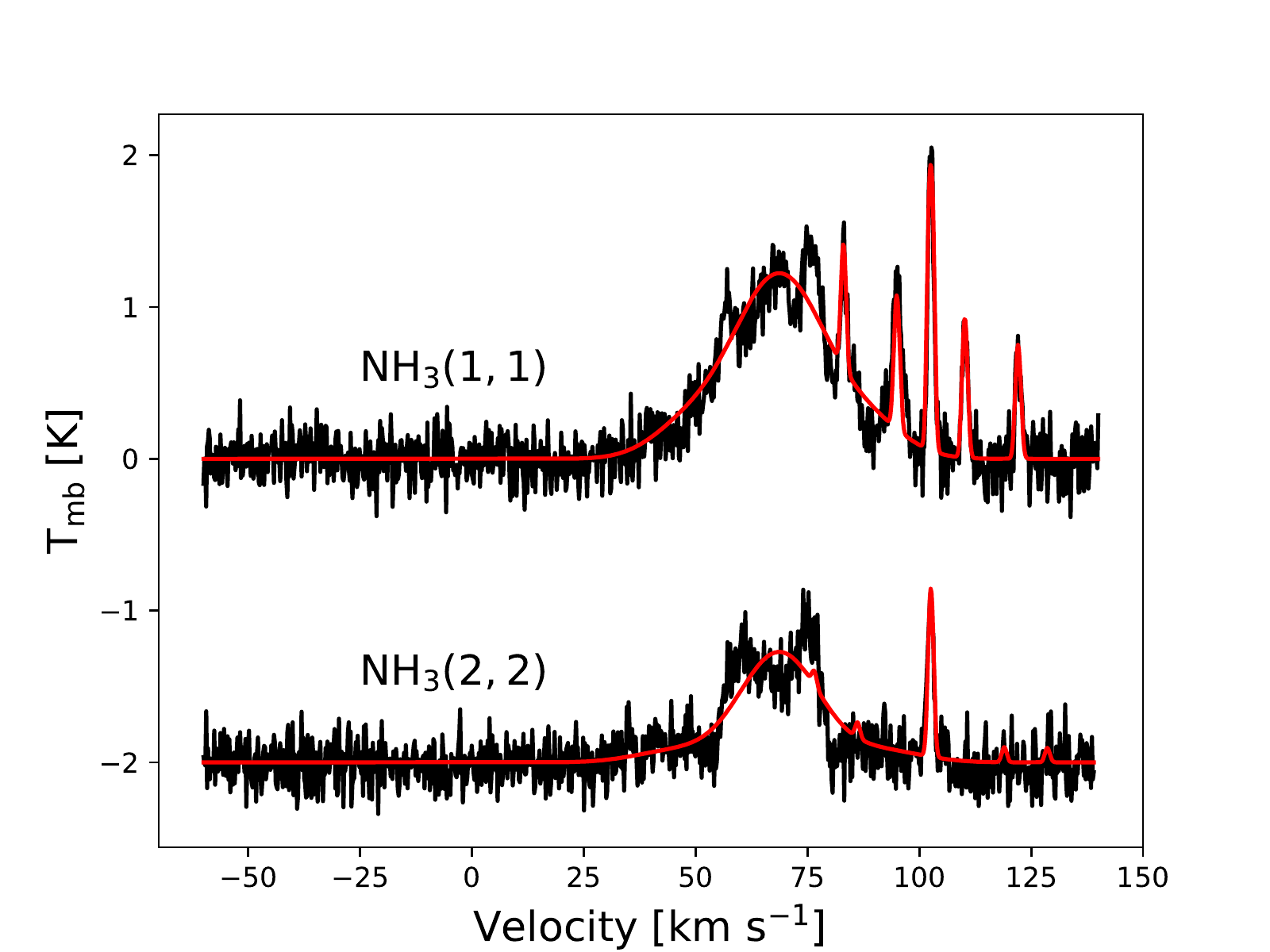}
\caption{$\mathrm{NH_{3}}$(1,1) and (2,2) spectra are shown in black, and best-fit models for the source G23.33-0.30 (near $v=70 \ \mathrm{km \ s^{-1}}$) and another source near $v=100 \ \mathrm{km \ s^{-1}}$ are overplotted in red. This source is unique for its large best-fit $\sigma$, which is the result of two velocity components that have been fit by a single-component model, as well as G23.33-0.30's intrinsically large line widths.}
\label{fig:G23_ex}
\end{figure}

The model degeneracy for small values of $\tau$ is best illustrated by Figure~\ref{fig:ntoterr_vs_err_fillfrac_subplots}, which again shows the error in the column density plotted against the column density, but now with the symbol color representing the filling fraction and error in the filling fraction. All of the fits with low column densities and low column density errors have their filling fractions pegged at the upper limit of $\phi=1$. These fits are not plotted on the right of Figure~\ref{fig:ntoterr_vs_err_fillfrac_subplots} because they do not have meaningful errors. Fits with low column densities and large errors on their column density have moderate values for the filling fraction, but these values are completely unconstrained. Thus, the small number of fits with large errors on $\phi$ have unreliable values for $N_{\rm{tot}}$ but can still provide the beam-averaged column density ($\phi N_{\rm{tot}}$). Another potential issue in our model is our LTE assumption. In LTE, the amplitude ratios of the two inner satellite lines and the two outer satellite lines of the $\mathrm{NH_{3}}$(1,1) transition are unity. The departure from LTE hyperfine line amplitude ratios is referred to as the hyperfine intensity anomaly \citep[HIA; see ][ and references therein]{2015ApJ...806...74C}. The HIA is ubiquitous in high-mass SFRs; thus, a significant number of our $\mathrm{NH_{3}}$(1,1) spectra exhibit this effect. The LTE $\mathrm{NH_{3}}$ model attempted to fit these spectra assuming that the amplitude ratios of the inner and outer satellite pairs were unity; thus, two of the satellite lines were fit with Gaussians that were larger than the expected amplitude, while the other two satellite lines were fit with Gaussians that were smaller than the expected amplitude. Although the cause of the HIA is not well understood, \citet{1977ApJ...214L..67M} proposed that the HIA is the result of selective trapping in the hyperfine structure of the $\mathrm{NH_{3}}$(2,1)-(1,1) transition. This has the effect of shifting photons from the left outer to the right outer satellite line, while simultaneously shifting photons from the right inner to the left inner satellite line. Because this mechanism is not expected to change the average amplitude of the inner or outer satellite line, we do not expect that this anomaly will have a large affect on the accuracy of our fits that assume LTE. We refer readers to \citet{1985A&A...144...13S} for a more detailed discussion of the HIA. Because this is a preliminary analysis of the RAMPS data, users of the dataset in its current form are cautioned to make use of the errors on the parameter values and the reduced $\chi^{2}$ of the fits to determine the reliability of the best-fit parameters.

\begin{figure}[!hbtp]
\centering
\includegraphics[scale=0.55,trim={2cm 1.5cm 1cm 0.5cm}]{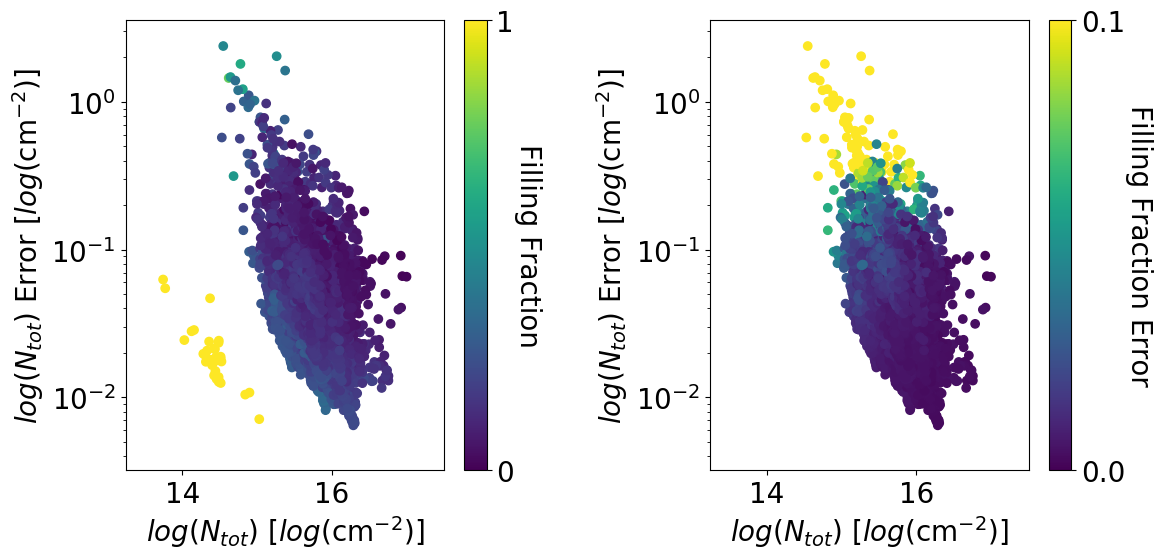}
\caption{Plot of the error in the column density against the column density. Left: the color mapping shows the filling fraction. Right: the color mapping shows the error in the filling fraction. Fits with low column densities and column density errors have their filling fractions pegged at the upper limit. Fits with low column densities and large errors on column densities have large errors on their filling fractions. This likely represents a degeneracy in the model between the column density and filling fraction parameters.}
\label{fig:ntoterr_vs_err_fillfrac_subplots}
\end{figure}

\subsection{$\mathrm{H_{2}O}$ Masers}
\label{subsec:H2O}

For the preliminary analysis of the RAMPS $\mathrm{H_{2}O}$ data, we focused on determining maser positions by calculating their integrated intensity-weighted positions. We began by creating integrated intensity maps of the $\mathrm{H_{2}O}$ data cubes using a similar method to that we used for the $\mathrm{NH_{3}}$ data. To help separate closely spaced masers when determining positions, we created integrated intensity maps of the brightest maser line in each spectrum by calculating the integral in a 1 $\mathrm{km \ s^{-1}}$ window around the brightest channel in each spectrum. For overlapping masers, we created an integrated intensity map for both masers and used these to find the positions of each maser. To reduce the effect of noise on the integrated intensity, we utilized a similar masking method as that used for the $\mathrm{NH_{3}}$ integrated intensity maps. This method masks all channels with intensity less than 3$\sigma$, as well as those channels that were not contiguous with at least two other masked channels. We then summed over the channels that were unmasked. 

Our method of locating $\mathrm{H_{2}O}$ masers produced a few suspected false detections. These false detections generally showed significant signal in only one pixel or in a few pixels that were contiguous in the Galactic latitude direction. This is a result of the coarser sampling in the Galactic latitude direction of our maps, which also degrades the spatial resolution of the $\mathrm{H_{2}O}$ maps in that direction. We suspected that these were false detections because they had low intensities and showed a relatively uniform velocity distribution, which are attributes we expect from random noise. The suspected false detections did not usually exhibit significant signal in two pixels that were contiguous in the Galactic longitude direction. To mask these pixels, we required that each unmasked pixel have an unmasked neighbor in the Galactic longitude direction for the $\mathrm{H_{2}O}$ integrated intensity maps. 

After creating the integrated intensity maps, we calculated the integrated intensity-weighted position of each maser using the equations
\begin{equation}
l = \frac{\sum\limits_{i} l_{i} I_{i}}{\sum\limits_{i} I_{i}}
\end{equation}
\begin{equation}
b = \frac{\sum\limits_{i} b_{i} I_{i}}{\sum\limits_{i} I_{i}}
\end{equation}
where $l$ is the calculated Galactic longitude, $b$ is the calculated Galactic latitude, $l_i$ is the Galactic longitude of a pixel, $b_i$ is the Galactic latitude of a pixel, and $I_i$ is the integrated intensity of a pixel. We estimated the error in the positions using 
\begin{equation}
\sigma_{l} = \sqrt{\sum\limits_{i} (l_i - l)^2 (\frac{\sigma_{I_i}}{\sum\limits_{i} I_i})^2 + \sigma_{\rm{pointing}}^2}
\end{equation}
\begin{equation}
\sigma_{b} = \sqrt{\sum\limits_{i} (b_i - b)^2 (\frac{\sigma_{I_i}}{\sum\limits_{i} I_i})^2 + \sigma_{\rm{pointing}}^2}
\end{equation}
where $\sigma_l$ is the error in the calculated Galactic longitude, $\sigma_b$ is the error in the calculated Galactic latitude, $\sigma_{I_i}$ is the error in the integrated intensity, and $\sigma_{\rm{pointing}}$ is the 1D error in the pointing of the telescope. The error in the integrated intensity ($\sigma_{I_i}$) is given by 
\begin{equation}
\sigma_{I_i} = I_i \sqrt{N}
\end{equation}
where $N$ is the number of unmasked channels used to calculate $I_i$. The error in the pointing of the telescope is estimated from the average wind speed during our observing period (GBT Dynamic Scheduling Project Note 18.1).  A conservative estimate of a 5 $\mathrm{m \ s^{-1}}$ wind speed during the daytime at a height of 10 m above the ground results in $\sigma_{\rm{pointing}} \approx 4.2\arcsec$. The error in the telescope pointing dominates the error in the maser positions, such that $\sigma_l \approx \sigma_b \approx \sigma_{\rm{pointing}}$.

The 22.235 GHz $\mathrm{H_{2}O}$ maser transition is associated with both SFRs and AGB stars \citep{1981ARA&A..19..231R}. As part of our preliminary analysis, we estimated the associated environment of all of the $\mathrm{H_{2}O}$ masers detected with the RAMPS pilot survey. If an $\mathrm{H_{2}O}$ maser is spatially coincident with $\mathrm{NH_{3}}$ emission and there is $\mathrm{H_{2}O}$ maser emission within 30 $\mathrm{km \ s^{-1}}$ of the $\mathrm{NH_{3}}$ velocity, we inferred that the maser is associated with an SFR. Although the 30 $\mathrm{km \ s^{-1}}$ velocity criterion is somewhat arbitrary, masers with no emission near the clump velocity are unlikely to be associated with an SFR within the clump. For masers that are not associated with $\mathrm{NH_{3}}$ emission, we checked for the presence of a compact 24 $\mathrm{\mu m}$ source in MIPSGAL data \citep{2009PASP..121...76C}. This emission feature probably represents a large, red AGB star. Masers that are associated with neither SFRs nor AGB stars have an unknown association. 

In Table~\ref{table:masers} we present data for all 325 $\mathrm{H_{2}O}$ maser sites detected during the RAMPS pilot survey. Table~\ref{table:masers} gives the maser positions, the errors on the positions, the velocity and intensity of the brightest maser line, and the associated environment of each maser. We found that out of 325 detected masers, 185 ($57\pm4$\%) are associated with an SFR, 92 ($28\pm5$\%) are associated with an AGB star, and 48 ($15\pm5$\%) have an unknown association. Figures~\ref{fig:maser_Tb_hist} and \ref{fig:maser_vel_hist} show histograms of the maser intensities and velocities for the full sample of masers, and histograms of the masers separated by association, respectively. Figure~\ref{fig:maser_Tb_hist} shows that the slopes of the maser intensity distributions look roughly like a power law past the peak of each distribution. The sharp cutoff at $\sim1$ K in each distribution represents our completeness limit. This number is expected, since the typical 1$\sigma$ rms in our $\mathrm{H_{2}O}$ spectra is $\sim 0.4$ K and we require that each detected maser has an intensity of at least three times the noise. While the intensity distributions of the various maser groups look similar, the distribution associated with SFRs shows several more masers with intensities in excess of 100 K. Figure~\ref{fig:maser_vel_hist} shows that we have detected masers predominantly at positive velocities, particularly the masers associated with SFRs. On the other hand, both masers associated with AGB stars and those with an unknown association have a much broader spread in their velocity distributions. 

\begin{deluxetable}{ccccccc}
\tabletypesize{\footnotesize}
\tablewidth{0pt}
\tablecolumns{7}

\tablecaption{RAMPS $\mathrm{H_{2}O}$ Masers\label{table:masers}} 
\tablehead{\colhead{$l$} &
		  \colhead{$b$}&
		  \colhead{Error on $l$}&
		  \colhead{Error on $b$}&
		  \colhead{Velocity}&
		  \colhead{Intensity}&
		  \colhead{Association} \\
		  \colhead{(deg)} &
		  \colhead{(deg)}&
		  \colhead{(arcsec)}&
		  \colhead{(arcsec)}&
		  \colhead{($\mathrm{km \ s^{-1}}$)}&
		  \colhead{(K)}&
		  \colhead{}}
\startdata
9.621 & 0.194 & 4.2 & 4.2 & 6.1 & 230.4 & SFR\\
9.651 & -0.06 & 4.2 & 4.2 & 49.5 & 10.6 & AGB\\
9.731 & -0.142 & 4.2 & 4.2 & -15.5 & 2.7 & AGB\\
9.777 & -0.021 & 4.2 & 4.21 & 34.1 & 1.8 & AGB\\
9.829 & -0.2 & 4.2 & 4.21 & 18.5 & 3.9 & SFR\\
9.912 & -0.348 & 4.2 & 4.21 & 11.5 & 2.9 & AGB\\
9.92 & -0.125 & 4.2 & 4.21 & 116.9 & 1.6 & AGB\\
9.961 & -0.369 & 4.2 & 4.2 & -13.3 & 3.6 & SFR\\
9.986 & -0.029 & 4.2 & 4.2 & 48.1 & 13.3 & SFR\\
10.001 & -0.193 & 4.22 & 4.21 & -62.2 & 3.1 & U\\
10.004 & -0.193 & 4.2 & 4.2 & -58.7 & 4.2 & U\\
10.02 & -0.393 & 4.2 & 4.2 & 9.2 & 74.3 & SFR\\
10.054 & -0.077 & 4.2 & 4.2 & 68.7 & 2.1 & U\\
10.072 & -0.095 & 4.2 & 4.2 & 26.1 & 2.6 & AGB\\
10.072 & -0.095 & 4.2 & 4.2 & 22.9 & 4.0 & AGB\\
10.249 & -0.111 & 4.2 & 4.2 & 8.1 & 2.1 & SFR\\
10.271 & -0.138 & 4.22 & 4.2 & 49.7 & 1.7 & SFR\\
10.285 & -0.117 & 4.2 & 4.21 & 13.0 & 4.2 & SFR\\
10.287 & -0.125 & 4.2 & 4.2 & 15.3 & 13.8 & SFR\\
10.334 & -0.148 & 4.2 & 4.2 & 16.8 & 3.2 & SFR\\
10.341 & -0.143 & 4.2 & 4.2 & 8.5 & 17.5 & SFR\\
10.385 & -0.014 & 4.2 & 4.2 & 46.0 & 3.8 & AGB\\
10.444 & -0.019 & 4.2 & 4.2 & 71.8 & 31.4 & SFR\\
10.472 & 0.027 & 4.2 & 4.2 & 59.3 & 163.0 & SFR\\
10.472 & 0.027 & 4.2 & 4.2 & 88.6 & 98.4 & SFR \\
22.538 & -0.032 & 4.2 & 4.2 & 35.6 & 2.4 & U\\
22.569 & -0.388 & 4.2 & 4.2 & 62.4 & 5.2 & AGB\\
22.595 & 0.211 & 4.2 & 4.2 & 45.5 & 1.6 & U\\
22.739 & 0.229 & 4.2 & 4.2 & -83.8 & 4.7 & AGB\\
22.74 & -0.242 & 4.21 & 4.21 & 105.1 & 1.0 & SFR\\
22.823 & -0.165 & 4.2 & 4.2 & 39.6 & 11.5 & AGB\\
22.827 & -0.031 & 4.2 & 4.2 & 62.1 & 4.1 & AGB\\
22.896 & -0.001 & 4.2 & 4.2 & 64.2 & 5.1 & SFR\\
22.896 & -0.001 & 4.2 & 4.2 & 62.7 & 6.4 & SFR\\
22.909 & 0.074 & 4.2 & 4.2 & 23.8 & 6.8 & AGB\\
22.974 & -0.378 & 4.2 & 4.2 & 75.4 & 11.8 & SFR\\
23.011 & -0.397 & 4.2 & 4.2 & 48.7 & 11.8 & SFR\\
23.013 & -0.397 & 4.27 & 4.2 & 72.9 & 0.9 & SFR\\
23.035 & -0.279 & 4.2 & 4.21 & 106.3 & 1.8 & U\\
23.044 & 0.198 & 4.2 & 4.2 & 107.9 & 1.5 & SFR\\
23.047 & -0.148 & 4.21 & 4.21 & 84.1 & 2.5 & AGB\\
23.127 & -0.147 & 4.2 & 4.2 & 119.7 & 10.4 & AGB\\
23.165 & -0.383 & 4.2 & 4.2 & 57.4 & 1.7 & AGB\\
23.209 & -0.377 & 4.2 & 4.2 & 79.1 & 930.8 & SFR\\
23.268 & 0.077 & 4.2 & 4.2 & 105.7 & 2.7 & SFR\\
23.298 & -0.254 & 4.2 & 4.21 & 103.9 & 1.8 & SFR\\
23.352 & -0.14 & 4.2 & 4.21 & 94.3 & 1.9 & SFR\\
23.395 & -0.221 & 4.2 & 4.2 & 101.6 & 3.1 & SFR\\
23.415 & -0.108 & 4.2 & 4.2 & 55.6 & 10.0 & SFR\\
23.416 & -0.239 & 4.2 & 4.22 & 55.5 & 1.7 & SFR\\
23.419 & -0.239 & 4.21 & 4.21 & 76.3 & 1.7 & SFR\\
23.436 & -0.185 & 4.2 & 4.2 & 107.2 & 3.4 & SFR\\
23.448 & -0.255 & 4.2 & 4.2 & 54.1 & 1.8 & U\\
23.454 & -0.2 & 4.2 & 4.2 & 61.7 & 3.6 & SFR\\
23.457 & -0.018 & 4.2 & 4.2 & 76.3 & 5.5 & SFR\\
23.484 & 0.096 & 4.2 & 4.2 & 83.9 & 3.7 & SFR\\
23.515 & -0.02 & 4.2 & 4.21 & 95.1 & 1.8 & SFR\\
23.518 & -0.049 & 4.2 & 4.2 & 106.8 & 8.0 & SFR\\
23.569 & -0.137 & 4.2 & 4.2 & 22.0 & 3.4 & AGB\\
23.629 & 0.031 & 4.2 & 4.21 & 42.8 & 2.1 & SFR\\
23.653 & -0.016 & 4.2 & 4.22 & 128.7 & 1.6 & U\\
23.704 & 0.184 & 4.2 & 4.21 & 110.2 & 1.8 & SFR\\
23.732 & 0.298 & 4.2 & 4.2 & 93.3 & 2.8 & AGB\\
23.742 & -0.158 & 4.2 & 4.2 & 81.6 & 10.8 & SFR\\
23.75 & 0.29 & 4.2 & 4.21 & 129.9 & 1.0 & U\\
23.818 & 0.383 & 4.2 & 4.2 & 80.7 & 13.1 & AGB\\
23.818 & 0.384 & 4.2 & 4.2 & 64.7 & 3.5 & AGB\\
23.845 & -0.124 & 4.2 & 4.2 & 80.2 & 1.4 & SFR\\
23.868 & -0.122 & 4.21 & 4.2 & 82.0 & 0.8 & SFR\\
23.868 & -0.127 & 4.21 & 4.21 & 82.0 & 1.0 & SFR\\
23.87 & -0.091 & 4.2 & 4.21 & 15.0 & 1.0 & U\\
23.885 & 0.06 & 4.2 & 4.2 & 44.2 & 3.3 & SFR\\
23.907 & 0.072 & 4.2 & 4.2 & 70.6 & 8.1 & AGB\\
23.93 & -0.063 & 4.2 & 4.2 & 47.0 & 6.4 & U\\
23.936 & -0.153 & 4.22 & 4.2 & 110.2 & 0.8 & AGB\\
23.95 & 0.154 & 4.2 & 4.2 & 82.2 & 33.4 & SFR\\
23.962 & 0.136 & 4.2 & 4.2 & 63.7 & 1.8 & SFR\\
23.966 & -0.11 & 4.2 & 4.2 & 75.6 & 11.8 & SFR\\
23.97 & -0.164 & 4.2 & 4.2 & 87.2 & 2.1 & AGB\\
23.991 & 0.25 & 4.2 & 4.2 & 102.5 & 3.5 & AGB\\
23.991 & 0.25 & 4.2 & 4.2 & 97.1 & 4.4 & AGB\\
23.996 & -0.099 & 4.2 & 4.2 & 47.0 & 5.8 & SFR\\
23.998 & 0.117 & 4.21 & 4.21 & 123.6 & 1.2 & SFR\\
24.0 & 0.127 & 4.2 & 4.2 & -27.5 & 1.5 & AGB\\
24.014 & 0.047 & 4.2 & 4.2 & 102.8 & 5.4 & SFR\\
24.022 & 0.145 & 4.2 & 4.2 & -57.7 & 8.1 & U\\
24.047 & -0.215 & 4.2 & 4.2 & 81.3 & 3.4 & SFR\\
24.113 & -0.172 & 4.2 & 4.2 & 78.9 & 2.5 & SFR\\
24.12 & 0.141 & 4.2 & 4.2 & 97.4 & 2.3 & SFR\\
24.152 & -0.009 & 4.2 & 4.21 & 24.4 & 1.2 & SFR\\
24.158 & 0.167 & 4.2 & 4.2 & 71.1 & 3.4 & SFR\\
24.162 & 0.163 & 4.21 & 4.2 & 98.3 & 0.9 & SFR\\
24.162 & -0.019 & 4.2 & 4.2 & 96.3 & 1.5 & SFR\\
24.188 & -0.033 & 4.21 & 4.21 & 61.8 & 0.8 & SFR\\
24.232 & 0.298 & 4.21 & 4.21 & -54.2 & 1.6 & AGB\\
24.3 & -0.148 & 4.2 & 4.21 & 105.1 & 1.0 & SFR\\
24.328 & 0.148 & 4.2 & 4.2 & 70.2 & 33.4 & SFR\\
24.347 & 0.039 & 4.2 & 4.2 & 120.9 & 3.1 & SFR\\
24.377 & -0.157 & 4.2 & 4.2 & 50.0 & 2.8 & SFR\\
24.405 & 0.159 & 4.2 & 4.2 & 81.4 & 2.3 & U\\
24.417 & 0.103 & 4.2 & 4.2 & 118.4 & 11.5 & SFR\\
24.426 & 0.298 & 4.2 & 4.2 & 50.9 & 6.8 & AGB\\
24.436 & 0.25 & 4.2 & 4.21 & 119.4 & 2.1 & SFR\\
24.461 & 0.199 & 4.2 & 4.2 & 125.7 & 9.4 & SFR\\
24.488 & -0.037 & 4.2 & 4.2 & 74.0 & 6.1 & SFR\\
24.496 & -0.038 & 4.2 & 4.2 & 104.7 & 10.9 & SFR\\
24.503 & -0.218 & 4.2 & 4.2 & 31.6 & 4.9 & SFR\\
27.5 & 0.196 & 4.2 & 4.2 & -43.4 & 3.3 & U\\
27.591 & 0.085 & 4.2 & 4.2 & 36.0 & 9.2 & AGB\\
27.623 & 0.149 & 4.2 & 4.2 & 92.1 & 2.3 & AGB\\
27.64 & 0.07 & 4.2 & 4.21 & 102.8 & 1.6 & SFR\\
27.647 & 0.092 & 4.2 & 4.21 & 42.9 & 1.8 & U\\
27.664 & 0.125 & 4.2 & 4.2 & 98.9 & 3.4 & SFR\\
27.696 & 0.195 & 4.2 & 4.2 & 60.2 & 12.1 & AGB\\
27.716 & -0.257 & 4.2 & 4.2 & 63.6 & 2.7 & AGB\\
27.725 & 0.034 & 4.2 & 4.21 & 104.4 & 2.2 & SFR\\
27.742 & 0.181 & 4.2 & 4.2 & 84.8 & 2.0 & SFR\\
27.783 & 0.056 & 4.2 & 4.2 & 106.5 & 18.5 & SFR\\
27.785 & 0.056 & 4.2 & 4.2 & 101.8 & 26.3 & SFR\\
27.786 & -0.26 & 4.2 & 4.2 & 74.7 & 3.3 & SFR\\
27.801 & -0.062 & 4.2 & 4.2 & 23.8 & 2.4 & AGB\\
27.802 & -0.063 & 4.21 & 4.22 & 26.9 & 0.6 & AGB\\
27.872 & -0.238 & 4.2 & 4.2 & 38.9 & 108.0 & AGB\\
27.915 & -0.151 & 4.2 & 4.2 & 40.6 & 4.1 & AGB\\
27.924 & -0.03 & 4.2 & 4.2 & 47.7 & 3.9 & SFR\\
27.927 & 0.243 & 4.2 & 4.2 & 46.6 & 64.9 & AGB\\
27.962 & 0.066 & 4.2 & 4.2 & 98.2 & 4.9 & SFR\\
28.009 & -0.038 & 4.2 & 4.2 & 78.6 & 4.1 & SFR\\
28.033 & 0.242 & 4.21 & 4.2 & 75.9 & 1.9 & AGB\\
28.088 & 0.038 & 4.21 & 4.2 & 52.4 & 2.3 & U\\
28.146 & -0.004 & 4.2 & 4.2 & 89.9 & 11.2 & SFR\\
28.156 & -0.185 & 4.2 & 4.2 & 115.6 & 3.4 & AGB\\
28.178 & 0.01 & 4.2 & 4.2 & 98.9 & 9.0 & SFR\\
28.2 & -0.048 & 4.2 & 4.2 & 95.1 & 25.3 & SFR\\
28.227 & 0.359 & 4.2 & 4.2 & 59.8 & 17.4 & SFR\\
28.27 & -0.185 & 4.2 & 4.22 & 85.8 & 1.6 & SFR\\
28.299 & -0.193 & 4.2 & 4.2 & 76.4 & 4.6 & SFR\\
28.328 & 0.159 & 4.2 & 4.2 & 31.9 & 45.7 & SFR\\
28.329 & 0.064 & 4.21 & 4.2 & 88.6 & 1.4 & SFR\\
28.332 & 0.113 & 4.2 & 4.2 & 39.6 & 2.2 & U\\
28.342 & 0.144 & 4.2 & 4.2 & -1.6 & 4.5 & U\\
28.348 & 0.063 & 4.2 & 4.2 & 80.8 & 3.5 & SFR\\
28.354 & 0.143 & 4.2 & 4.2 & 75.0 & 6.1 & SFR\\
28.371 & -0.046 & 4.2 & 4.2 & 23.0 & 5.4 & AGB\\
28.371 & -0.046 & 4.2 & 4.2 & 19.8 & 5.4 & AGB\\
28.397 & 0.079 & 4.2 & 4.2 & 78.4 & 243.5 & SFR\\
28.398 & -0.305 & 4.2 & 4.2 & 36.5 & 2.6 & SFR\\
28.406 & 0.064 & 4.2 & 4.2 & 20.5 & 60.3 & U\\
28.453 & 0.128 & 4.2 & 4.21 & 84.2 & 4.0 & AGB\\
28.454 & 0.126 & 4.2 & 4.2 & 81.4 & 2.4 & AGB\\
28.516 & 0.008 & 4.2 & 4.2 & 88.1 & 1.2 & AGB\\
28.521 & 0.121 & 4.2 & 4.2 & 34.7 & 1.3 & U\\
28.532 & -0.151 & 4.2 & 4.2 & 84.0 & 4.3 & SFR\\
28.586 & -0.226 & 4.2 & 4.2 & 84.9 & 2.2 & SFR\\
28.605 & -0.341 & 4.21 & 4.21 & 30.2 & 1.2 & U\\
28.611 & -0.027 & 4.2 & 4.21 & 42.5 & 1.2 & AGB\\
28.617 & 0.298 & 4.2 & 4.21 & 4.3 & 2.2 & AGB\\
28.722 & 0.169 & 4.2 & 4.2 & 91.6 & 2.6 & AGB\\
28.732 & 0.175 & 4.2 & 4.2 & 88.9 & 3.7 & U\\
28.784 & 0.234 & 4.2 & 4.2 & 115.0 & 1.7 & SFR\\
28.785 & 0.234 & 4.2 & 4.2 & 112.0 & 2.2 & SFR\\
28.801 & 0.174 & 4.2 & 4.2 & 104.1 & 2.0 & SFR\\
28.803 & 0.175 & 4.2 & 4.21 & 62.7 & 1.5 & SFR\\
28.805 & 0.202 & 4.2 & 4.2 & 34.0 & 3.8 & SFR\\
28.814 & 0.361 & 4.2 & 4.2 & 86.6 & 22.8 & SFR\\
28.826 & -0.156 & 4.2 & 4.2 & 55.5 & 0.6 & AGB\\
28.833 & -0.254 & 4.2 & 4.2 & 77.9 & 47.5 & SFR\\
28.862 & 0.064 & 4.2 & 4.2 & 105.1 & 458.6 & SFR\\
28.884 & 0.258 & 4.2 & 4.2 & 51.3 & 9.9 & AGB\\
28.884 & -0.022 & 4.2 & 4.2 & 114.5 & 4.4 & SFR\\
28.903 & 0.29 & 4.21 & 4.2 & 8.2 & 1.7 & AGB\\
28.959 & -0.203 & 4.2 & 4.2 & 95.2 & 11.6 & SFR\\
28.963 & 0.388 & 4.2 & 4.2 & 17.1 & 14.0 & AGB\\
28.982 & 0.067 & 4.2 & 4.2 & 68.9 & 2.2 & SFR\\
29.036 & -0.13 & 4.2 & 4.2 & 96.6 & 3.7 & U\\
29.121 & 0.029 & 4.2 & 4.21 & 98.7 & 1.1 & SFR\\
29.163 & 0.017 & 4.2 & 4.2 & 78.9 & 2.8 & SFR\\
29.245 & -0.111 & 4.22 & 4.2 & 44.3 & 1.1 & AGB\\
29.256 & -0.257 & 4.2 & 4.2 & 115.8 & 6.3 & AGB\\
29.256 & -0.257 & 4.2 & 4.2 & 117.9 & 6.0 & AGB\\
29.261 & -0.301 & 4.2 & 4.21 & 73.3 & 1.6 & SFR\\
29.274 & -0.007 & 4.2 & 4.2 & -18.5 & 1.0 & AGB\\
29.289 & -0.277 & 4.2 & 4.2 & 52.1 & 4.9 & AGB\\
29.306 & -0.213 & 4.2 & 4.2 & 59.1 & 4.1 & AGB\\
29.32 & -0.164 & 4.2 & 4.21 & 45.4 & 1.2 & SFR\\
29.417 & 0.139 & 4.22 & 4.2 & 81.8 & 0.3 & U\\
29.431 & 0.157 & 4.2 & 4.21 & 105.9 & 1.0 & AGB\\
29.433 & 0.111 & 4.2 & 4.2 & 63.1 & 0.9 & AGB\\
29.475 & -0.181 & 4.2 & 4.2 & 107.0 & 1.8 & SFR\\
29.492 & 0.151 & 4.2 & 4.2 & 65.4 & 1.6 & SFR\\
29.497 & 0.183 & 4.2 & 4.2 & 40.6 & 3.9 & AGB\\
29.515 & -0.193 & 4.2 & 4.2 & 44.7 & 2.0 & AGB\\
29.579 & 0.133 & 4.2 & 4.2 & 38.2 & 8.6 & AGB\\
29.648 & 0.414 & 4.2 & 4.21 & 28.9 & 3.4 & AGB\\
29.729 & -0.05 & 4.2 & 4.2 & 80.9 & 2.4 & AGB\\
29.782 & -0.342 & 4.21 & 4.2 & 39.7 & 2.2 & U\\
29.784 & -0.334 & 4.2 & 4.2 & 40.1 & 16.1 & AGB\\
29.827 & -0.202 & 4.2 & 4.21 & 84.1 & 1.8 & SFR\\
29.889 & -0.019 & 4.2 & 4.2 & 103.6 & 4.7 & SFR\\
29.918 & -0.044 & 4.2 & 4.2 & 40.1 & 32.8 & SFR\\
29.929 & -0.06 & 4.2 & 4.2 & 100.1 & 1.0 & SFR\\
29.917 & -0.044 & 4.2 & 4.2 & 36.2 & 12.8 & U\\
29.929 & -0.06 & 4.2 & 4.2 & 100.3 & 1.6 & SFR\\
29.942 & -0.05 & 4.2 & 4.21 & 99.3 & 0.9 & SFR\\
29.958 & -0.016 & 4.2 & 4.2 & 95.2 & 102.7 & SFR\\
29.958 & -0.016 & 4.2 & 4.2 & 93.9 & 217.7 & SFR\\
29.979 & -0.048 & 4.2 & 4.2 & 79.3 & 2.0 & SFR\\
29.975 & -0.05 & 4.2 & 4.2 & 101.4 & 2.1 & SFR\\
29.981 & -0.049 & 4.2 & 4.2 & 105.5 & 1.5 & SFR\\
29.979 & -0.049 & 4.2 & 4.2 & 99.0 & 1.7 & SFR\\
29.985 & 0.109 & 4.2 & 4.2 & 118.3 & 17.7 & AGB\\
29.999 & -0.146 & 4.2 & 4.2 & 96.2 & 1.7 & SFR\\
30.003 & -0.264 & 4.2 & 4.2 & 105.3 & 13.1 & SFR\\
30.08 & -0.139 & 4.21 & 4.2 & 124.3 & 1.2 & AGB\\
30.137 & -0.233 & 4.2 & 4.2 & 64.6 & 6.6 & AGB\\
30.141 & -0.126 & 4.2 & 4.2 & 7.4 & 1.8 & AGB\\
30.231 & -0.144 & 4.2 & 4.2 & -13.4 & 2.5 & AGB\\
30.233 & -0.137 & 4.2 & 4.2 & -10.0 & 6.9 & AGB\\
30.244 & -0.084 & 4.2 & 4.2 & 113.3 & 14.3 & AGB\\
30.264 & 0.046 & 4.22 & 4.2 & 105.9 & 1.3 & AGB\\
30.316 & 0.071 & 4.2 & 4.2 & 45.9 & 17.2 & SFR\\
30.342 & -0.118 & 4.2 & 4.2 & 94.3 & 2.1 & SFR\\
30.349 & 0.391 & 4.2 & 4.2 & 64.8 & 20.5 & SFR\\
30.354 & 0.428 & 4.22 & 4.2 & 76.0 & 2.0 & SFR\\
30.363 & 0.108 & 4.22 & 4.2 & 56.0 & 2.2 & U\\
30.395 & 0.135 & 4.2 & 4.2 & 32.8 & 6.6 & AGB\\
30.401 & -0.292 & 4.2 & 4.2 & 101.5 & 5.2 & SFR\\
30.401 & -0.296 & 4.2 & 4.2 & 89.7 & 1.4 & SFR\\
30.42 & -0.233 & 4.2 & 4.2 & 104.7 & 9.1 & SFR\\
30.465 & 0.034 & 4.2 & 4.2 & 97.6 & 3.1 & SFR\\
30.487 & -0.021 & 4.2 & 4.2 & -43.5 & 8.0 & U\\
30.51 & -0.074 & 4.2 & 4.21 & -10.2 & 1.8 & U\\
30.516 & 0.029 & 4.2 & 4.2 & 70.0 & 1.4 & U\\
30.539 & 0.019 & 4.21 & 4.2 & -39.9 & 1.0 & U\\
30.591 & -0.042 & 4.2 & 4.2 & 40.8 & 1.6 & AGB\\
30.608 & 0.171 & 4.2 & 4.2 & 105.1 & 15.9 & SFR\\
30.641 & 0.328 & 4.2 & 4.2 & 135.3 & 3.3 & AGB\\
30.65 & -0.122 & 4.2 & 4.2 & 104.0 & 1.0 & SFR\\
30.688 & -0.233 & 4.2 & 4.2 & 27.7 & 6.2 & AGB\\
30.688 & -0.233 & 4.2 & 4.2 & 30.5 & 8.0 & AGB\\
30.688 & -0.257 & 4.2 & 4.21 & 100.5 & 2.9 & SFR\\
30.695 & -0.07 & 4.2 & 4.2 & 83.7 & 0.3 & SFR\\
30.704 & -0.068 & 4.2 & 4.2 & 93.0 & 24.1 & SFR\\
30.715 & 0.427 & 4.2 & 4.2 & 52.4 & 4.2 & AGB\\
30.72 & -0.083 & 4.2 & 4.2 & 88.9 & 2.9 & SFR\\
30.726 & 0.141 & 4.2 & 4.2 & 40.0 & 4.1 & SFR\\
30.744 & -0.061 & 4.2 & 4.2 & 43.1 & 26.4 & U\\
30.762 & -0.257 & 4.2 & 4.2 & 101.7 & 5.0 & U\\
30.744 & -0.061 & 4.2 & 4.2 & 43.2 & 27.3 & U\\
30.762 & -0.053 & 4.2 & 4.2 & 89.1 & 5.9 & SFR\\
30.763 & -0.053 & 4.2 & 4.2 & 116.4 & 2.5 & SFR\\
30.77 & -0.116 & 4.2 & 4.21 & 94.1 & 1.6 & SFR\\
30.77 & -0.117 & 4.2 & 4.2 & 96.3 & 2.5 & SFR\\
30.785 & 0.229 & 4.2 & 4.2 & 21.7 & 7.8 & U\\
30.786 & 0.203 & 4.2 & 4.2 & 86.9 & 24.3 & SFR\\
30.787 & 0.203 & 4.2 & 4.2 & 83.7 & 24.6 & SFR\\
30.816 & -0.033 & 4.2 & 4.2 & 99.9 & 3.5 & SFR\\
30.817 & -0.058 & 4.2 & 4.2 & 99.8 & 221.6 & SFR\\
30.818 & 0.272 & 4.2 & 4.2 & 103.9 & 8.1 & SFR\\
30.821 & 0.059 & 4.2 & 4.2 & 40.9 & 12.1 & SFR\\
30.821 & 0.055 & 4.2 & 4.2 & 35.6 & 26.7 & SFR\\
30.822 & -0.155 & 4.2 & 4.2 & 110.2 & 96.5 & SFR\\
30.848 & 0.121 & 4.21 & 4.2 & 41.7 & 3.2 & U\\
30.884 & 0.203 & 4.2 & 4.2 & 108.9 & 13.3 & AGB\\
30.895 & 0.161 & 4.21 & 4.2 & 105.5 & 4.0 & SFR\\
30.897 & 0.106 & 4.2 & 4.2 & 30.9 & 13.1 & U\\
30.912 & 0.093 & 4.2 & 4.21 & 95.9 & 4.7 & SFR\\
30.94 & -0.157 & 4.2 & 4.2 & 79.7 & 14.2 & AGB\\
30.944 & 0.03 & 4.2 & 4.2 & -54.5 & 19.2 & AGB\\
30.956 & 0.086 & 4.2 & 4.2 & 39.7 & 37.2 & SFR\\
30.956 & 0.086 & 4.2 & 4.2 & 35.5 & 40.7 & SFR\\
31.0 & -0.11 & 4.2 & 4.2 & 22.5 & 3.3 & U\\
31.012 & -0.22 & 4.2 & 4.2 & 119.9 & 49.7 & AGB\\
31.045 & 0.36 & 4.2 & 4.2 & 82.0 & 2.6 & SFR\\
31.053 & 0.109 & 4.2 & 4.2 & 13.8 & 4.6 & AGB\\
31.06 & 0.093 & 4.2 & 4.2 & 17.7 & 8.1 & U\\
31.096 & -0.118 & 4.2 & 4.2 & -18.1 & 10.8 & U\\
31.12 & 0.026 & 4.2 & 4.2 & 33.8 & 2.4 & SFR\\
31.15 & 0.268 & 4.2 & 4.2 & 90.9 & 18.6 & SFR\\
31.156 & 0.049 & 4.22 & 4.2 & 18.2 & 2.1 & SFR\\
31.167 & -0.022 & 4.2 & 4.2 & 50.5 & 7.9 & U\\
31.209 & -0.094 & 4.2 & 4.2 & 17.4 & 3.0 & U\\
31.222 & 0.019 & 4.2 & 4.2 & 69.4 & 8.4 & SFR\\
31.223 & -0.038 & 4.2 & 4.21 & 41.5 & 2.5 & SFR\\
31.241 & -0.126 & 4.2 & 4.2 & 80.0 & 4.6 & SFR\\
31.242 & -0.111 & 4.2 & 4.2 & 25.9 & 175.6 & SFR\\
31.277 & 0.064 & 4.2 & 4.2 & 108.8 & 37.7 & SFR\\
31.279 & 0.06 & 4.2 & 4.2 & 105.7 & 1.9 & SFR\\
31.286 & 0.125 & 4.2 & 4.2 & 29.4 & 9.1 & U\\
31.396 & -0.257 & 4.2 & 4.2 & 84.3 & 3.4 & SFR\\
31.398 & 0.243 & 4.2 & 4.21 & 49.2 & 3.3 & AGB\\
31.41 & 0.309 & 4.2 & 4.2 & 97.6 & 96.5 & SFR\\
31.435 & -0.001 & 4.2 & 4.2 & 51.0 & 2.7 & AGB\\
31.443 & -0.064 & 4.2 & 4.2 & 82.1 & 14.5 & AGB\\
31.445 & -0.229 & 4.21 & 4.2 & 33.1 & 1.6 & U\\
31.469 & 0.189 & 4.2 & 4.2 & 107.8 & 4.9 & SFR\\
31.494 & 0.178 & 4.2 & 4.2 & 105.1 & 8.3 & SFR\\
37.546 & -0.109 & 4.2 & 4.2 & 40.1 & 3.0 & SFR\\
37.562 & -0.321 & 4.2 & 4.2 & 96.8 & 7.7 & SFR\\
37.594 & -0.125 & 4.2 & 4.2 & -33.4 & 4.9 & AGB\\
37.677 & -0.11 & 4.2 & 4.2 & 64.5 & 2.2 & SFR\\
37.736 & -0.112 & 4.2 & 4.2 & 52.1 & 71.7 & SFR\\
37.736 & -0.113 & 4.2 & 4.2 & 41.9 & 68.0 & SFR\\
37.754 & -0.188 & 4.2 & 4.2 & 60.1 & 7.9 & SFR\\
37.766 & -0.219 & 4.2 & 4.2 & 67.0 & 5.7 & SFR\\
37.766 & -0.219 & 4.2 & 4.2 & 68.7 & 5.5 & SFR\\
37.906 & -0.34 & 4.2 & 4.2 & 71.0 & 2.9 & U\\
37.982 & 0.046 & 4.2 & 4.2 & 28.1 & 2.3 & AGB\\
38.042 & -0.297 & 4.2 & 4.2 & 59.8 & 3.8 & SFR\\
38.103 & -0.126 & 4.2 & 4.2 & 31.2 & 20.8 & SFR\\
38.247 & -0.148 & 4.2 & 4.2 & 26.8 & 2.7 & U\\
38.252 & -0.147 & 4.2 & 4.2 & 39.5 & 5.7 & U\\
38.259 & -0.075 & 4.2 & 4.2 & 9.4 & 45.5 & AGB\\
45.068 & 0.132 & 4.2 & 4.2 & 62.6 & 37.7 & SFR\\
45.111 & 0.127 & 4.2 & 4.2 & 54.1 & 3.5 & SFR\\
45.164 & 0.091 & 4.2 & 4.2 & 60.5 & 3.3 & AGB\\
45.279 & -0.141 & 4.2 & 4.2 & 42.1 & 21.1 & AGB\\
45.425 & 0.082 & 4.2 & 4.2 & 56.4 & 22.8 & SFR\\
45.441 & -0.001 & 4.2 & 4.2 & 46.4 & 1.7 & SFR\\
45.467 & 0.044 & 4.2 & 4.2 & 72.2 & 15.9 & SFR\\
45.48 & 0.054 & 4.2 & 4.2 & 54.7 & 17.1 & SFR\\
45.452 & 0.064 & 4.2 & 4.2 & 48.2 & 2.4 & SFR\\
45.479 & 0.134 & 4.2 & 4.2 & 66.7 & 7.8 & SFR\\
45.494 & 0.125 & 4.2 & 4.2 & 74.0 & 4.6 & SFR\\
47.001 & 0.219 & 4.2 & 4.2 & 4.3 & 3.3 & U\\
\enddata
\tablecomments{Starting from the leftmost column, we present the Galactic longitude, the Galactic latitude, the error in the Galactic longitude, the error in the Galactic latitude, the velocity of the brightest channel in the spectrum, the intensity of the brightest channel in the spectrum, and the associated environment. We classify $\mathrm{H_{2}O}$ masers as being associated with a star-forming region (SFR), an asymptotic giant branch (AGB) star, or an unknown environment.}

\end{deluxetable}

\begin{figure}[!hbtp]
\centering
\includegraphics[scale=0.8]{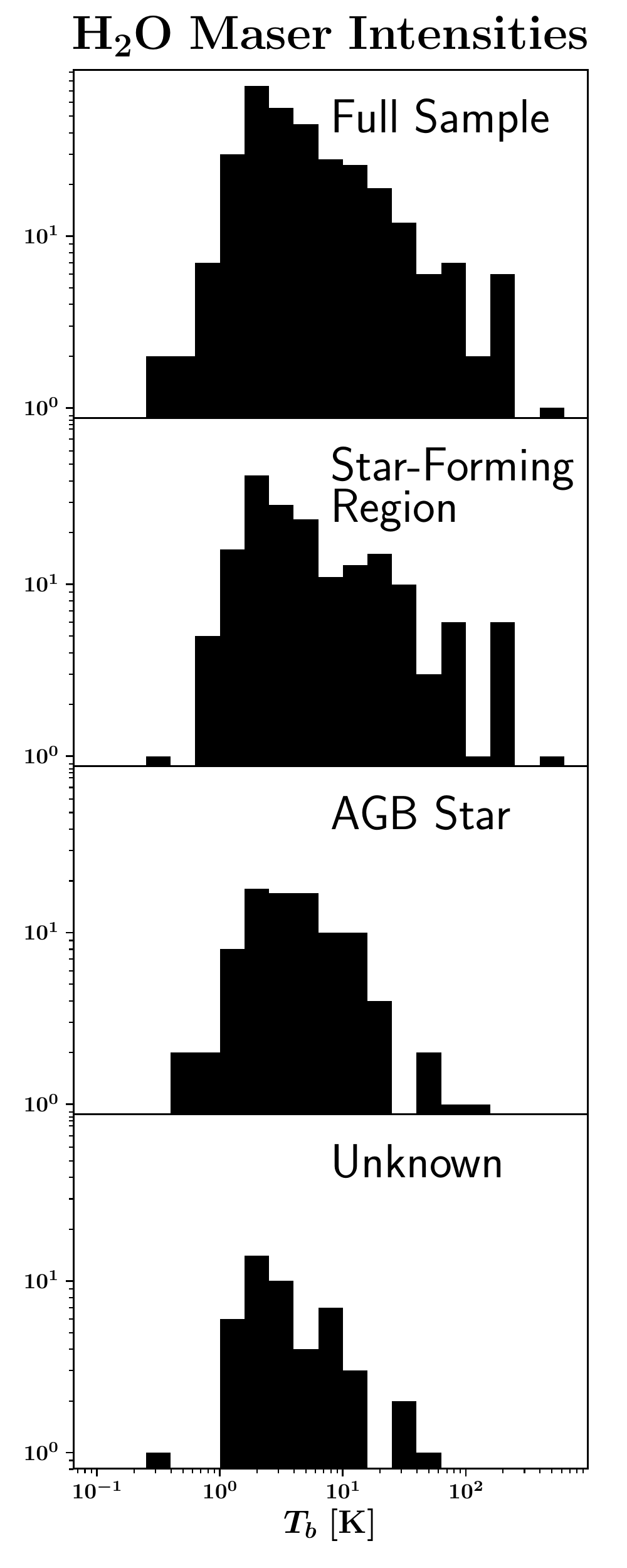}
\caption{From top to bottom, histograms of maser intensities for the full sample of masers, the masers associated with SFRs, the masers associated with AGB stars, and the masers with an unknown association, respectively. The histograms qualitatively look similar, with a roughly power-law slope at large intensities and a sharp cutoff at $\sim 1$ K.}
\label{fig:maser_Tb_hist}
\end{figure}

\begin{figure}[!hbtp]
\centering
\includegraphics[scale=0.8]{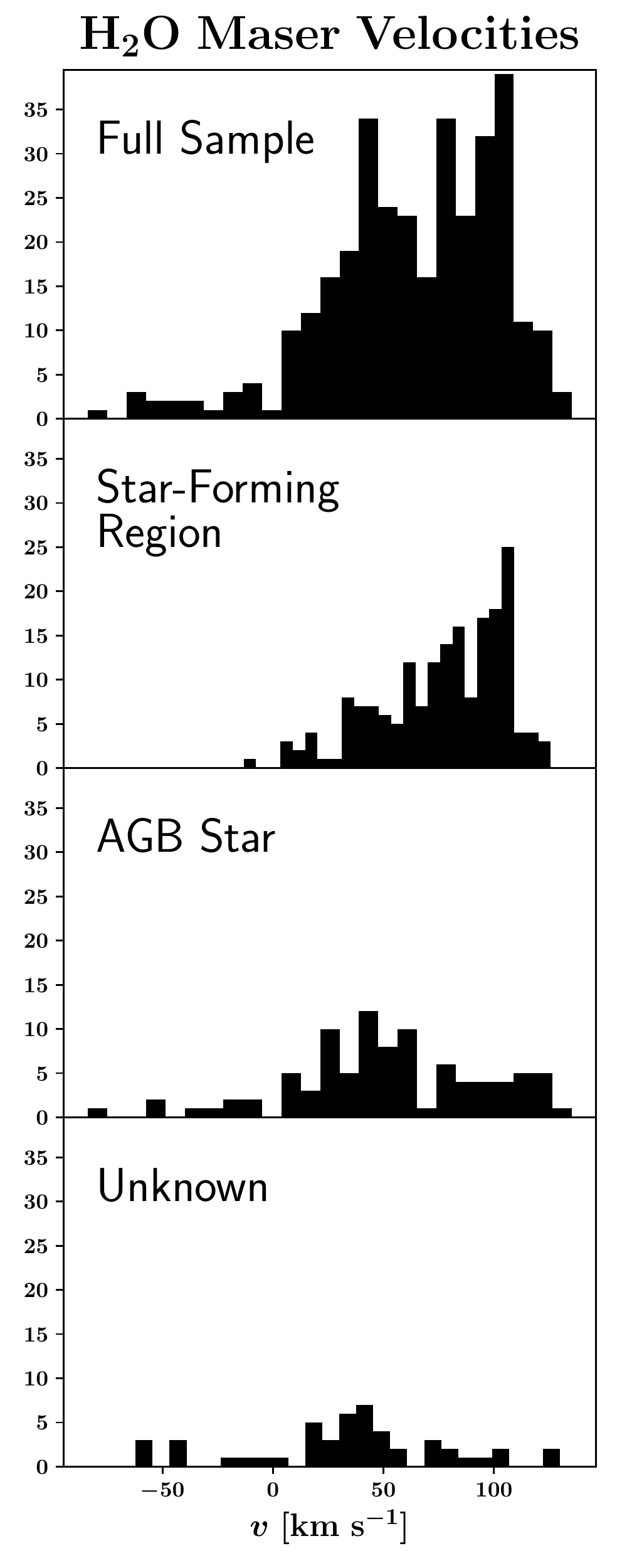}
\caption{From top to bottom, histograms of maser velocities for the full sample of masers, the masers associated with SFRs, the masers associated with AGB stars, and the masers with an unknown association, respectively. The distribution of masers associated with SFRs is more concentrated at positive velocities than all other distributions. This difference may indicate that some of the masers associated with AGB stars are in the Galactic halo.}
\label{fig:maser_vel_hist}
\end{figure}

To compare the distributions in a quantitative manner, we performed a two-sample Kolmogorov$-$Smirnov (K-S) test. The two-sample K-S test determines how different two samples are, and the K-S statistic, which can have a value between 0 and 1, is a measure of this difference. If the K-S statistic of a particular test equals 0, this indicates that we cannot reject the hypothesis that the two samples are derived from the same distribution. A larger K-S statistic implies that the two samples are less likely drawn from the same distribution. For each pair of samples, we also calculated the $p$-value, which is the probability that two samples are derived from the same distribution. Table~\ref{tab:statistics} shows the results of the statistical tests. The K-S test and the $p$-test both show that the three intensity distributions are only moderately different from each other. On the other hand, the differences in the velocity distributions are significant. The $p$-values for these tests show that the velocities of the masers associated with SFRs are almost certainly drawn from a different distribution than the velocities of both the masers associated with AGB stars and those with an unknown association. The differences in velocity distribution between the unknown and AGB categories are more moderate. The difference in the velocity distributions is likely due to differing spatial distributions within the Galaxy. Masers associated with SFRs are found only where there is molecular gas; thus, these masers are excited primarily within the midplane of the Galaxy and follow roughly circular orbits. Consequently, the SFRs that are in the first quadrant of the Galaxy and less than $\sim8$ kpc from the Galactic center have positive velocities. Unlike SFRs, AGB stars can be found in both the Galactic plane and the Galactic halo. Stars in the Galactic halo can have a wide range in $v_{LSR}$; thus, we expect to detect some masers with large negative velocities. The results of the statistical tests could indicate that masers with an unknown association are more likely associated with AGB stars, although some of these masers could be associated with an SFR, but exhibit a greater than 30 $\mathrm{km \ s^{-1}}$ velocity offset from the source's systemic velocity.

\begin{deluxetable}{lcc}
\tabletypesize{\footnotesize}
\tablewidth{0pt}
\tablecolumns{3}

\tablecaption{Comparison of $\mathrm{H_{2}O}$ Maser Distributions\label{tab:statistics}} 
\tablehead{\colhead{Distributions} &
		  \colhead{K-S Statistic}&
		  \colhead{$P$-value}}
\startdata
Intensity $-$ SFR and AGB & 0.15 & 0.14\\
Intensity $-$ SFR and U & 0.19 & 0.12\\
Intensity $-$ AGB and U & 0.17 & 0.30\\
Velocity $-$ SFR and AGB & 0.37 & $8.2\times10^{-8}$\\
Velocity $-$ SFR and U & 0.53 & $3.0\times10^{-10}$\\
Velocity $-$ AGB and U & 0.21 & 0.10\\
\enddata

\end{deluxetable}

\section{Comparison with Other Surveys}
\label{sec:comparison}

Having established the capabilities of our survey and presented a preliminary analysis of the RAMPS dataset, we now compare RAMPS to previous Galactic plane surveys. First, we will compare our detection threshold for clumps to that of the BGPS, a 1 mm dust continuum survey. Due to spatial filtering, BGPS is biased toward compact, and presumably dense, sources, which makes it a good continuum survey to compare with RAMPS. Figure~\ref{fig:L23_bolocam} presents the RAMPS L23 $\mathrm{NH_{3}(1,1)}$ integrated intensity map overlaid with 3~$\sigma$ BGPS 1 mm dust emission contours. RAMPS detects most of the clumps detected by BGPS, indicating that our sensitivity is sufficient to observe a large sample of molecular clumps. In Figure~\ref{fig:L23_bolocam} there are a few clumps detected by RAMPS that are not detected at a significance of $3\sigma$ by BGPS, as well as a few clumps that are detected by BGPS that do not meet the significance threshold we set for the RAMPS integrated intensity maps. While part of this difference is due to the fact that we require a $5\sigma$ detection of a line to meet our significance threshold, there may also be differences between the gas and dust that lead to different emission properties. Although investigating differences between the gas and dust emission of molecular clumps is interesting and important, these clumps are faint enough that we did not attempt to fit these spectra for the rotational temperature and column density. We intend to perform a robust comparison between the RAMPS dataset and dust continuum data in a future project.

\begin{figure}[!hbtp]
\centering
\includegraphics[scale=0.7]{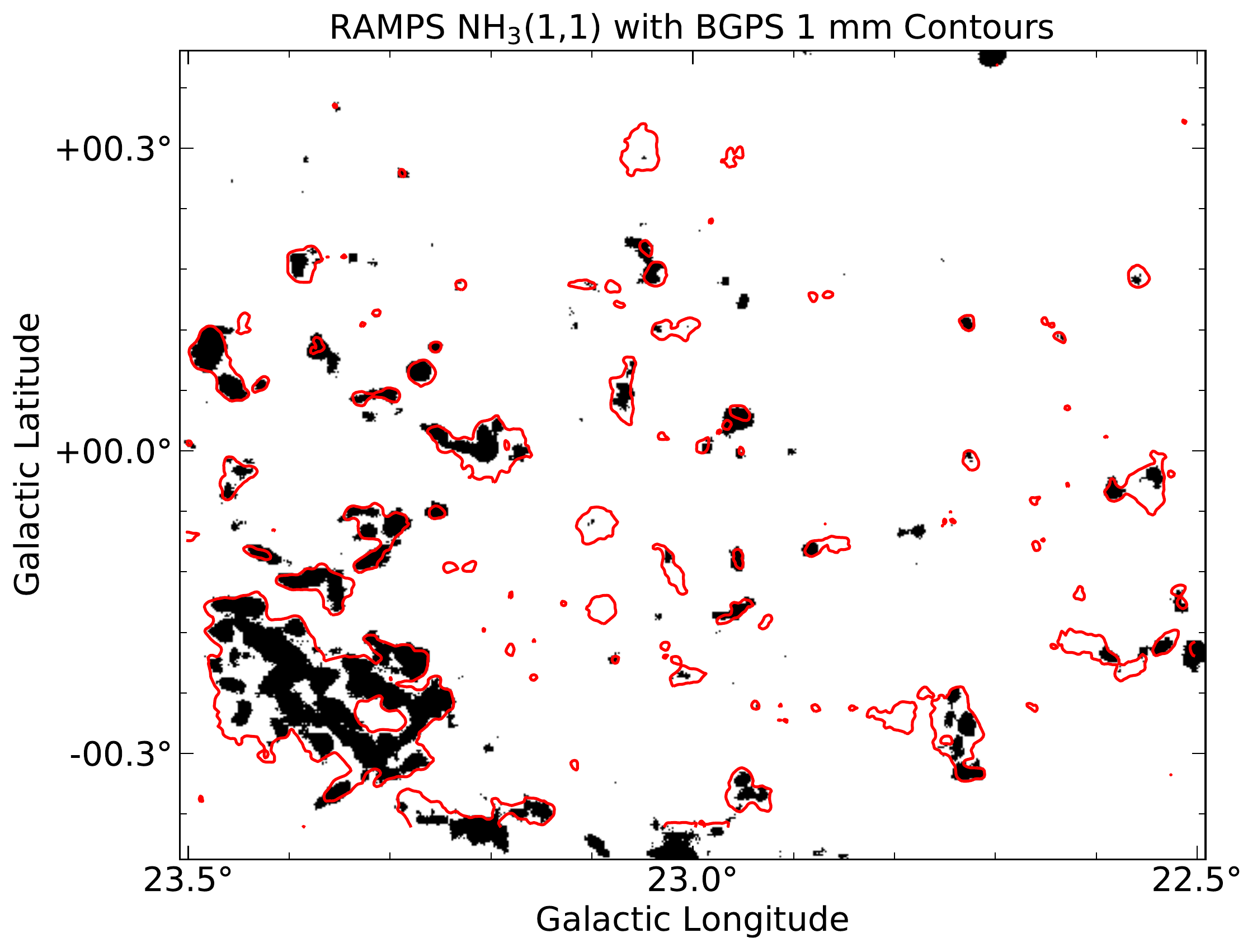}
\caption{L23 $\mathrm{NH_{3}(1,1)}$ integrated intensity map overlaid with BGPS 1 mm dust emission contours. The $\mathrm{NH_{3}}$ emission is in black for visibility. The contour level is at 140 mJy beam$^{-1}$, which is approximately three times the noise in the BGPS map. RAMPS detects most of the BGPS sources in this field.}
\label{fig:L23_bolocam}
\end{figure}

RAMPS is more sensitive than the previous large, blind $\mathrm{NH_{3}}$ survey, HOPS \citep{2011MNRAS.416.1764W,P12}. HOPS is a 100 deg$^{2}$ molecular line survey primarily targeting $\mathrm{NH_{3}(1,1)-(3,3)}$ and $\mathrm{H_{2}O}$ using the 22 m Mopra telescope. Figure~\ref{fig:L23-24_HOPS} compares the RAMPS and HOPS $\mathrm{NH_{3}(1,1)}$ integrated intensity maps of the L23 and L24 fields. RAMPS is clearly more sensitive than HOPS and has much better angular resolution. One consequence of better spatial resolution is that small clumps, which are severely beam diluted in the large Mopra beam, are better resolved by the GBT beam and are thus easier to detect. The finer angular resolution also resolves the larger clump complexes into their constituent clumps. The GBT beam resolves many of the clumps throughout the map, often revealing structure in the maps of temperature, column density, line width, and velocity. Probing clumps at this scale is crucial for understanding how the onset of high-mass star formation affects the evolution of the surrounding clump.

\begin{figure}[!hbtp]
\centering
\includegraphics[scale=1.2,trim={1cm 4cm 1cm 4cm},clip]{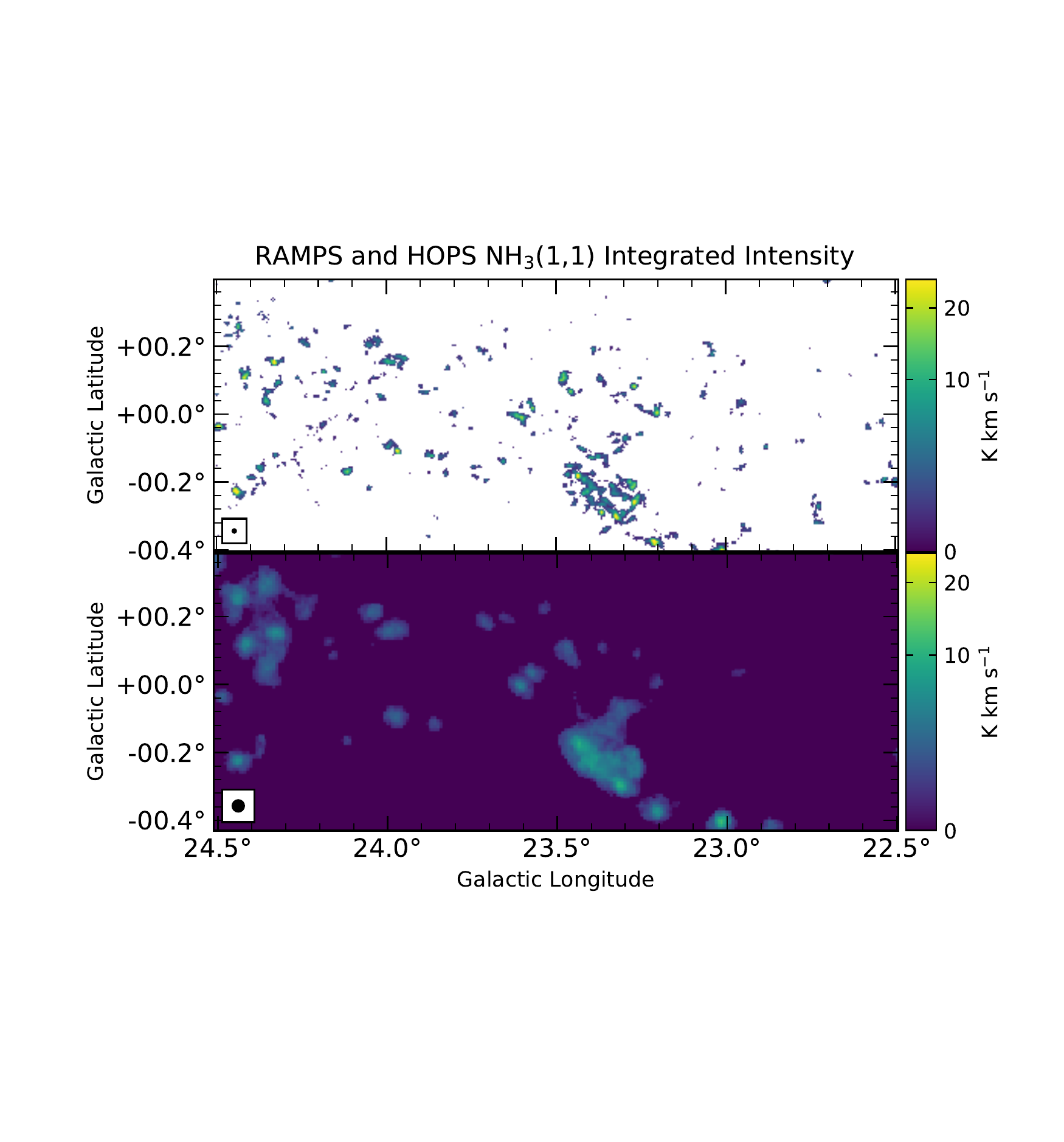}
\caption{Comparison between RAMPS and HOPS $\mathrm{NH_{3}(1,1)}$ integrated intensity maps. Top: RAMPS $\mathrm{NH_{3}(1,1)}$ integrated intensity map of the L23-24 fields. Bottom: HOPS $\mathrm{NH_{3}(1,1)}$ integrated intensity map of the same region. The beam size for each survey is shown in a box at the lower left corner of each map. Given that the GBT is much larger than the 22 m Mopra telescope, the RAMPS maps have much better spatial resolution and can be used to distinguish clumps smaller than the Mopra beam. Consequently, RAMPS detects many smaller clumps and resolves the large clump complexes into the individual clumps of which they are composed.}
\label{fig:L23-24_HOPS}
\end{figure}

RAMPS is now the most sensitive large, blind survey of $\mathrm{H_{2}O}$ masers to date; thus, it is important to compare it to HOPS, the previous large $\mathrm{H_{2}O}$ maser survey. HOPS detected 540 sites of maser emission in a 100 deg$^2$ survey, or 5.4 masers/deg$^2$. The RAMPS pilot survey detected 325 masers in 6.5 deg$^2$, or 50 masers/deg$^2$. Since the two survey regions have only moderate overlap, it is not meaningful to compare these two numbers directly. For a direct comparison, Figure~\ref{fig:HOPS_maser} shows the L23 and L24 fields, which were observed by both surveys. RAMPS $\mathrm{NH_{3}(1,1)}$ integrated intensity is in black, with the colored symbols overlaid showing the positions of masers. The orange stars represent the $\mathrm{H_{2}O}$ masers detected by HOPS, while the other symbols represent masers detected by RAMPS. The RAMPS masers are further separated by their associated environment, with the blue squares representing masers associated with SFRs, the green triangles representing masers associated with AGB stars, and the pink circles representing masers associated with an unknown environment. While HOPS detected 15 masers in this region, RAMPS has detected 82, demonstrating that RAMPS offers a significant leap in sensitivity. We note that HOPS detected two masers in this region that RAMPS does not detect. A possible explanation for this difference is maser variability. Considering that maser intensities can vary \citep{1992ARA&A..30...75E}, it is possible that these masers were brighter during the HOPS observations but faded to intensities below the detection limit during the more sensitive RAMPS observations. Given that the GBT is $\sim20\times$ more sensitive to point sources than Mopra, the variability would need to be large to explain the nondetections.

\begin{figure}[!hbtp]
\centering
\includegraphics[scale=0.73,trim={0 2cm 0 3cm},clip]{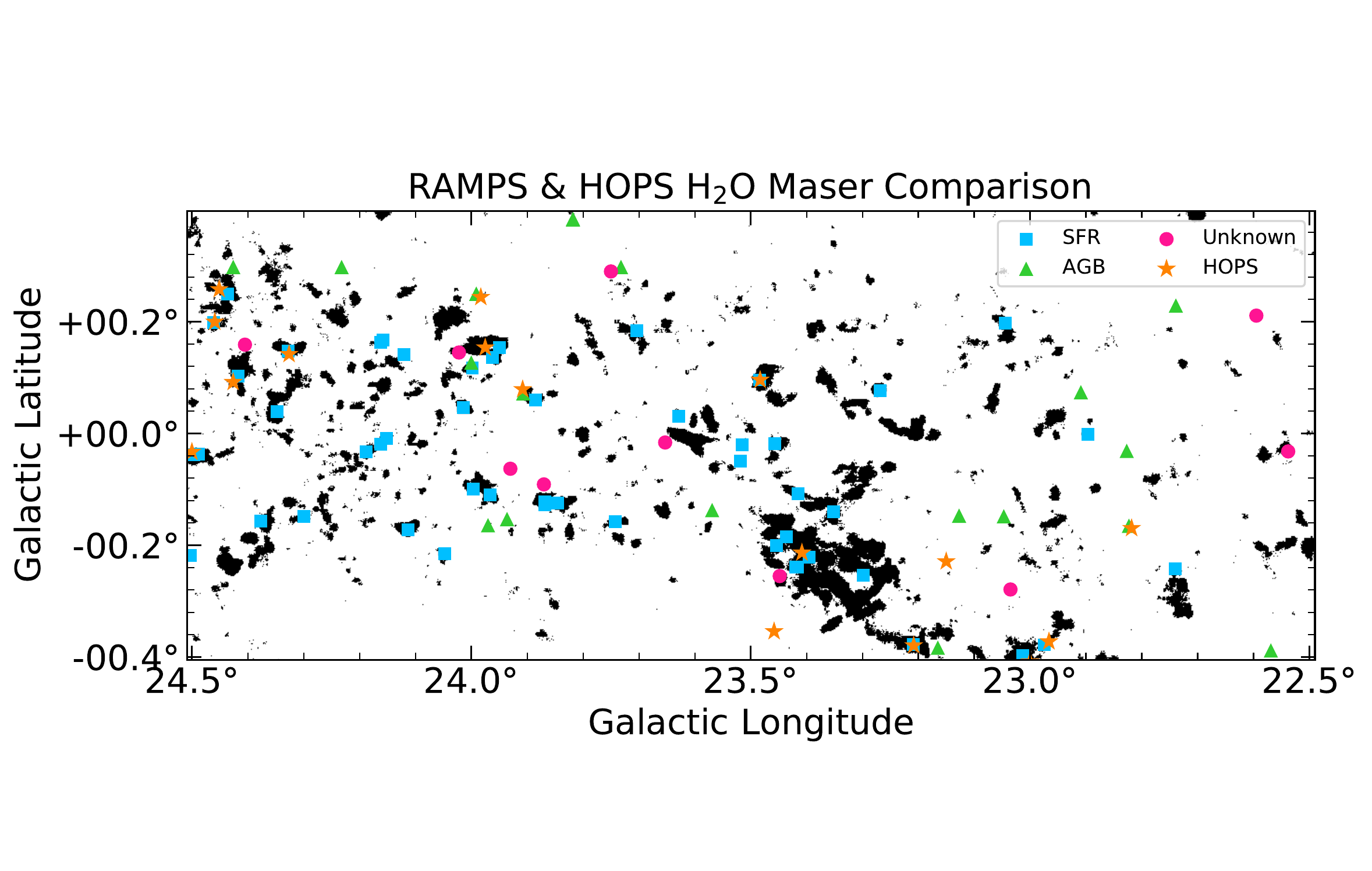}
\caption{Comparison between RAMPS and HOPS $\mathrm{H_{2}O}$ maser detections. RAMPS $\mathrm{NH_{3}(1,1)}$ integrated intensity is in black for visibility, while maser positions are overlaid with colored symbols. Because masers appear as point sources to the GBT beam, the symbol sizes do not represent the physical size of the masers. The HOPS masers are represented with orange stars, while the rest of the symbols represent masers detected only by RAMPS. The RAMPS masers are further separated by their associated environment, with blue squares representing SFRs, the green triangles representing AGB stars, and the pink circles representing an unknown environment.}
\label{fig:HOPS_maser}
\end{figure}

We now investigate whether the better sensitivity of RAMPS changes the detection rate relative to that deduced by HOPS. We find that $57\pm4$\% of RAMPS masers are associated with SFRs, while $28\pm5$\% are associated with AGB stars. On the other hand, \citet{2014MNRAS.442.2240W} found that $69\pm2$\% of HOPS masers are associated with SFRs and only $19\pm4$\% are associated with AGB stars. There is some discrepancy between these detection rates, but it is uncertain why it arises. If RAMPS and HOPS observed the same region of the Galaxy, this difference would likely point toward differing flux distributions for the two maser populations. In reality, HOPS observed much more of the Galactic center than RAMPS has. A possible explanation for their larger detection rate of masers associated with SFRs is a longitudinal variation in the relative occurrence of these masers. Another possible explanation is our differing classification schemes. Further investigations of the RAMPS $\mathrm{H_{2}O}$ maser data will constitute future research.

\section{Conclusion}
\label{sec:conclusion}

RAMPS is an ongoing molecular line survey in the first quadrant of the Galactic midplane. In this paper, we have reported on the pilot survey, which mapped approximately 6.5 square degrees of the RAMPS survey region. RAMPS is a significant improvement on previous large molecular line surveys owing to advancements in instrumentation on the GBT. While the GBT provides excellent sensitivity and spatial resolution, the KFPA receiver array and the VEGAS spectrometer make a large $K$-band survey possible. The KFPA's seven receivers can map large areas in a relatively short amount of time, while VEGAS is able to observe simultaneously a large number of spectral lines over a wide frequency range. This combination gives RAMPS a distinct advantage in fast mapping at $K$-band frequencies. 

An important consequence of the new instrumentation is our ability to map simultaneously a suite of useful lines, namely, the $\mathrm{NH_{3}}$ inversion transitions, $\mathrm{NH_{3}}$(1,1)$-$(5,5), and the 22.235 GHz $\mathrm{H_{2}O}$ maser line. Not only do the $\mathrm{NH_{3}}$ inversion lines trace the dense molecular clumps where high-mass stars can form, but they also provide robust estimates of the gas temperature and column density. Furthermore, measured line widths allow us to determine the virial state of molecular clumps, while their velocities can help determine their distances. Among other things, $\mathrm{H_{2}O}$ masers can be used as an indicator of active star-formation; thus, an $\mathrm{H_{2}O}$ maser associated with $\mathrm{NH_{3}}$ can help indicate whether stars are forming within a molecular clump. RAMPS is a leap forward in large surveys of $\mathrm{NH_{3}}$ and $\mathrm{H_{2}O}$ masers; thus, the RAMPS dataset is an important step toward a better understanding of high-mass star formation. 

We have presented integrated intensity maps of $\mathrm{NH_{3}(1,1)}$ and $\mathrm{NH_{3}(2,2)}$, $\mathrm{H_{2}O}$ positions, and associations for six fields within the Galactic plane. In addition, we have presented representative maps of $\mathrm{NH_{3}}$ velocity, $\mathrm{NH_{3}}$ rotational temperature, total $\mathrm{NH_{3}}$ column density, and $\mathrm{NH_{3}}$ line width for the L23 and L24 fields. The data cubes and maps for the entire RAMPS pilot survey are now available on the RAMPS website (see footnote \ref{foot:ramps}). With the successful results from the pilot survey, we have shown that RAMPS works as expected. Following the pilot survey, RAMPS has been awarded additional observing time on the GBT to extend the survey. We plan to release RAMPS data publicly after calibration and verification. We anticipate that the full RAMPS dataset will support numerous scientific investigations in the future. 

\section*{Acknowledgments} 

We would like to acknowledge Toney Minter for his excellent support in the formulation of the project and the execution of the observations, as well as Joe Masters for help with the GBT Mapping Pipeline. For this research we employed several useful software packages, which we also acknowledge below. In particular we thank Adam Ginsburg for his help with the PySpecKit $\mathrm{NH_{3}}$ line fitting. We would also like to thank the referee for the helpful comments and suggestions we received. RAMPS is funded by the National Science Foundation under grant AST-1616635.

\software{gbtgridder (https://github.com/nrao/gbtgridder), GBT Mapping Pipeline (Masters et al. 2011), PySpecKit (Ginsburg et al. http://doi.org/10.5281/zenodo.12490), APLpy (Robitaille and Bressert, 2012)}

\bibliography{RAMPS_Pilot_Paper_arxiv}{}
\end{document}